\documentclass[fleqn,usenatbib,useAMS]{mnras}

\pdfoutput=1

\usepackage{graphicx}	% Including figure files
\usepackage{multicol}        % Multi-column entries in tables
\usepackage{bm}		% Bold maths symbols, including upright Greek
\usepackage{pdflscape}	% Landscape pages
\usepackage[usenames,dvipsnames]{color}

\newcommand{\Sref}[1]{Section \ref{#1}}
\newcommand{\Tref}[1]{Table \ref{#1}}
\newcommand{\Aref}[1]{Appendix \ref{#1}}
\sfcode`\.=1001\sfcode`\?=1001\sfcode`\!=1001
\newcommand{\Fref}[1]{\ifhmode \ifnum\spacefactor=1001 Figure \ref{#1}\else Fig.\ \ref{#1}\fi \else Figure \ref{#1}\fi}
\newcommand{\Eref}[1]{\ifhmode \ifnum\spacefactor=1001 Equation (\ref{#1})\else equation (\ref{#1})\fi \else Equation (\ref{#1})\fi}

\newcommand{\auv}{$\alpha_\text{UV}$}

\newcommand{\kms}{\ensuremath{\textrm{km\,s}^{-1}}}

\newcommand{\lya}{\ensuremath{\textrm{Ly}\alpha}}
\newcommand{\lyb}{\ensuremath{\textrm{Ly}\beta}}
\newcommand{\lyc}{\ensuremath{\textrm{Ly}\gamma}}
\newcommand{\zem}{\ensuremath{z_\textrm{\scriptsize em}}}
\newcommand{\zab}{\ensuremath{z_\textrm{\scriptsize abs}}}
\newcommand{\mNHI}{N_\text{H\kern 0.2em\textsc{i}}}
\newcommand{\NHI}{\ensuremath{N_\textsc{h\scriptsize{\,i}}}}
\newcommand{\NDI}{\ensuremath{N_\textsc{d\scriptsize{\,i}}}}
\newcommand{\NCIV}{\ensuremath{N_\textsc{c\scriptsize{\,iv}}}}
\newcommand{\nH}{\ensuremath{n_\textsc{h}}}

\newcommand{\lNHI}{\ensuremath{\log_{10}(N_\textsc{h\scriptsize{\,i}}/\textrm{cm}^{-2})}}
\newcommand{\lNDI}{\ensuremath{\log_{10}(N_\textsc{d\scriptsize{\,i}}/\textrm{cm}^{-2})}}

\newcommand{\lnH}{\ensuremath{\log_{10}(n_\textsc{h}/\textrm{cm}^{-3})}}
\newcommand{\tran}[3]{\ensuremath{\textrm{#1\,{\scshape{#2}}}\,\lambda\,#3}}
\newcommand{\doublet}[4]{\ensuremath{\textrm{#1\,{\scshape{#2}}}\,\lambda\lambda#3#4}}
\newcommand{\MH}[1]{\ensuremath{<ft[\textrm{M}/\textrm{H}\right]}}

\newcommand{\popler}{\ensuremath{\textsc{uves\_popler}}}
\newcommand{\HI}   {{\rm H}{\sc \,i}}
\newcommand{\DI}   {{\rm D}{\sc \,i}}
\newcommand{\msun}{$M_\odot$}

\newcommand{\lmetal}{\ensuremath{\log_{10} (Z / Z_\odot)}}

\newcommand{\nHI}{$n_\mathrm{H\kern  0.1em \scriptsize \textsc{i}}$}
\newcommand{\mnHI}{n_\mathrm{H\kern 0.1em\scriptsize \textsc{i}}}
\newcommand{\Hy}{\rm H}
\newcommand{\He}{\rm He}
\newcommand{\D}{\rm D}
\newcommand{\hethree}{$^3$\rm He}
\newcommand{\hefour}{$^4$\rm He}

\newcommand{\liseven}{$^7$\rm Li}

\newcommand{\FeII} {{\rm Fe}{\sc \,ii}}
\newcommand{\CII}  {{\rm C}{\sc \,ii}}

\newcommand{\CIII} {{\rm C}{\sc \,iii}}
\newcommand{\CIV}  {{\rm C}{\sc \,iv}}
\newcommand{\AlII} {{\rm Al}{\sc \,ii}}
\newcommand{\AlIII}{{\rm Al}{\sc \,iii}}
\newcommand{\SiII} {{\rm Si}{\sc \,ii}}
\newcommand{\SiIII}{{\rm Si}{\sc \,iii}}
\newcommand{\SiIV} {{\rm Si}{\sc \,iv}}

\usepackage[T1]{fontenc}
\usepackage{ae,aecompl}

\usepackage{newtxtext,newtxmath}
\DeclareRobustCommand{\VAN}[3]{#2}
\let\VANthebibliography\thebibliography
\def\thebibliography{\DeclareRobustCommand{\VAN}[3]{##3}\VANthebibliography}

\title[Three new near-pristine absorption clouds]{Discovery of three new near-pristine absorption clouds at \boldmath{$z=2.6$}--4.4}

\author[Robert et al.]{P. Fr\'{e}d\'{e}ric Robert,$^{1}$
Michael T. Murphy,$^1$\thanks{E-mail: mmurphy@swin.edu.au (MTM)}
John M. O'Meara,$^{2,3}$
Neil H. M. Crighton,$^1$\newauthor
Michele Fumagalli$^{4,5,6}$
\\
$^{1}$Centre for Astrophysics and Supercomputing, Swinburne University of Technology, Hawthorn, Victoria 3122, Australia\\
$^{2}$Department of Chemistry \& Physics, Saint Michael's College, One Winooski Park, Colchester, VT 05439, USA\\
$^{3}$W.\ M.\ Keck Observatory 65-1120 Mamalahoa Highway, Kamuela, HI 96743, USA\\
$^{4}$Institute for Computational Cosmology and Centre for Extragalactic Astronomy, Durham University, South Road, Durham DH1 3LE, UK\\
$^{5}$Dipartimento di Fisica G.\ Occhialini, Universit\`a degli Studi di Milano Bicocca, Piazza della Scienza 3, 20126 Milano, Italy\\
$^{6}$INAF Osservatorio Astronomico di Trieste, via G.\ Tiepolo 11, Trieste, Italy
}

\date{Accepted 2022 May 31. Received 2022 April 24; in original form 2021 June 4}

\pubyear{2022}

\volume{{\rm in press}}

\begin{document}
\label{firstpage}
\pagerange{\pageref{firstpage}--\pageref{lastpage}}
\maketitle

% Single paragraph, not more than 250 words (200 for Letters), no references.
\begin{abstract}
We report the discovery of three new ``near-pristine'' Lyman Limit Systems (LLSs), with metallicities $\approx$1/1000 solar, at redshifts 2.6, 3.8 and 4.0, with a targeted survey at the Keck Observatory. High resolution echelle spectra of eight candidates yielded precise column densities of hydrogen and weak, but clearly detected, metal lines in seven LLSs; we previously reported the one remaining, apparently metal-free LLS, to have metallicity $<$1/10000 solar. Robust photoionisation modelling provides metallicities $\textrm{[Si/H]} = -3.05$ to $-2.94$, with 0.26\,dex uncertainties (95\% confidence) for three LLSs, and $\textrm{[Si/H]} \ga -2.5$ for the remaining four. Previous simulations suggest that near-pristine LLSs could be the remnants of PopIII supernovae, so comparing their detailed metal abundances with nucleosynthetic yields from supernovae models is an important goal. Unfortunately, at most two different metals were detected in each new system, despite their neutral hydrogen column densities ($10^{19.2\textrm{--}19.4}\,\textrm{cm}^{-2}$) being two orders of magnitude larger than the two previous, serendipitously discovered near-pristine LLSs. Nevertheless, the success of this first targeted survey for near-pristine systems demonstrates the prospect that a much larger, future survey could identify clear observational signatures of PopIII stars. With a well-understood selection function, such a survey would also yield the number density of near-pristine absorbers which, via comparison to future simulations, could reveal the origin(s) of these rare systems.
\end{abstract}

\begin{keywords}
line: profiles -- galaxies: haloes -- intergalactic medium -- quasars: absorption lines.
\end{keywords}

%%%%%%%%%%%%%%%%%%%%%%%%%%%%%%%%%%%%%%%%%%%%%%%%%%

%%%%%%%%%%%%%%%%% BODY OF PAPER %%%%%%%%%%%%%%%%%%

\section{Introduction}
\label{s:intro}

In the current standard model of cosmology, the first light elements were created through Big Bang Nucleosynthesis: hydrogen (\Hy), helium (\He), deuterium (\D), helium-$3$ (\hethree), helium-$4$ (\hefour), and a small amount of lithium-$7$ (\liseven). The first stars, referred to as PopIII stars, were made of this pristine material and, through stellar nucleosynthesis, created the elements heavier than helium: metals. These metals were then dispersed in the surrounding environment of the PopIII stars after they exploded as supernovae. The gas in the newly metal-polluted environment was used in the formation of the second generation of stars, PopII. This mechanism is commonly accepted, and within cosmological simulations it is possible to follow the initial collapse of pristine material leading up to the creation of PopIII stars, their death, and the following formation of PopII stars \citep[e.g.][]{2004ARA&A..42...79B, 2009Natur.459...49B}. Yet, since we lack observational constraints, the chemical and physical properties of PopIII stars are not currently known. One obvious method to constrain their properties would be direct observations, but it is thought that most of the PopIII stars died within the first $\sim 0.5$ billion years of the Universe. They are therefore too distant, too faint and too short-lived to be directly observed with current telescopes.

One way to infer the characteristics of the PopIII stars is to study element abundances in very old metal-poor stars found in the Galactic halo. Stars with extremely low iron abundances, $\textrm{[Fe/H]} \sim -7$ to $-3$, are believed to have formed from the remnants of PopIII stars \citep{2002Natur.419..904C,2005Natur.434..871F,2007ApJ...660L.117F,2014Natur.506..463K}. Their abundance patterns should then reflect the environment in which they formed and could be compared with theoretical supernova yields of PopIII stars. However, that comparison is complicated by the interpretation of the observed stars' elemental abundance patterns. Indeed, metal-poor stars can have very different abundance patterns to each other \citep{2013ApJ...762...26Y}. Moreover, these patterns could also be different from the time when the star formed, as its stellar interior may have been polluted by its environment and its nuclear burning. Therefore, linking the metal-poor stars of the Milky Way to the yields of PopIII stars is a difficult process. A complementary approach would be to study the remnants of PopIII stars: gas clouds that have been polluted by metals ejected during the death of PopIII stars as supernovae. Depending on the mass ranges and explosion energies of the PopIII progenitors, these remnants could remain in the intergalactic medium (IGM), or the circumgalactic environment (CGM) of a galaxy \citep[e.g.][]{2008ApJ...682...49W,2012ApJ...745...50W,2020MNRAS.497.2839L}.

Some Lyman limit systems (LLSs) may be excellent candidates for such PopIII remnants. The physical properties of LLSs -- usually defined as quasar absorbers with neutral hydrogen column densities $17.2 \le \lNHI < 20.3$ -- have been investigated in several surveys by differents groups at $z \leq 1$ \citep[e.g.][]{2013ApJ...770..138L,2016ApJ...833..283L,2016ApJ...831...95W} and $z \geq 2$ \citep[e.g.][]{1990ApJS...74...37S,2013ApJ...775...78F,2015ApJS..221....2P,2015ApJ...812...58C,2016MNRAS.455.4100F,2016ApJ...833..283L}. These studies have shown that LLSs are relatively metal-poor, with the metallicity distribution spanning $-2.0 \leq \lmetal \leq 0.4$ at low redshift ($\zab \leq 2$), and $-4.0 \leq \lmetal \leq 1$ at high redshift ($\zab \geq 2$). The high redshift regime is the most relevant for the LLSs studied in this paper. Along with these observational surveys, simulations \citep[e.g.][]{2005MNRAS.363....2K,2006MNRAS.368....2D,2009Natur.457..451D,2011MNRAS.418.1796F,2011MNRAS.412L.118F,2012MNRAS.421.2809V} have shown that the prime candidates for the metal-poor streams fuelling the formation of galaxies are LLSs, therefore making them potential candidates for circumgalactic PopIII remnants. Finally, three clouds with no apparent associated metal lines and metallicity upper limits of $\lmetal \leq -4$\ have been discovered: two at $z \sim 3$, serendipitously, by \citet{2011Sci...334.1245F} (LLS1134 and LLS0958B), and the third, recently, at $z \sim 4.4$ (LLS1723) from a dedicated search by \citet{2019MNRAS.483.2736R}. The very low metallicity of these systems, combined with the lack of detectable metal lines, imply that they could be intergalactic PopIII remnants. Another possibility, emphasised in \citet{2016MNRAS.455.4100F}, is that they are entirely pristine, having experienced no pollution from nearby galaxies for $\sim$2 billion years.

Even LLSs with the lowest column densities [i.e.\ $\lNHI \approx 17.3$], and in the lowest part of the metallicity distribution at high redshift ($\lmetal \leq -3$), still display detectable metal lines. Indeed, \citet{2016MNRAS.457L..44C} discovered the lowest metallicity LLS with metal-line detections, LLS1249, with $\lmetal = -3.41 \pm 0.26$. This very low metallicity falls in the range expected for PopIII remnants \citep{2012ApJ...745...50W}. With detections of C and Si, this LLS presents an abundance ratio of $\textrm{[C/Si]} = -0.26 \pm 0.17$, consistent with both predictions of nucleosynthesis models of PopIII stars \citep[e.g.][]{2014ApJ...791..116C,2010ApJ...724..341H}, and PopII stars \citep{2008ApJ...674..644C}. Its size, density, and temperature suggest that LLS1249 is more likely to be found in the IGM rather than in the CGM, although these parameters have large uncertainties and so the evidence for this is correspondingly weak. Following this first discovery of a ``near-pristine'' LLS, \citet{2016ApJ...833..283L} also reported a similar system (LLS0958A) and complimentary integral field spectroscopy by \citet{2016MNRAS.462.1978F} mapped its galactic environment, along with another absorber in the same sightline, LLS0958B, which is one of the two apparently metal-free LLSs discovered by \citet{2011Sci...334.1245F}. Interestingly, the environment of LLS0958A showed no nearby galaxies, potentially indicating an intergalactic environment. However, the environment of LLS0958B showed 5 \lya-emitting galaxies at a similar redshift, with 3 appearing to be aligned in projection, suggesting a filamentary structure akin to a cold stream in the vicinity of galactic haloes. In the context of a simple expectation that the lowest metallicity absorbers arise furthest from galaxies, it is perhaps surprising that the opposite was found in this study, though it remains possible that LLS0958A arises in gas closely associated with galaxies below the detection limits of \citet{2016MNRAS.462.1978F}.

Hence, given this somewhat counter-intuitive situation from a very small sample, more ``near-pristine'' LLSs need to be discovered to refine our understanding of their origins. While they appear, so far, to be rare, previous searches have not targeted them directly; instead, a dedicated search is required and now possible with existing LLS surveys. We reported the first such search in \citet{2019MNRAS.483.2736R}, in which we established a list of very-metal poor candidates and focussed on the discovery and properties of the apparently metal-free absorber, LLS1723. Here we study the rest of the sample, i.e.\ 7 systems. They all show detectable metal lines and could therefore be near-pristine LLSs and candidates for PopIII remnants. Importantly, the sample was deliberately biased towards LLSs with higher hydrogen column densities (\NHI) than the first near-pristine absorber discovered by \citet{2016MNRAS.457L..44C} (LLS1249) so that other metallic species, in addition to Si and C, could be detectable even at very low metallicities $\lmetal \leq -3$. This should improve prospects for distinguishing between PopIII and PopII nucleosynthetic scenarios.

This paper is structured as follows. In \Sref{s:observations}, we summarise briefly how the sample of very metal-poor candidates was selected and describe our Keck observing campaign. In \Sref{s:general}, we summarise our general approach to analysing all the candidates and, in \Sref{s:interesting}, we describe the absorption line features of each LLS and their physical properties obtained through photoionisation modelling. This allows us to accurately reassess their metallicity, and check the efficiency of our dedicated search. Indeed, among the 7 very metal-poor candidates, we ultimately find that 3 are near-pristine, i.e.\ have metallicities $\lmetal \la -3$. Their origins are discussed in \Sref{s:discussion}, and we also present suggestions for a strategy to improve future dedicated searches for very metal-poor LLSs. Note that, unless otherwise stated, all column density measurements in this paper are quoted with 1$\sigma$ uncertainties, all column density upper limits are 2$\sigma$, and all metallicity and abundance ratio estimates (including limits) from our photoionisation analyses are quoted at 95\% confidence.

\section{Target selection, observations, and data reduction}
\label{s:observations}

In \citet{2019MNRAS.483.2736R}, we described a dedicated search for very metal-poor LLSs. Briefly, we targeted 8 LLSs, for which no previous high-resolution ($R>30000$) spectra existed. We observed these 8 targets with Keck/HIRES during three runs in 2016--2017, obtaining signal-to-noise ratios (SNRs) $\ga$ 20 per $\approx$2.3\,\kms\ pixel in the continuum near the expected wavelength of the strongest metal absorption lines (generally \tran{Si}{ii}{1260}). The targets were selected to satisfy the following criteria: (i) their lower resolution quasar spectra, from \citet{2015ApJ...812...58C} and \citet{2015ApJS..221....2P}, showed either no metal lines at the LLS redshifts or only very weak (usually not very clear) detections; (ii) their metallicity estimates from these lower resolution spectra were consistent with $\lmetal \leq -3$; and (iii) their \HI\ column density estimates were higher than for LLS1249 discovered by \citet{2016MNRAS.457L..44C}, i.e.\ $\lNHI \geq 17.36$. The third criterion was introduced to attempt to identify near-pristine LLSs in which a larger variety of metallic species could be detected. \Tref{t:log_obs_lls} lists the metallicity estimates obtained by \citet{2015ApJ...812...58C} from their Magellan/MagE spectra, and by \citet{2016MNRAS.455.4100F} for their Keck/ESI and Magellan/MIKE spectra of the others, along with the literature \NHI\ estimates. One of the systems listed, LLS1723, appeared to be free of metal lines in the HIRES spectra and indicated a significantly lower metallicity than the others; this LLS was studied in \citet{2019MNRAS.483.2736R}.

\begin{table*}
\caption{Literature information used to select the 8 very metal-poor candidate LLSs targeted in our survey. LLS1153 and LLS1304 are from the sample of \citet{2015ApJ...812...58C} which provided ionic column densities and metallicity estimates for 17 LLSs at $z \geq 2$ observed with the Magellan/MagE spectrograph. The other 6 LLSs are from the survey of \citet{2015ApJS..221....2P} which provided ionic column densities for 157 LLSs at $z \geq 2$ observed with Keck/ESI or Magellan/MIKE. Their metallicities were estimated in a subsequent study described in \citet{2016MNRAS.455.4100F}.}
\label{t:log_obs_lls}
%\addtolength{\tabcolsep}{-3.6pt}
\footnotesize
%\begin{tabular*}{=\textwidth}{@{\extracolsep{\fill} }ccccccc}
\begin{tabular}{ccccccc}
\hline
Quasar name                                & LLS name         & Spectrograph  & QSO redshift     &    LLS redshift       &  \lNHI                            &   \lmetal  \\\hline
SDSS J010619.20$+$004823.3                 &  LLS0106         & MIKEb         &    4.430         &    4.17157            &   $19.05 \pm 0.20$                &  $\leq -3.19$     \\
SDSS J034402.80$-$065300.0                 &  LLS0344         & MIKEb         &    3.957         &    3.84280            &   $19.55 \pm 0.15$                &  $\leq -3.01$     \\
SDSS J095256.41$+$332939.0                 &  LLS0952         & ESI           &    3.396         &    3.26180            &   $19.90 \pm 0.20$                &  $\leq -3.00$     \\
SDSS J115321.68$+$101112.9                 &  LLS1153         & MagE          &    4.127         &    4.03800            &   $17.00 \leq$ \NHI $\leq 19.00$  &  $\leq -2.90$     \\
SDSS J115659.59$+$551308.1                 &  LLS1156         & ESI           &    3.110         &    2.61594            &   $19.10 \pm 0.30$                &  $\leq -3.10$     \\
SDSS J130452.57$+$023924.8                 &  LLS1304         & MagE          &    3.651         &    3.33690            &   $17.90 \leq$ \NHI $\leq 18.70$  &  $-2.81 \pm 0.15$ \\
SDSS J172323.20$+$224358.0                 &  LLS1723         & ESI           &    4.520         &    4.39100            &   $18.30 \pm 0.30$                &  $\leq -3.25$     \\
SDSS J224147.70$+$135203.0                 &  LLS2241         & MIKEb         &    4.470         &    3.65393            &   $20.20 \pm 0.20$                &  $\leq -3.45$     \\
 \hline
\end{tabular}
\end{table*}

The journal of our new Keck/HIRES observations of the 7 LLSs studied here is provided in \Tref{t:journal_obs}. HIRES was configured with the red cross-disperser, with a slit width of 1.148$^{\prime\prime}$ (C5 decker, with length 7$^{\prime\prime}$) to provide a resolving power of $R=37500$ for all quasars. The data reduction is described in detail in \citet{2019MNRAS.483.2736R}. Briefly, we performed the initial reduction steps with the \textsc{makee} pipeline: flat-fielding, order tracing, background subtraction and extraction of a 1D spectrum for each exposure. We wavelength-calibrated the data in the standard way by using spectra of a ThAr lamp but, unlike the case of J1723$+$2243, \textsc{makee} automatically identified a sufficient number of ThAr lines during the calibration process. We then used \popler\ \citep{Murphy:2016:UVESpopler} to merge together the extracted exposures, with the same process as \citet{2019MNRAS.482.3458M}; again see \citet{2019MNRAS.483.2736R} for more details of this procedure, and in particular how the continuum was manually estimated in the Lyman forest region of each spectrum.

\begin{table*}
\caption{Journal of HIRES observations. ``EA'' and ``XDA'' are the echelle and cross-disperser angles; $\lambda$ range is the wavelength coverage. The slit-width was 1.148$^{\prime\prime}$ for all exposures, providing a nominal resolving power $R=37500$. Notes:
\textsuperscript{a}Very poor weather most of the night; snow during preceding day.
\textsuperscript{b}Dome closed after 2733\,s of second exposure.
\textsuperscript{c}Exposure paused at 2425\,s because target transited and could not be tracked.
\textsuperscript{d}Stopped at 3315\,s because an incorrect echelle angle was used which did not cover the important \tran{Si}{ii}{1260} line.
\textsuperscript{e}Echelle angle was changed to focus on \tran{Si}{ii}{1260}.}
\label{t:journal_obs}
%\addtolength{\tabcolsep}{-3.6pt}
\footnotesize
\begin{tabular}{lllllll}
\hline
Quasar                     & EA   & XDA & $\lambda$ range & Date        & Exposure    & Seeing       \\
                           & {[\textdegree]} & {[\textdegree]}       & {[\AA]}             & {[UT]}      & {[s]}       & {[$\arcsec$]}\\
\hline
SDSS J010619.20$+$004823.3 & $-0.330$     &    $0.830$             & 4740--9360        & 2016-09-22    &   3600, 2733       & $\sim$0.8\textsuperscript{a,b} \\
%                           & $-0.330$     &    $0.830$             & 4740--9360        & 2016-09-22    &   2733       & $\sim$0.8\textsuperscript{a} \\
                           & $-0.300$     &    $0.700$             & 4680--9120        & 2017-08-19    &   3210       & $\sim$0.65--0.8 \smallskip\\
SDSS J034402.80$-$065300.0 & 0.200        &    0.400               &  4250--8700       & 2017-08-19    &   2$\times$2710       & $\sim$0.9 \\
%                           & 0.200        &    0.400               &  4250--8700       & 2017-08-19    &   2710       & $\sim$0.9 \\
                           & 0.400        &    0.200               &  4010--8500       & 2017-08-20    &   3000       & $\sim$0.65\textsuperscript{a} \smallskip\\ 
SDSS J095256.41$+$332939.0 & 0.000        &    1.878               &  3750--6600       & 2015-11-14    &   3000       & $\sim$0.8 \\
                           & 0.000        &    1.752               &  3600--6480       & 2015-11-14    &   2700       & $\sim$0.8 \smallskip\\
SDSS J115321.68$+$101112.9 & $-0.100$     &    0.400               &  4340--8690       & 2017-02-18    &   2$\times$3600       & $\sim$0.7--0.9 (mostly $\sim$0.8)\\
%                           & $-0.100$     &    0.400               &  4340--8690       & 2017-02-18    &   3600       & $\sim$0.7--0.9 (mostly $\sim$0.8)\\
                           & $-0.100$     &    0.400               &  4340--8690       & 2017-02-18    &   3600       & $\sim$0.7--0.8 (mostly $\sim$0.75) \smallskip\\
SDSS J115659.59$+$551308.1 & 0.000        &    0.010               &  3640--8100       & 2017-02-17    &   3600       & $\sim$0.9--1.2 (mostly $\sim$1.0) \\
%                           & 0.000        &    0.010               &  3640--8100       & 2017-02-17    &   3600       & $\sim$1.0--1.3 (mostly $\sim$1.1) \\
                           & 0.000        &    0.010               &  3640--8100       & 2017-02-17    &   2$\times$3600       & $\sim$1.0--1.3 (mostly $\sim$1.1) \smallskip\\
SDSS J130452.57$+$023924.8 & 0.190        &    0.000               &  4000--8490       & 2017-02-17    &   2$\times$3300       & $\sim$1.0--1.3 (mostly $\sim$1.1) \\
%                           & 0.190        &    0.000               &  4000--8490       & 2017-02-17    &   3300       & $\sim$1.0--1.3 (mostly $\sim$1.1) \\
                           & 0.190        &    0.000               &  4000--8490       & 2017-02-18    &   3300, 3300       & $\sim$0.75, $\sim$0.75--1.0 (mostly $\sim$0.9) \\
%                           & 0.190        &    0.000               &  4000--8490       & 2017-02-18    &   3300       & $\sim$0.75--1.0 (mostly $\sim$0.9) \\
                           & 0.190        &    0.000               &  4000--8490       & 2017-06-13    &   3600       & $\sim$0.7--1.1 (mostly $\sim$0.7) \\
                           & 0.190        &    0.000               &  4000--8490       & 2017-06-13    &   3600       & $\sim$0.8--1.0 (mostly $\sim$0.8) \smallskip\\
PSS J172323.10$+$224357.14 & 0.080        &    0.900               &  4786--9230       & 2017-06-13    &   3750, 3315\textsuperscript{c}\textsuperscript{d}       & $\sim$0.7--0.8 \\
%                           & 0.080        &    0.900               &  4786--9230       & 2017-06-13    &   3315\textsuperscript{c}\textsuperscript{d} & $\sim$0.7--0.8 \\
                           & $-0.150$\textsuperscript{e} &  0.900  &  4786--9217       & 2017-06-13    &   2$\times$3300       & $\sim$0.75--0.8 \\
%                           & $-0.150$     &    0.900               &  4786--9217       & 2017-06-13    &   3300       & $\sim$0.7--0.8 \\
                           & $-0.150$     &    0.900               &  4786--9217       & 2017-06-13    &   3400, 3600       & $\sim$0.8--1.0, $\sim$0.9--1.5 (mostly $\leq$1.2) \smallskip\\
%                           & $-0.150$     &    0.900               &  4786--9217       & 2017-06-13    &   3600       & $\sim$0.9--1.5 (mostly $\leq$1.2) \smallskip\\
SDSS J224147.70$+$135203.0 & $-0.250$     &    0.730               &  4680--9130       & 2017-08-19    &   2$\times$3210       & $\sim$0.6--0.7 \\
%                           & $-0.250$     &    0.730               &  4680--9130       & 2017-08-19    &   3210       & $\sim$0.65 \\
\hline
\end{tabular}
\end{table*}

For the 7 LLSs studied here, our reassessment of their absorption features and metallicities with the newly acquired HIRES spectra showed that 4 have metallicity consistent with $\lmetal \la -3$. As stated above, the analysis of LLS1723 is the focus of \citet{2019MNRAS.483.2736R} due to its apparent lack of metal lines. The three others -- LLS0344, LLS1153, and LLS1156 -- do show weak metal absorption lines and have $\lmetal \la -3$: they are new examples of ``near-pristine'' LLSs. Therefore, they are the focus of \Sref{s:interesting} where the details of the absorption line properties and photoionisation analysis are provided. The absorption feature properties and metallicity estimates of the remaining 4 LLSs are reported briefly in \Aref{s:higher}. Nonetheless, these higher-metallicity systems are important for considering an improved selection strategy for low metallicity LLS candidates in future; we discuss this in \Sref{s:strategy}.

\section{General analysis approach for all absorbers}
\label{s:general}

Here we introduce the general analysis approach used for all the very metal-poor candidates. We start with the study of their absorption features to derive the metal and \HI\ column densities, followed by the photoionisation analysis which returns the physical properties of the absorbers -- most importantly, the metallicity.

\subsection{Spectral analysis}
\label{s:spectral_general}

The starting point for all absorbers is to identify the metal lines detected at, or very near, the LLS redshift determined by previous authors in \Tref{t:log_obs_lls}. They are generally the strongest metal transitions detectable for LLSs at $z \geq 2$  and correspond to the different ionisation stages of carbon and silicon: \CII, \CIII, \CIV, \SiII, \SiIII\ and \SiIV. At the resolution of Keck/HIRES, the detection of these transitions is unambiguous, especially if using the strongest available transition (highest oscillator strength) for an ion; e.g. \tran{Si}{ii}{1260}. Therefore, the determination of the absorption redshift, a key component of the LLS's model, is robust.

The LLSs are modeled using the \textsc{vpfit} software (version 9.5) \citep{2014ascl.soft08015C} which fits ideal Voigt profiles to absorption lines and minimises $\chi^2$ between the data and model to return their best-fit column densities $N$, Doppler parameters $b$ (line widths), absorption redshifts \zab\ and the associated uncertainties. All the LLSs analysed in this paper show detectable metal absorption lines and the fitting process for them is rather straightforward. For instance, the LLS redshift, \zab, is determined from the metal lines and then fixed for the fit to the \HI\ lines. The main purpose of the Voigt profile modelling is to measure the total column density of each ion, integrated across all fitted velocity components. The same velocity components are fitted to all singly-ionised species simultaneously. We then assess whether the same components can be fitted to the higher-ionisation species, using changes in $\chi^2$ per degree of freedom as a rough guide. This aids our assessment of whether the low and high ion absorption likely arises in the same gas phase or not, which informs our approach to the photoionisation model of each LLS. However, it should be emphasised that fitting the high ions with the same or different velocity structure as the low ions does not appreciably change the total column density returned for any species.

The total \HI\ line column density \NHI\ and representative $b$\ parameter is then determined for the LLS. But unlike metal absorption lines, it is challenging to fit ideal Voigt profiles to the different lines of the Lyman series: e.g.\ at $z \geq 2$, the \lya\ forest is very thick and several unrelated systems can contribute to the absorption features seen; the \lya\ line associated with a LLS can be on the flat part of the curve-of-growth and not have strong damping wings; and the Lyman series lines can be saturated. Due to these possible issues, we use a simple one-component Voigt profile fit for \HI\ and follow one of the options below:
\begin{itemize}
\item When the presence of any weak damping wings in the profile of its \lya\ absorption line cannot be confidently identified due to Lyman forest absorption, then $\lNHI \la 19$. The low order lines such as \lya\ or \lyb\ can only provide an upper limit on \NHI. However, establishing a \emph{lower} limit is much more important as this provides the highest possible (i.e.\ most conservative) metallicity upper limit. The common approach is to use the flux level bluewards of the Lyman limit to establish a lower-limit on \NHI\ which provides a metallicity upper limit \citep[e.g.][]{2010ApJ...718..392P,2013ApJ...765..137O,2015ApJS..221....2P}. This has the advantage that the measured \NHI\ is independent of $b$. We used this method for LLS0106 and LLS1304, and our analysis of LLS1723 in \citet{2019MNRAS.483.2736R} provides a detailed example of this approach.
\item When a LLS's \lya\ profile has clear damping wings, then $\lNHI \geq 19$. \NHI\ is easily measured, and strongly constrained, by the observed transmission near the \lya\ line and its damping wings. We used this method for LLS0952, LLS1156 and LLS2241. Again, the estimated $b$ value does not have a significant impact on the \NHI\ measurement.
\end{itemize}

Finally, we note that two of our target LLSs have apparent \DI\ \lya\ detections at the expected $-82$\,\kms\ offset from their associated \HI\ lines. In both cases, the \HI\ \lya\ shows only weak damping wings [i.e.\ $\lNHI \leq 19$], so the \DI\ lines provide an opportunity to refine the \NHI\ measurement: combined with a precise determination of the [D/H] ratio in low metallicity LLSs and damped \lya\ systems (DLAs), taken from \citet{2014ApJ...781...31C}, the estimate of \NDI\ provides a more precise, and likely more accurate, estimate of \NHI\ (with $\sim$0.1\,dex uncertainty) which is consistent with the value determined form the \HI\ \lya\ line itself (with $\sim$0.3\,dex uncertainty). However, the \NDI\ and $b$-parameter estimates in our two LLSs are not accurate enough for a competitive constraint on the primordial [D/H] value: other transitions of the \DI\ Lyman series are not detected in our LLSs, and the \DI\ \lya\ lines are at-least partially blended with other, unrelated \HI\ lines, so they will not provide an accuracy in [D/H] near $\sim$0.02\,dex which is typical for the best measurements \citep[e.g.][]{2018ApJ...855..102C}.

\subsection{Estimating metallicities with photoionisation analysis}
\label{s:cloudy_general}

In this paper, we measure metallicity using silicon as the probe, and so we equate $\lmetal$ with [Si/H], i.e.\ the silicon abundance in the absorber measured relative to the solar values. Nonetheless, one can not assume that the observed ions are fully representative of the hydrogen and metal content of the LLSs. To convert the measured ionic column densities into atomic ones, ionisation corrections are derived with the photoionisation simulation software \textsc{Cloudy} (version 13.03)\footnote{See \urlstyle{rm}\url{https://www.nublado.org/}.} \citep{2013RMxAA..49..137F} which performs radiative transfer calculations for an ideal gas cloud representative of a LLS.

We follow closely the method described in \citet{2015MNRAS.446...18C} so provide only a brief summary here. The \textsc{Cloudy} models are generated assuming that a LLS is a single phase slab with a constant density and illuminated on one side by an ionising spectrum, the HM12 UV background of \citet{2012ApJ...746..125H} set at the LLS's redshift\footnote{The earlier ``HM05'' spectrum \citep{1996ApJ...461...20H} included background radiation from quasars and galaxies, while the HM12 spectrum also includes a contribution from ionising photons escaping galaxies close to the gas slab under consideration. There is considerable discussion in the literature about which background spectrum is most appropriate for modelling LLSs \citep[as summarised in, e.g.,][]{2015MNRAS.446...18C,2016MNRAS.455.4100F,2019ApJ...887....5L}. \citet{2019ApJ...872...81W} found that metallicities derived through \textsc{Cloudy} modelling with the HM12 background were $\approx$0.2--0.4\,dex higher than those derived using the HM05. However, in their work, the difference reduced strongly as a function of \NHI, and for $\lNHI > 19$ there is no evidence for a systematic difference. Similar results were also inferred by \citet{2019ApJ...883...78P}. We use the HM12 background in this work for consistency with \citet{2016MNRAS.457L..44C} and to ensure our metallicity estimates remain conservative for the 3 near-pristine systems we identify with $\lNHI > 19$, i.e.\ we do not spuriously identify higher-metallicity systems as having $\lmetal \la -3$.}. We generate a large grid of models as a function of \NHI, the hydrogen volume density \nH, [Si/H] and a free parameter \auv\ which tunes the hardness of the ionising spectrum, effectively representing the uncertainty in its shape \citep[see][]{2015MNRAS.446...18C}. The column densities of the different ions are then calculated by interpolation on the grid.

To more efficiently use our computing resources, for each LLS we first generated a coarse \textsc{Cloudy} grid covering $-4.2 < \lnH < -1$, $15.5<\lNHI<20.5$, $-6 < $[Si/H]$ < -2$ (all with steps of $0.5$\,dex) and set \auv\ to 0, i.e.\ the un-tilted HM12 UV background. This allows us to quickly reassess the LLS's metallicity, filtering out those inconsistent with $\lmetal \leq -3$. For those with metallicities near, or clearly lower than this threshold, grids with finer steps (requiring more computing time) were then generated. The details of each grid are provided in \Sref{s:interesting} for each very metal-poor LLS. For all those grids, \auv\ was varied from $-2.50$ to 1.50 in steps of $0.4$ dex.

Still following the approach of \citet{2015MNRAS.446...18C}, a Markov Chain Monte Carlo (MCMC) technique is used to find the metallicity probability distribution of each LLS. Briefly, the predictions of \textsc{Cloudy} for the ionic column densities are compared to the measurements, and we construct a likelihood function $\ln\mathcal{L}$ with the \textsc{Cloudy} grid parameters: \NHI, \nH, [Si/H] and \auv. We maximise $\ln\mathcal{L}$ to determine the posterior distributions of its parameters using a MCMC sampler provided by the \textsc{emcee} code \citep{2013PASP..125..306F}. During the sampling of the coarse and fine grids, we impose a flat prior on the density \nH\ of $-3.5 < \lnH < -1.0$. Briefly, this range of densities represents the likely extremes that characterise LLSs arising in intergalactic and circumgalactic environments; see \citet{2016MNRAS.455.4100F} (figure 3) and \citet{2019MNRAS.483.2736R} for a detailed discussion. While solar abundances are assumed for the metals when constructing the \textsc{Cloudy} grids, non-solar patterns can be allowed \textsl{a posteriori} by allowing deviations over a range set via a flat prior. This is generally done in the sampling of the fine grids when a metal ion is detected, but it is also possible to infer an upper limit on a ratio from the non-detection of a metal species. As the metallicity is determined using the $\alpha$\ element Si, a deviation from solar abundance pattern for an element X is expressed as [X/Si] and this ratio is allowed to vary, such as $-2 \leq \textrm{[X/Si]} \leq 2$.

From the analysis with the coarse grids, 3 LLSs appeared to be near-pristine candidates, i.e.\ with detected metal lines and metallicities consistent with $\textrm{[Si/H]} \leq -3$: LLS0344, LL1153, LL1156. The analysis of these LLSs is described below in \Sref{s:interesting},
%For each LLS we (i) list the metal absorption lines identified and show their absorption profiles; (ii) present the H{\sc \,i} profile analysis; and (iii) present the photoionisation modeling results for the coarse and fine grids of models.
%The analysis of the remaining 4 LLSs, with metallicities $\textrm{[Si/H]} \ga -2.5$, is summarised in \Aref{s:higher}.
while the properties of the remaining 4 LLSs are provided in \Aref{s:higher}.
\Tref{t:logN} summarises the metal and \HI\ column density measurements and upper limits for each near-pristine LLS, together with the final metallicity estimate from the photoionisation analysis.
%Note that the column density values listed represent the sum of all components used in a Voigt profile fit.
%We provide full details of the fitted model parameters in the Supporting Information online.

\begin{table}
\caption{Column densities and metallicity measurements for the 3 near-pristine LLSs. The neutral hydrogen and metallic ion column density measurements are provided for all detected lines with 1$\sigma$ uncertainties, or 2$\sigma$ upper limits for non-detections. The metal column density values correspond to the sum of all velocity components across the absorption profile. The upper limits were derived using the apparent optical depth method \citep{1991ApJ...379..245S}. The last row provides the fiducial metallicities, [Si/H], inferred from the photoionisation analyses, with 95\% confidence intervals.}
\label{t:logN}
\begin{center}
\begin{tabular}{l|ccc}\hline 
    Ion  & \multicolumn{3}{c}{$\log_{10} (N/\mathrm{cm^{-2}})$} \\
 \hline
  & LLS0344 & LLS1153 & LLS1156 \\
 \ion{Si}{ii}                       & $ 12.48 \pm 0.02 $ & $ 12.22 \pm 0.03 $ & $12.35 \pm 0.15 $  \\
 \ion{Si}{iv}                 & $ 12.88 \pm 0.04 $ & $ 12.46 \pm 0.04 $ & -  \\
 \ion{C}{ii}                      & $ 13.37 \pm 0.03 $ & -                  & -  \\
 \ion{C}{iv}                  & $ 13.70 \pm 0.02 $ & $ 13.37 \pm 0.04 $ & $14.83 \pm 0.09 $  \\
 \ion{Fe}{ii}                      & $ \leq 12.85     $ & -                  & -  \\
 \ion{Al}{ii}                      & $ \leq 11.45     $ & $ \leq 11.06     $ & $11.45 \pm 0.09 $  \\
 \ion{Al}{iii}               & -                  & -                  & $ \leq 11.55 $  \\
 \DI                                       & $14.63 \pm 0.10  $ & $14.80 \pm 0.10  $ & - \\  
 \HI                                       & $19.23 \pm 0.10  $ & $19.40 \pm 0.10  $ & $19.30 \pm 0.10 $ \\
 \hline
  [Si/H]                                   & $-3.00 \pm 0.26  $ & $ -3.05 \pm 0.26 $ & $-2.94 \pm 0.26 $ \\
\hline
\end{tabular}
\end{center}
\end{table}

\section{Absorption line and photoionisation analysis of the new near-pristine absorbers}
\label{s:interesting}

\subsection{LLS0344}\label{s:lls0344}

\subsubsection{Metal line column densities}

LLS0344 was first identified in \citet{2015ApJS..221....2P} towards the $\zem=3.957$ quasar SDSS J034402.80$-$065300.0 (hereafter J0344$-$0653), based on absorption features at redshift $\zab=3.843$, using a MIKE spectrum. At this redshift \NHI\ was estimated at $\lNHI=19.55 \pm 0.15$ along with $b=30$\,\kms. These values were determined by visual inspection of theoretical Voigt profiles superimposed on the \lya\ line. At $\zab=3.843$ they identified \tran{C}{ii}{1334}, \doublet{C}{iv}{1548/1550}, \tran{Si}{ii}{1260}, \doublet{Si}{iv}{1393/1402}, \tran{O}{i}{1302}, and obtained column density upper limits for \tran{Fe}{ii}{1608}\ and \tran{Al}{ii}{1670}. The column densities were measured using the apparent optical depth method \citep{1991ApJ...379..245S}.

\begin{figure}
\centering
\includegraphics[width=0.95\columnwidth]{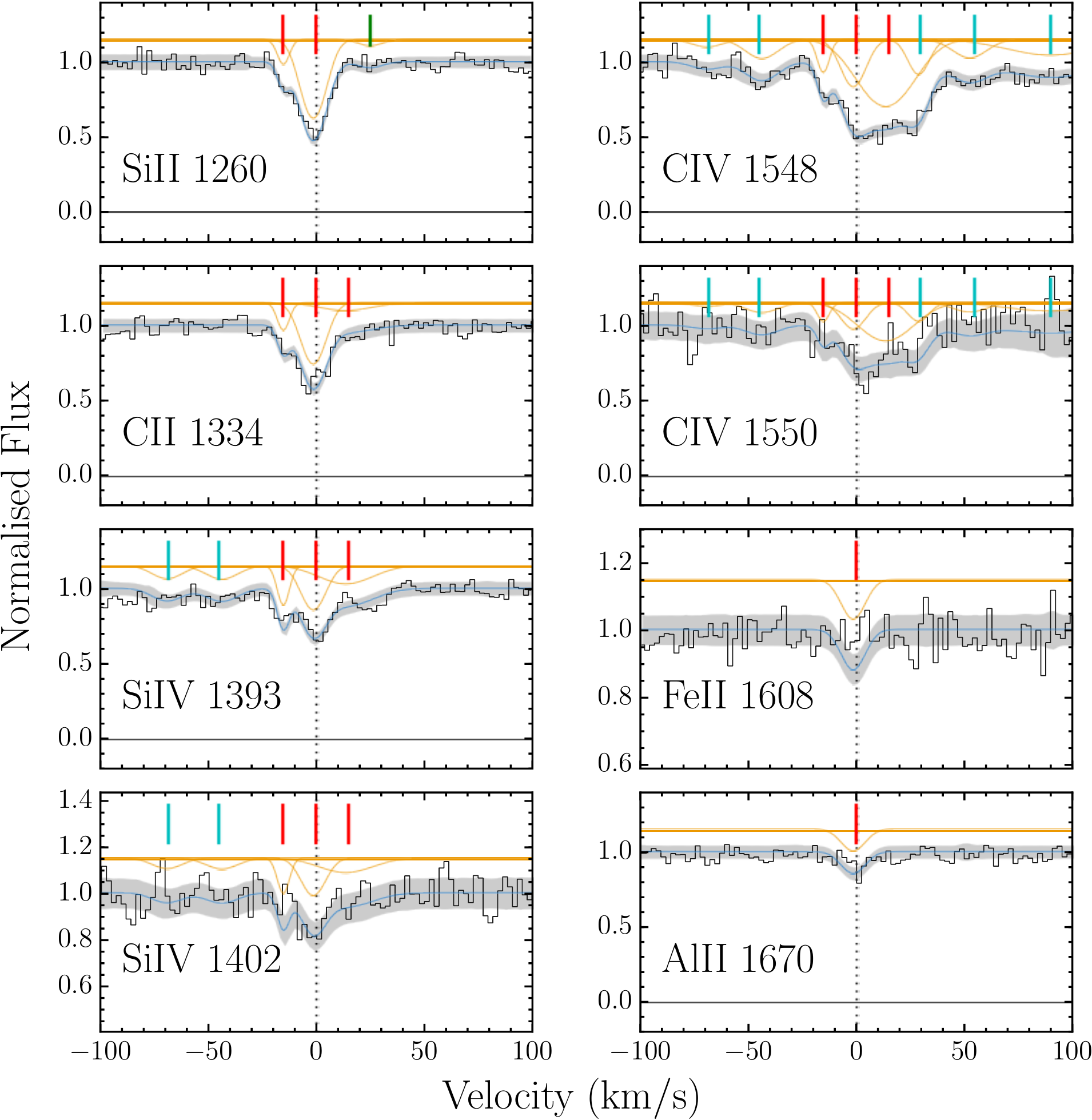}
\caption{Regions of the J0344$-$0653 HIRES spectrum (black histogram) where the strongest metal absorption lines of LLS0344 are expected, centered at $\zab=3.8428323$. The solid blue line in each panel represents the combined Voigt profile fit, with a total column density given in \Tref{t:logN}. The solid orange lines represent the Voigt profiles of the different velocity components contributing to the total absorption profile. These are indicated by vertical red and blue ticks. The red ticks are the main contributions to the total column density in each ion, while those represented by the cyan ticks contribute only weakly; in this absorber, the latter are only detected in the high ions \CIV\ and \SiIV. Note that the orange Voigt profile fits have been vertically shifted for clarity. The grey shading shows the 1$\sigma$ uncertainty in the flux. For \tran{Fe}{ii}{1608}\ and \tran{Al}{ii}{1670}, the column densities used for the Voigt profile fit correspond to a 2$\sigma$ upper limit derived using the apparent optical depth method. For \tran{Si}{ii}{1260}, in the top--left panel, the Voigt profile at $+25$\,\kms\ corresponds to the \tran{Fe}{ii}{1260} transition associated with this absorber.}
\label{f:lls0344_metals}
\end{figure}

Guided by the initial results above, we search the HIRES spectrum for metal absorption lines and we list the detections and upper limits in \Tref{t:logN}. \Fref{f:lls0344_metals} depicts the strongest (highest oscillator strength) transitions of the most abundant metal species expected at $\zab = 3.843$. For a detection, the Voigt profile fit depicted by a solid blue line corresponds to the combination of the profiles of each velocity component (solid orange lines).
%, and listed in the Supporting Information online.
For an upper limit, the profile illustrates the suitability of the 2$\sigma$ upper limit derived through the use of the apparent optical depth method.

The absorption profiles in \Fref{f:lls0344_metals} show that the majority of the total optical depth in each ion is concentrated in three velocity components between $-20$ and $+20$\,\kms\ (red tick marks) for both the low ions (\CII\ and \SiII) and high ions (\CIV\ and \SiIV). All three components can be fitted simultaneously, with the same redshifts and $b$ parameters, in all detected transitions, although the weakest, broadest component is not statistically required in the fit to \tran{Si}{ii}{1260}. There are additional velocity components detected over a larger velocity extent in \CIV\ and \SiIV\ (cyan ticks). However, these are minor contributors to the total column densities in those ions ($\approx$11\% for \SiIV, $\approx$36\% for \CIV). Given these observations, we assume for simplicity in our photoionisation analysis that all the metal ion column density arises in the same phase as the \HI\ detected in this system. Note that if, in reality, the high ion gas does not contribute to the \HI\ column density, our simple assumption will result in a slight over-estimate of the metallicity. That is, our assumption is conservative in that it will not result in a spurious near-pristine identification.

\subsubsection{\textbf{\HI}\ column density}

\begin{figure}
\centering
\includegraphics[width=0.95\columnwidth]{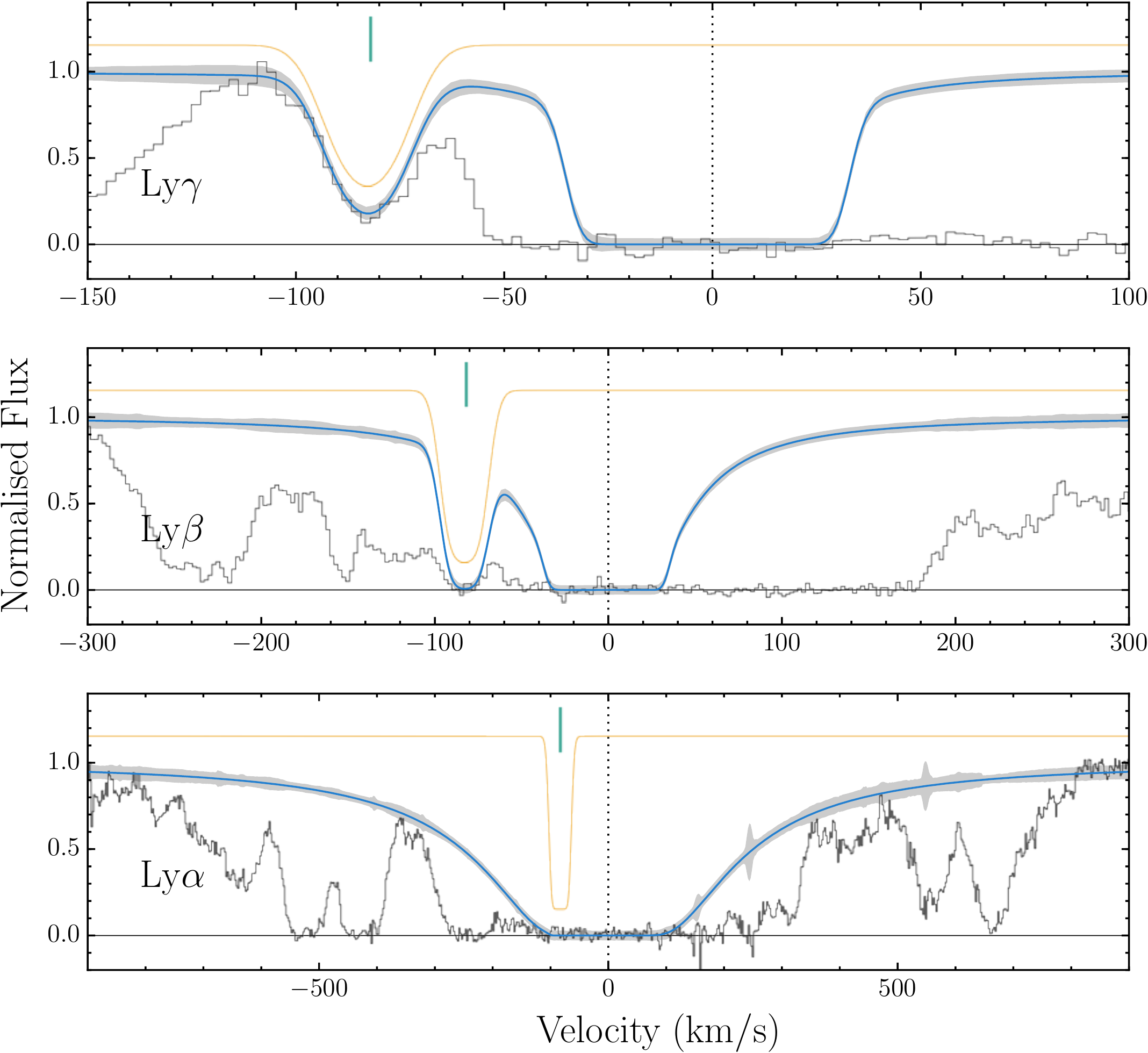}
\caption{Continuum-normalised flux for Lyman series absorption lines of LLS0344 in the J0344$-$0653 HIRES spectrum (black histogram). The zero velocity redshift is set at $\zab = 3.843$. The blue solid lines correspond to our fiducial \HI\ model with $\lNHI=19.23 \pm 0.10$, $\zab = 3.843$ and $b = 10$\,\kms. The solid orange line in each panel represents the Voigt profile fit of a \DI\ line with $\lNDI = 14.63 \pm 0.10$ and $b = 10$\,\kms. It contributes to the total \HI\ absorption profile and the vertical green ticks indicate its position, at the $-82$\,\kms\ isotopic shift from the associated \HI\ line. The grey shading shows the 1$\sigma$ uncertainty in the flux.}
\label{f:lls0344_deut}
\end{figure}

Our fiducial \HI\ model is depicted in \Fref{f:lls0344_deut} with $\lNHI=19.23 \pm 0.10$\ and $b = 10$\,\kms. Having established the redshift of LLS0344, we checked the corresponding Lyman series lines to assess \NHI. The model of \citet{2015ApJS..221....2P} with $\lNHI=19.55 \pm 0.15$\ appeared reasonable when compared with our HIRES spectrum. However, as \Fref{f:lls0344_deut} shows, there appears to be additional, lower column density Lyman absorption lines at almost all velocities across these profiles. This makes the presence, and extent, of the damping wings difficult to discern. It may also appear possible that \NHI\ is substantially lower than \citet{2015ApJS..221....2P}'s estimate, which would weaken the evidence that this system is a near-pristine LLS.

However, the detection of \DI\ at the expected $-82$\,\kms\ isotopic shift from the associated \HI\ \lyc\ line provides an alternative, more reliable estimate of \NHI\ in this system. The top panel of \Fref{f:lls0344_deut} shows this \DI\ detection and that there is coincident, saturated absorption at the expected position of the \lyb\ \DI\ line as well. This indicates that the majority -- if not all -- of the absorption at $-82$\,\kms\ in \lyc\ is most likely due to \DI. Indeed, \Fref{f:lls0344_deut} shows a single Voigt profile overlaying the data for these two \DI\ lines, at the same redshift as the \HI\ line, with \NDI\ and $b$ adjusted in \textsc{vpfit} to visually match the data. This yields $\lNDI = 14.63 \pm 0.10$ and $b = 10$\,\kms. Unfortunately, the higher order \DI\ lines are severely blended with Lyman forest absorption so they could not be used to further corroborate the \DI\ analysis. Nevertheless, using the D/H ratio measurement of \citet{2014ApJ...781...31C}, $ \lNDI = - 4.597 \pm 0.006$, our \NDI\ estimate implies that $\lNHI=19.23 \pm 0.10$. As the bottom panel of \Fref{f:lls0344_deut} shows, overlaying a Voigt profile with this column density on the data (with the same $b$ parameter of $10$\,\kms\ as for \DI) provides a self-consistent result: the model does not fall below the limits imposed by the flux peaks at $-350$, $-320$, $-150$, $+360$ and $+470$\,\kms. The uncertainty of $0.1$ dex in \NDI\ (and therefore \NHI) was determined visually to approximately match the scatter in the HIRES flux values and from the continuum placement. The latter is uncertain by as much as $\pm 15$--20\% and changing the continuum level by this amount requires an adjustment of the \NDI\ of $\sim$0.05 dex, well within the $0.1$ dex uncertainty ascribed to \NHI.

\subsubsection{Photoionisation modeling results}

\begin{figure}
\includegraphics[width=0.95\columnwidth]{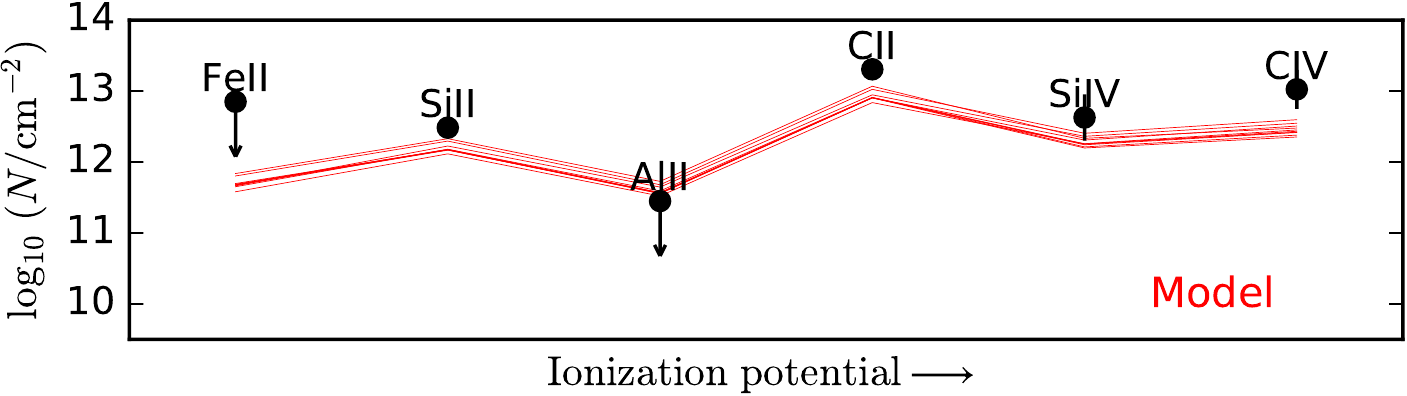}
\caption{\label{f:minimal_model_lls0344} Minimal photoionisation model of LLS0344. The observed column densities in \Tref{t:logN} are shown as data points, with 1$\sigma$ uncertainties and 2$\sigma$ upper limits, and the red lines represent \textsc{Cloudy} column density model values for 20 MCMC samples drawn at random.}
\end{figure}

Following \Sref{s:cloudy_general}, we start the photoionisation modeling of LLS0344 with the coarse \textsc{cloudy} grid and a `minimal model': a minimal set of free parameters in the MCMC analysis to match the observing ionic column densities in \Tref{t:logN}. In this case, the minimal model considers that the singly and triply ionised species are produced by the same phase. \Fref{f:minimal_model_lls0344} shows \textsc{cloudy}'s' predictions for the metal column densities. The main result derived by the MCMC sampling algorithm is the metallicity estimate of $\textrm{[Si/H]} = -3.29 \pm 0.26$. However, the predicted values are larger than the measured ones for \AlII, and smaller for \CII, \CIV, and \SiIV. Retaining the assumption that these ions are produced by the same phase, these mismatches may correspond to a deviation from a solar abundance pattern for C and Al and/or uncertainties related to the shape of the HM12 UV background. Reducing the difference between the measured column densities and the predictions of \textsc{cloudy} is interesting as it may provide a measurement of [C/Si] and an upper limit on [Al/Si].

\begin{figure}
\includegraphics[width=0.95\columnwidth]{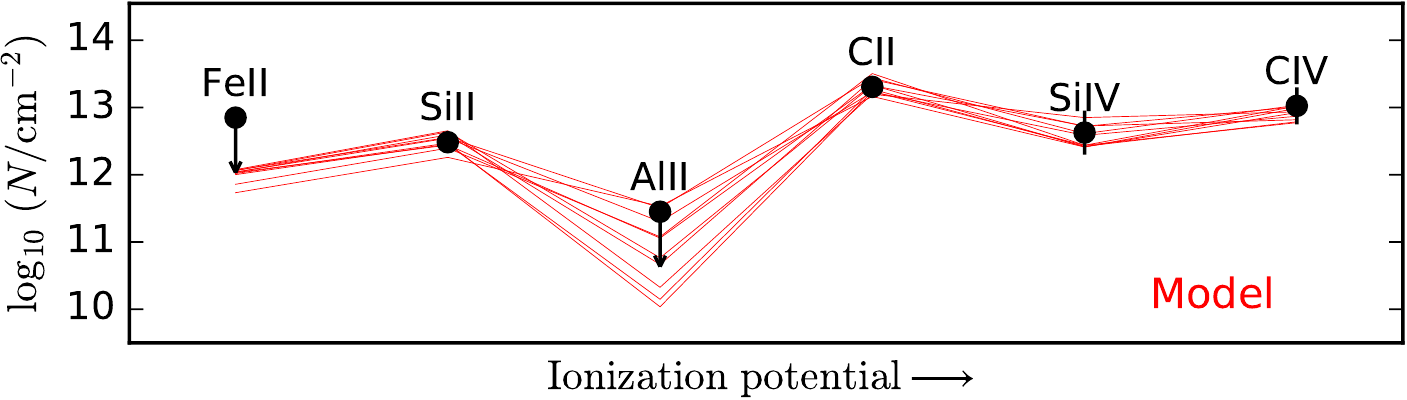}
\caption{\label{f:fiducial_model_lls0344}Same as \Fref{f:minimal_model_lls0344} but for the fiducial photoionisation model of LLS0344.}
\end{figure}

\begin{figure}
\includegraphics[width=0.95\columnwidth]{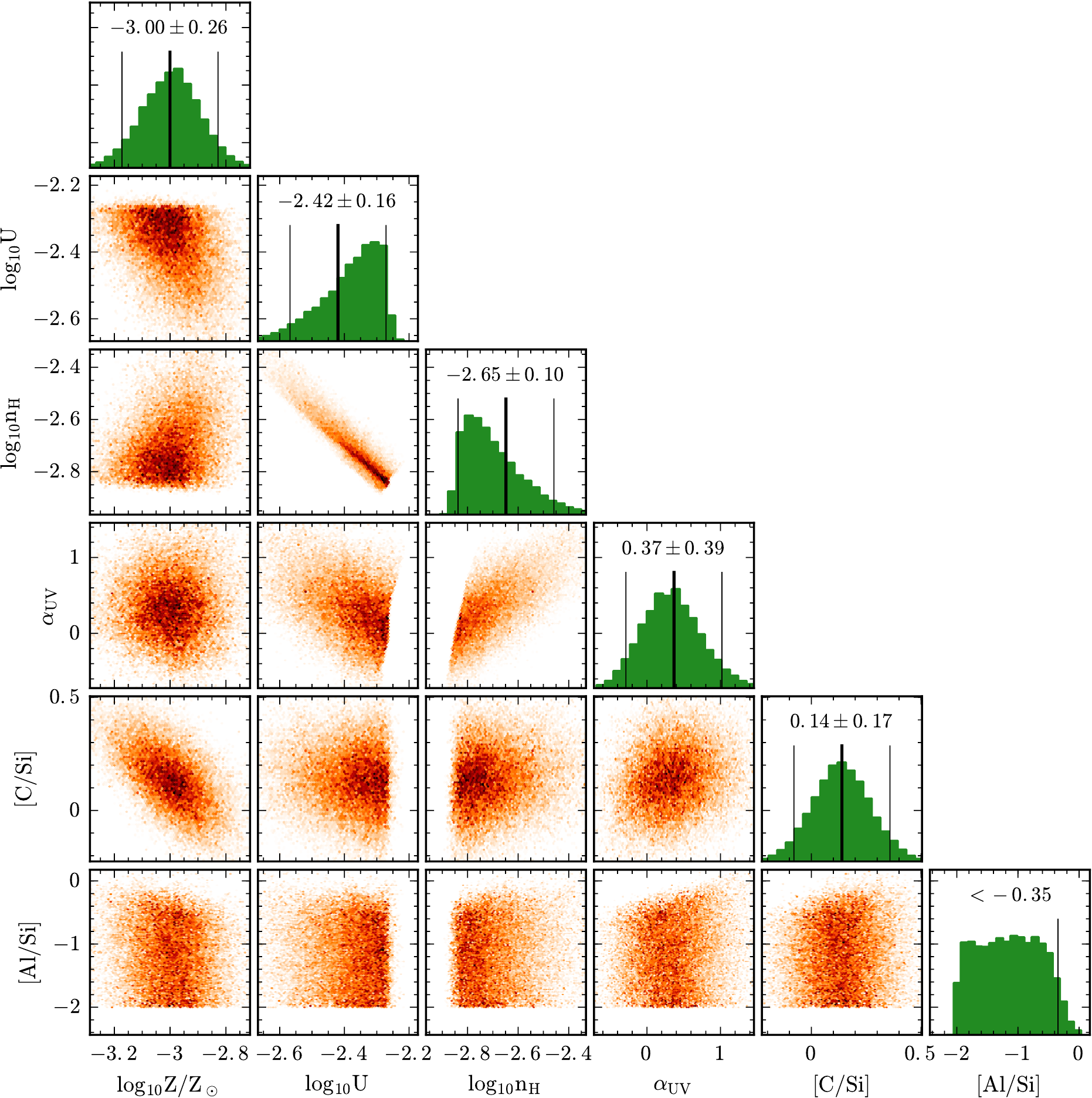}
\caption{\label{f:distribution_0344_fiducial} Fiducial photoionisation model results for LLS0344. Each panel shows the MCMC distributions of parameter pairs involving metallicity ($Z/Z_\odot$), the ionisation parameter ($U$), the volume density of hydrogen (\nH), the [C/Si] and [Al/Si] ratios, and the slope of the UV background (\auv). The panel at the top of each column shows the distribution of the corresponding parameter, together with its mean value or limit (95\% confidence).}
\end{figure}

Given this low initial metallicity estimate and the possibility of non-solar abundance patterns for C and Al, we then follow \Sref{s:cloudy_general} and generate a finer grid of \textsc{Cloudy}'s models for LLS0344. The grid covers $-4.2 < \lnH < -1$, $18.5<\lNHI<19.5$, and $-6< \textrm{[Si/H]} < -2$, all with respective steps of $0.2$\,dex. As stated earlier, we introduce a free parameter $-2.50< $\auv$ <1.50$ with steps of $0.4$ dex to account for uncertainties related to the shape of the HM12 UV background. We also apply a Gaussian prior on \auv, centred on 0 with a standard deviation of $\sigma = 0.5$. \Fref{f:fiducial_model_lls0344} shows \textsc{cloudy}'s predictions compared to the measured values. The distributions of [Si/H], \nH, \auv, [C/Si] and [Al/Si] derived by the MCMC sampling algorithm are shown in \Fref{f:distribution_0344_fiducial} for the fiducial model. For completeness, and comparison with models in other works, we also include the distribution of the ionisation parameter, $U$, defined as the ratio of the ionising photon density to \nH.

The main results from the fiducial model (\Fref{f:distribution_0344_fiducial}) are a metallicity measurement of $\textrm{[Si/H]} = -3.00 \pm 0.26$, a carbon-to-silicon abundance ratio of $\textrm{[C/Si]}=0.14 \pm 0.17$, and a upper limit on the aluminium-to-silicon abundance ratio of $\textrm{[Al/Si]}\leq -0.35$. As opposed to the minimal model (\Fref{f:minimal_model_lls0344}), there are no mismatches between the \textsc{cloudy} predictions and measured column densities in \Fref{f:fiducial_model_lls0344}, so it appears that this is a more reliable model. While the super-solar value of [C/Si] has only very marginal significance, the sub-solar upper-limit on [Al/Si] is robust to different assumptions about the photoionisation modelling. For example, assuming that the high ions (\CIV\ and \SiIV) are not produced by the same phase as the low ions (\CII, \SiII, \FeII\ and \AlII) and a HM12 UV background, $\textrm{[Al/Si]} \leq -0.14$. Further changing the UV background to ``HM05'' within \textsc{Cloudy}, a revised version of that originally published by \citet{1996ApJ...461...20H}, has little impact: we obtain $\textrm{[Al/Si]} \leq -0.18$. We discuss how these values for [C/Si] and [Al/Si] can be compared to the nucleosynthetic yields of PopIII stars in \Sref{s:origin}.

\subsection{LLS1153}
\label{s:lls1153}

\subsubsection{Metal line column densities}

LLS1153 was first identified in \citet{2015ApJ...812...58C} towards the $\zem=4.127$ quasar SDSS J115321.68$+$101112.9 (hereafter J1153$+$1011), based on absorption features at redshift $\zab=4.038$, using a Magellan/MagE spectrum. At this redshift, rather than explicitly estimating \NHI, they established a plausible range of values using the same approach explained in \Sref{s:spectral_general}: $17.7<\lNHI<19.0$. The upper limit derives from the lack of apparent damping wings for \lya, while the lack of flux bluewards of the Lyman limit imposes the lower limit. At $\zab=4.038$ they identified \doublet{C}{iv}{1548/1550}, \doublet{Si}{iv}{1393/1402}, and obtained column density upper limits for \tran{Si}{ii}{1260}, $\lambda$1304 and $\lambda$1526.

\begin{figure}
\centering
\includegraphics[width=0.95\columnwidth]{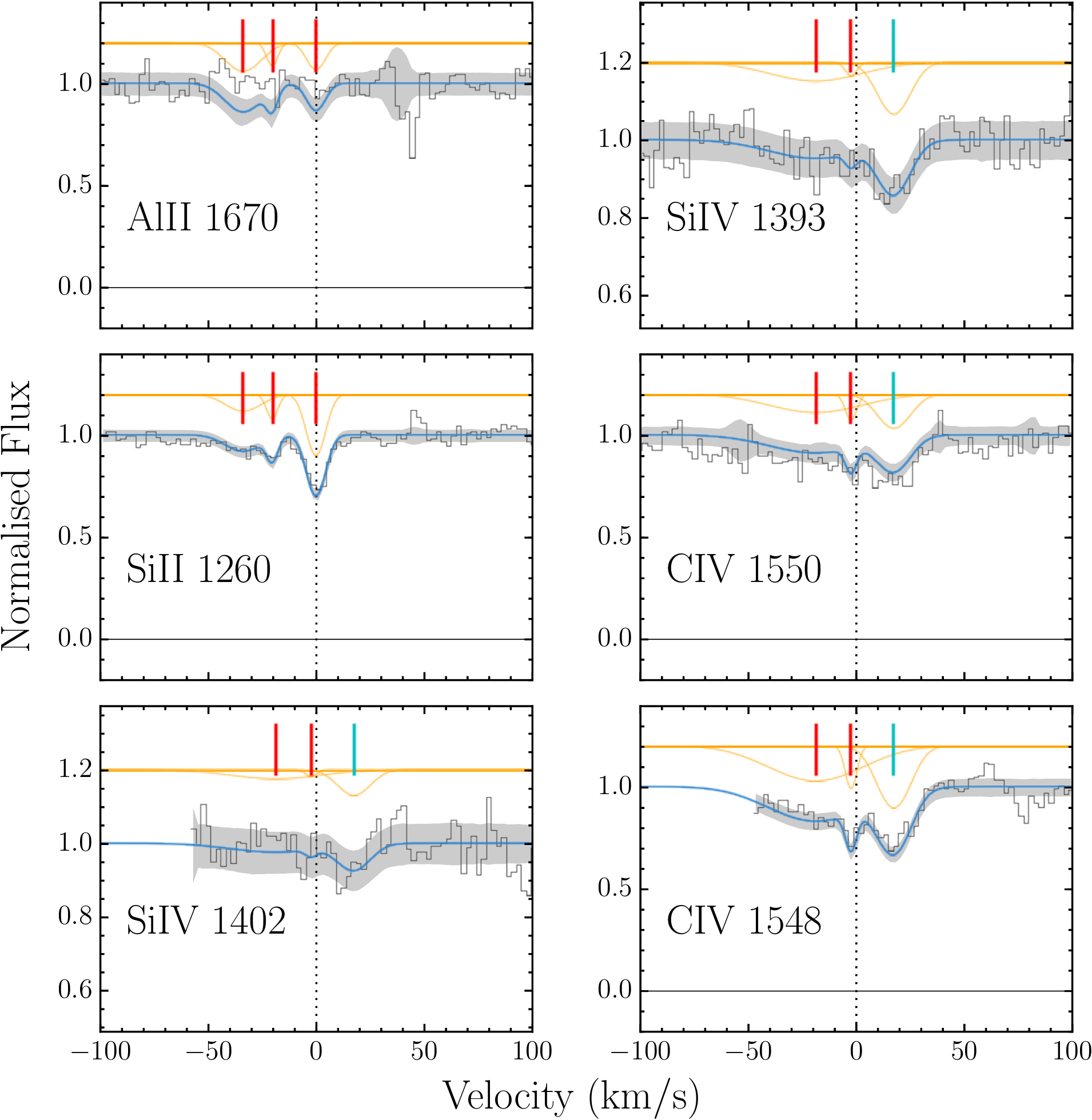}
\caption{Same as \Fref{f:lls0344_metals} but for LLS1153 centred at $\zab=4.038$ towards J1153$+$1011. For \tran{Al}{ii}{1670}, the profile represents the 2$\sigma$ upper limit on its total column density.}
\label{f:lls1153_metals}
\end{figure}

Guided by the initial results above, we searched the HIRES spectrum for metal absorption lines and we list the detections and non-detections (upper limits) in the second column of \Tref{t:logN}. \Fref{f:lls1153_metals} depicts the strongest (highest oscillator strength) transitions of the most abundant metal species expected at $\zab = 4.038$. \Fref{f:lls1153_metals} clearly shows that most of the column density of highly-ionised species resides in a component indicated by a cyan vertical tick, but that this is not present in the singly ionised species. Based on this difference, in our photoionisation modelling we assume that the low and high ions arise in different phases of LLS1153, and that only the low-ions are associated with the \HI\ detected in the Lyman lines. Below, we check the impact of including the triply ionised species on LLS1153's metallicity. In that case, the \CIV\ and \SiIV\ absorption in the two velocity components which are (seemingly) in common with the low ions are included in the model.

\subsubsection{\textbf{\HI}\ column density}

\begin{figure}
\centering
\includegraphics[width=0.95\columnwidth]{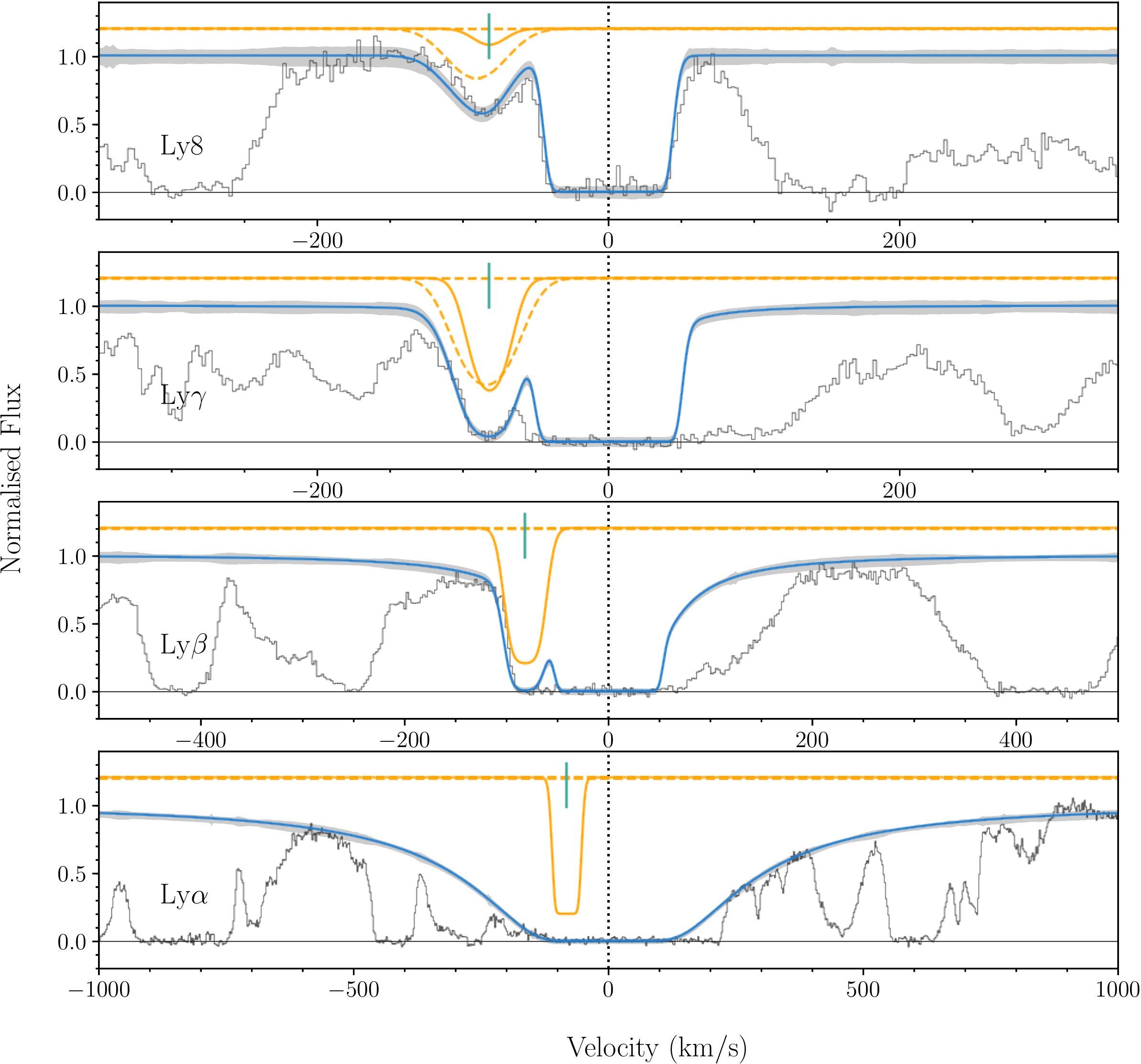}
\caption{Same as \Fref{f:lls0344_deut} but for LLS1153 centred at $\zab = 4.038$ towards J1153$+$1011. The model (blue line) has $\lNHI=19.40 \pm 0.10$ and $b = 15$\,\kms\, derived from the deuterium absorption (solid orange line) at $-82$\,\kms, with $\lNDI = 14.80 \pm 0.10$ (same $b$). The orange dashed lines in the top two panels represent \HI\ blends set at $z=3.0289$ for \lyc\ and $2.8243$ for Ly-8.}
\label{f:lls1153_hi}
\end{figure}

Our fiducial \HI\ model is depicted in \Fref{f:lls1153_hi} with $\lNHI=19.40 \pm 0.10$ and $b = 15$\,\kms. Having established the redshift of LLS1153 from its metal lines, we checked the corresponding Lyman series lines to assess its \NHI. With a hint of damping wings apparent for the \lya\ line in the third panel of \Fref{f:lls1153_hi}, it should be expected that $\lNHI \geq 19$. However, similar to LLS0344, the detection of \DI\ at the expected $-82$\,\kms\ isotopic shift from the associated \HI\ \lyb, \lyc\ and Ly-8 lines provides a more reliable estimate of \NHI. We emphasise again that such a method only provides estimates of the column density and width of the \DI\ lines and not a reliable [D/H] value; indeed, possibly several additional \HI\ blends would need to be comprehensively modeled to determine if a very precise fit of the \DI\ lines was even possible.

The \DI\ detection can be seen in the top panel of \Fref{f:lls1153_hi}, at the expected $-82$\,\kms\ isotopic shift from the associated \HI\ Ly-8 line. There are coincident, saturated absorption features at the same velocity in the \lyb\ and \lyc\ lines. For both Ly-8 and \lyc, part of the absorption at $-82$\,\kms\ is also due to unrelated \HI\ blends. Thus, in \Fref{f:lls1153_hi}, the two top panels each show the Voigt profile of an \HI\ blend to account for some of this additional absorption: $\zab=3.0289$ for \lyc\ and $2.8243$ for Ly-8. The third panel shows a single Voigt profile for \DI\ \lyb. An exact, detailed fit to the data is not sought in this case, particularly for the blends, as the main uncertainty in \NDI\ stems from the continuum level, amounting to $\sim$0.10\,dex (using the same approach as for LLS0344 in \Sref{s:lls0344}). Adjusting \NDI, $b$ and the blend parameters, within \textsc{vpfit} to visually match the data yields $\lNDI = 14.80 \pm 0.10$ and $b = 15$\,\kms. Using \citet{2014ApJ...781...31C}'s average measured [D/H] ratio, our \NDI\ estimate implies that $\lNHI=19.40 \pm 0.10$ and $b=15$\,\kms. As can be appreciated from \Fref{f:lls1153_hi}, this \HI\ model is reasonable as it does not produce too much absorption at (i) the transmission peaks in \lya\ at $-600$ and 400\,\kms; (ii) the sharp blue wing of the \DI\ \lyb\ $-100$\,\kms; (iii) the blue wing of \DI\ at $-100$\,\kms\ and sharp rise in flux from zero at $-50$\,\kms\ in \lyc; and (iv) the sharp rises in flux at $-110$ and $-40$\,\kms\ in Ly-8 and the transmission peak between them.

\subsubsection{Photoionisation modeling results}

We start the modeling of LLS1153 on the coarse \textsc{Cloudy} grids with minimal inputs to the photoionisation model. We use the column densities listed in the second column of \Tref{t:logN}\ and consider the bulk of \HI\ to be associated with the singly ionised species. The result is an initial metallicity estimate from the minimal model of $\textrm{[Si/H]} = -3.15 \pm 0.26$.

\begin{figure}
\includegraphics[width=0.95\columnwidth]{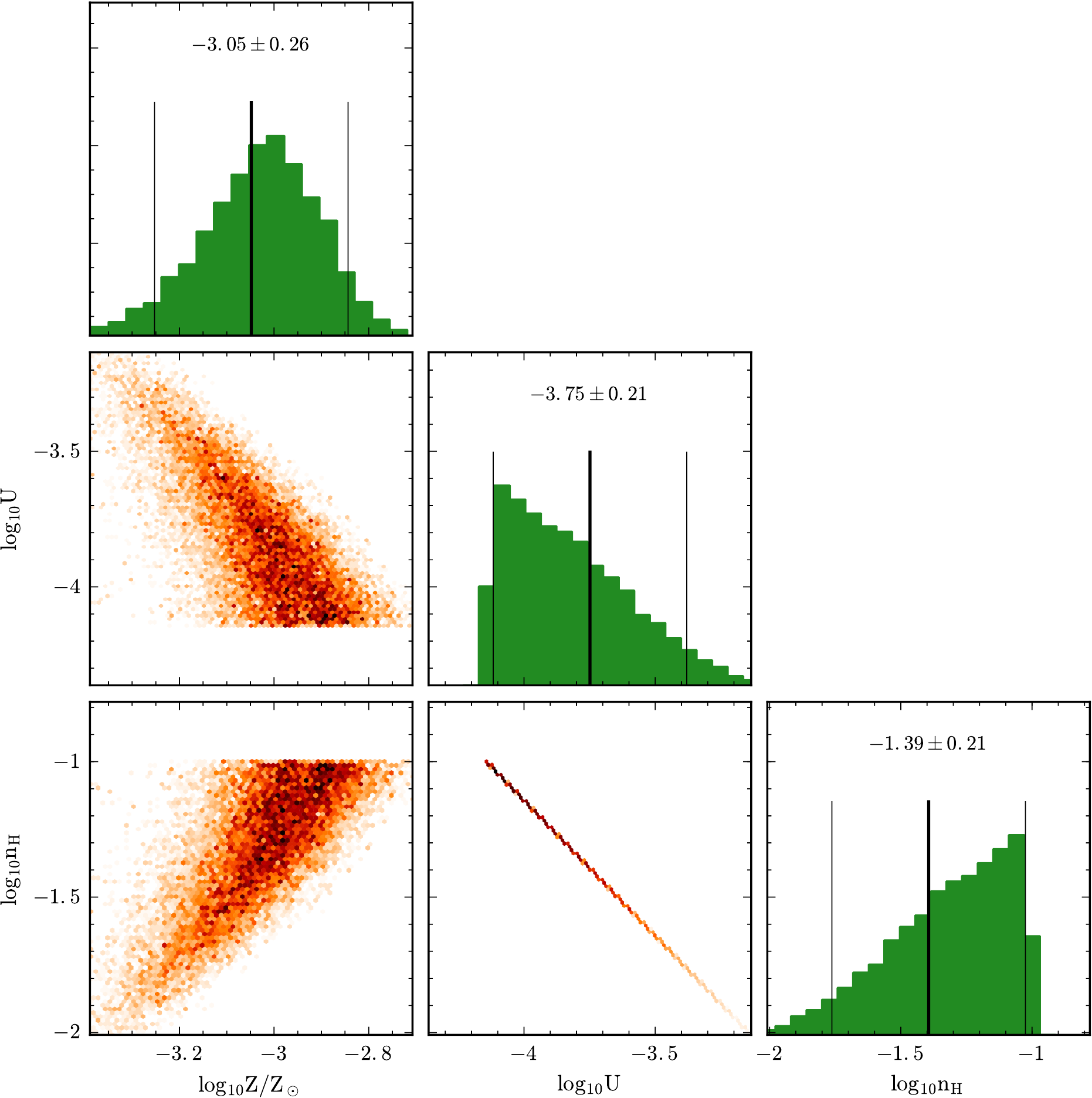}
\caption[Fiducial photoionisation model results for LLS1153.]{\label{f:finer_fiducial_distribution_1153} Same as \Fref{f:distribution_0344_fiducial} but for the fiducial photoionisation model of LLS1153.}
\end{figure}

Given this low initial metallicity estimate, we generate a finer grid of \textsc{Cloudy} models for LLS1153. However, given that we only have a column density measurement of one metal ion (\SiII), we do not consider varying the UV background slope or abundance ratios; in this case, the fiducial model has the same assumptions and inputs as the minimal model above. The distributions of $Z/Z_\odot$, $U$ and \nH\ derived by the MCMC sampling algorithm are shown in \Fref{f:finer_fiducial_distribution_1153}. The main result from the fiducial model is a metallicity measurement of $\textrm{[Si/H]} = -3.05 \pm 0.26$ (95\%-confidence). This is $0.1$\,dex higher than the initial estimate derived with the same model assumptions but a courser parameter sampling grid; even though this is well within the 95\%-confidence interval, it highlights the importance of using the finer grids.

\begin{figure}
\includegraphics[width=0.95\columnwidth]{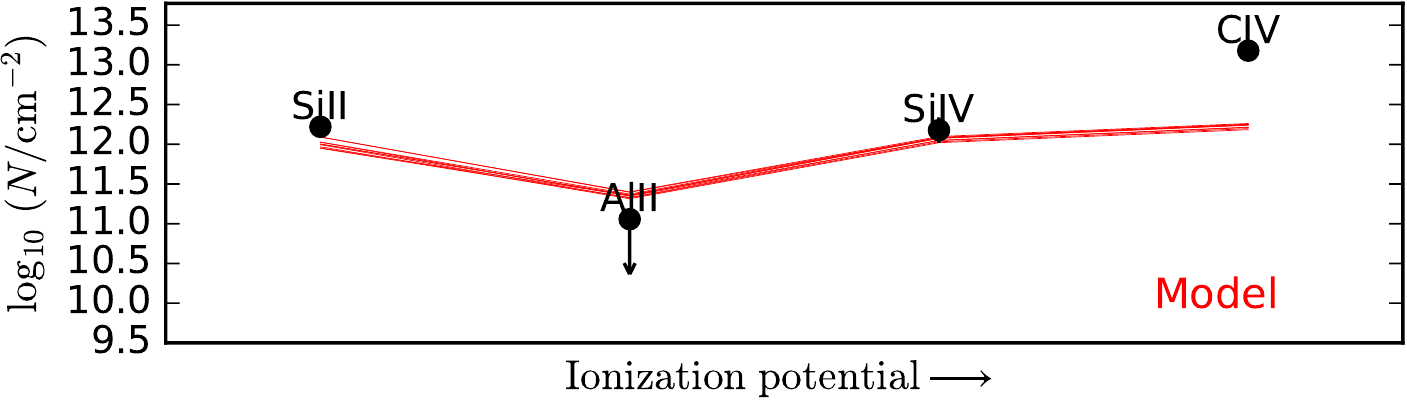}
\caption{\label{f:test_model_lls1153}Same as \Fref{f:minimal_model_lls0344} but for a test model of LLS1153 which includes the column densities of \CIV\ and \SiIV\ in the same phase as the singly ionised gas.}
\end{figure}

So far, the high ions (\CIV\ and \SiIV) were not considered to be part of the same phase as the low ions and \HI. However, from the metal-line velocity structures apparent in \Fref{f:lls1153_metals}, it appears possible that the \CIV\ and \SiIV\ absorption between $-40$ and $+10$\,\kms\ (marked by red ticks) arises in the two fitted low-ion components at $-20$ and $0$\,\kms. We therefore consider a test model for LLS1153 in which the two bluest (red tick) components are in the same phase as the low ions and are associated with the \HI\ absorption shown in \Fref{f:lls1153_hi}. \Fref{f:test_model_lls1153} compares the metallicity measurements with a sample of \textsc{cloudy}'s predictions from the MCMC algorithm for this test model. The metallicity measurement shifts to $\textrm{[Si/H]}  = -3.58 \pm 0.26$, significantly below our fiducial metallicity, showing that the latter is robustly below $\textrm{[Si/H]} = -3$. However, this test model clearly fails to match the measured column densities of the high ions, similar to the case of LLS0344's minimal model. Indeed, \NCIV\ is more than 1\,dex higher than the \textsc{cloudy} predictions, and causes a significant tension between the model and the upper limit on the \AlII\ column density.

It is therefore interesting to consider a further test model, similar to the fiducial model of LLS0344, in which we also allow UV background slope variations and non-solar abundance ratios for C and Al. We use the same Gaussian prior on \auv\ introduced for LLS0344; i.e.\ $\sigma=0.5$ centred on 0. The resulting metallicity is $\textrm{[Si/H]} = -3.28 \pm 0.26$, \auv$ = 0.13 \pm 0.48$ and abundance ratios $\textrm{[C/Si]}=0.89 \pm 0.17$ and $\textrm{[Al/Si]}\leq -0.01$. The latter is consistent with the solar value and \auv\ indicates consistency with the HM12 background. However, the low metallicity in this test model, combined with the overabundance of C, appears interesting at first. Nevertheless, we emphasise that the entire basis of this test model was to consider the same phase for the triply and singly ionised species, just to check the robustness of our fiducial metallicity $\textrm{[Si/H]} \leq -3$. Furthermore, only \CIV\ constrains the large value of [C/Si], i.e.\ it is simply a reflection of the assumption that \CIV\ arises in the same phase as the singly-ionised metals. In these regards, the fiducial model of LLS1153 represents the simplest, most robust set of assumptions, and it produces the highest (i.e.\ most conservative) metallicity measurement.

\subsection{LLS1156}
\label{s:lls1156}

\subsubsection{Metal lines}\label{para:lls1156_metal}

LLS1156 was first identified by \citet{2015ApJS..221....2P} towards the $\zem=3.110$ quasar SDSS J115659.59$+$551308.1 (hereafter J1156$+$5513) based on absorption features at redshift $\zab=2.616$ in a Keck/ESI spectrum. At this redshift \NHI\ was estimated at $\lNHI=19.10 \pm 0.30$ along with $b=30$\,\kms. This value of \NHI\ was determined by visual inspection of theoretical Voigt profiles superimposed on the \lya\ line, which has clear enough damping wings to provide reasonable constraints, even at ESI's low resolution of $R = 8000$ ($\textrm{FWHM} \sim 37$\,\kms). At $\zab=2.616$, \citet{2015ApJS..221....2P} identified \tran{Al}{ii}{1670}, \doublet{C}{iv}{1548/1550}\ and \tran{Si}{iv}{1393}, and obtained column density upper limits for \tran{Si}{ii}{1526}, \tran{Fe}{ii}{2344}\ and \tran{Al}{iii}{1854}. 

\begin{figure}
\centering
\includegraphics[width=0.95\columnwidth]{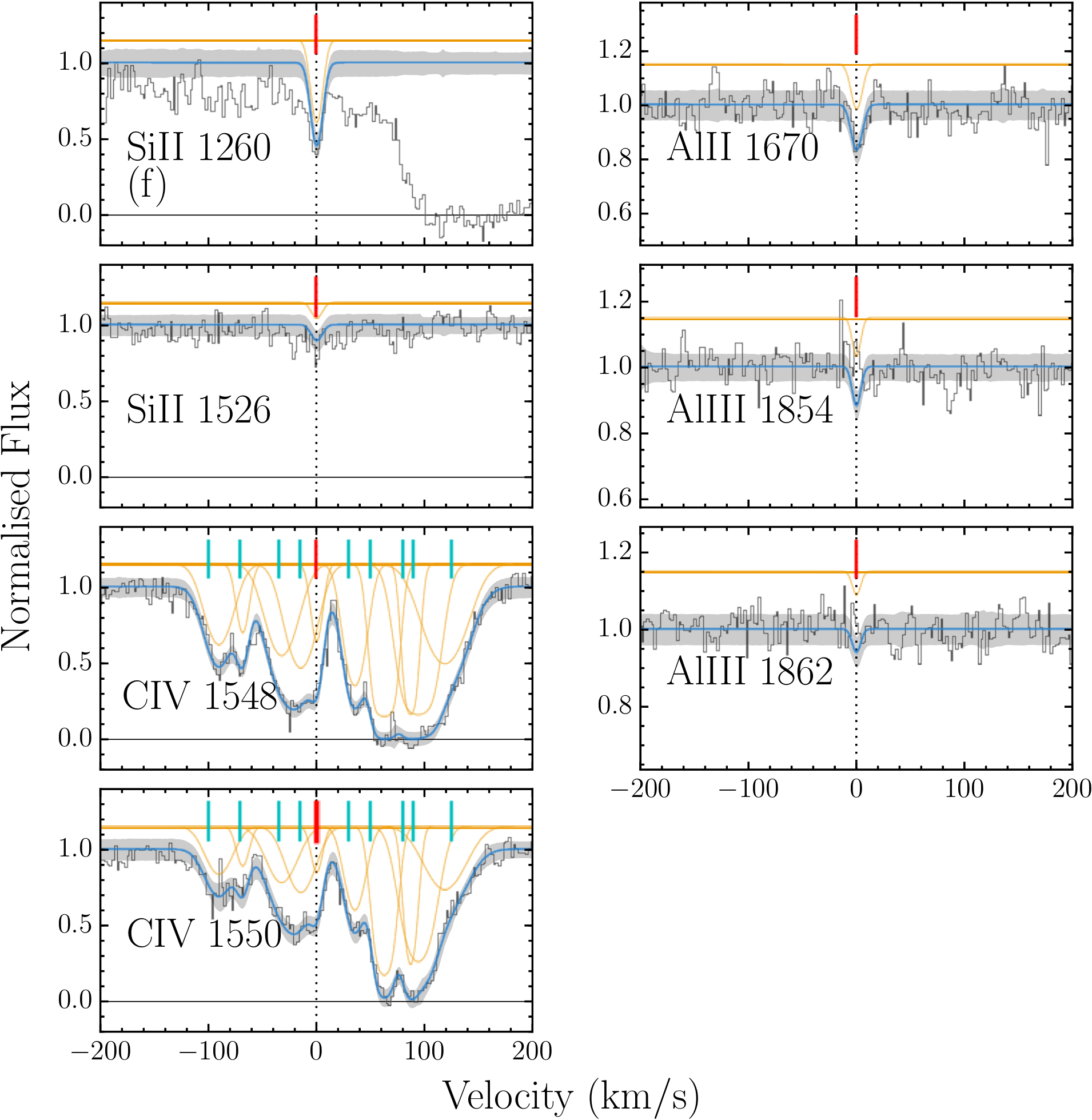}
\caption{Same as \Fref{f:lls0344_metals} but for LLS1156 centred at $\zab=2.616$ towards J1156$+$5513. The ``(f)'' label (in the \tran{Si}{ii}{1260} panel here) indicates that a transition falls in the Lyman forest of the quasar. For \doublet{Al}{iii}{1854/1862}, the column densities used for the Voigt profile fits correspond to a 2$\sigma$ upper limit on the column density.}
\label{f:lls1156_metals}
\end{figure}

Guided by the initial results above, we searched our HIRES spectrum for metal absorption lines and we list the detections and non-detections (upper limits) in the third column of \Tref{t:logN}. \Fref{f:lls1156_metals} depicts the strongest transitions of the most abundant metal species expected at $\zab = 2.616$. \Fref{f:lls1156_metals} clearly shows that the high-ion (\CIV) absorption has a broad, complex velocity structure, which is very different to the very simple, apparently single-component low-ion absorption (\SiII\ and \AlII). \tran{Si}{ii}{1260} falls in the Lyman forest of J1156$+$5513 and sits amongst an \HI\ structure which may affect our column density estimate. Indeed, some of the absorption we attribute to the \HI\ structure could, in fact, be additional \tran{Si}{ii}{1260} absorption; our column density may be a lower limit in that case. The weak detection of \tran{Al}{ii}{1670} will also allow us to constrain the [Al/Si] ratio in \Sref{para:lls1156}.

\subsubsection{\HI\ column density}\label{para:lls1156_hi}

\begin{figure}
\centering
\includegraphics[width=0.95\columnwidth]{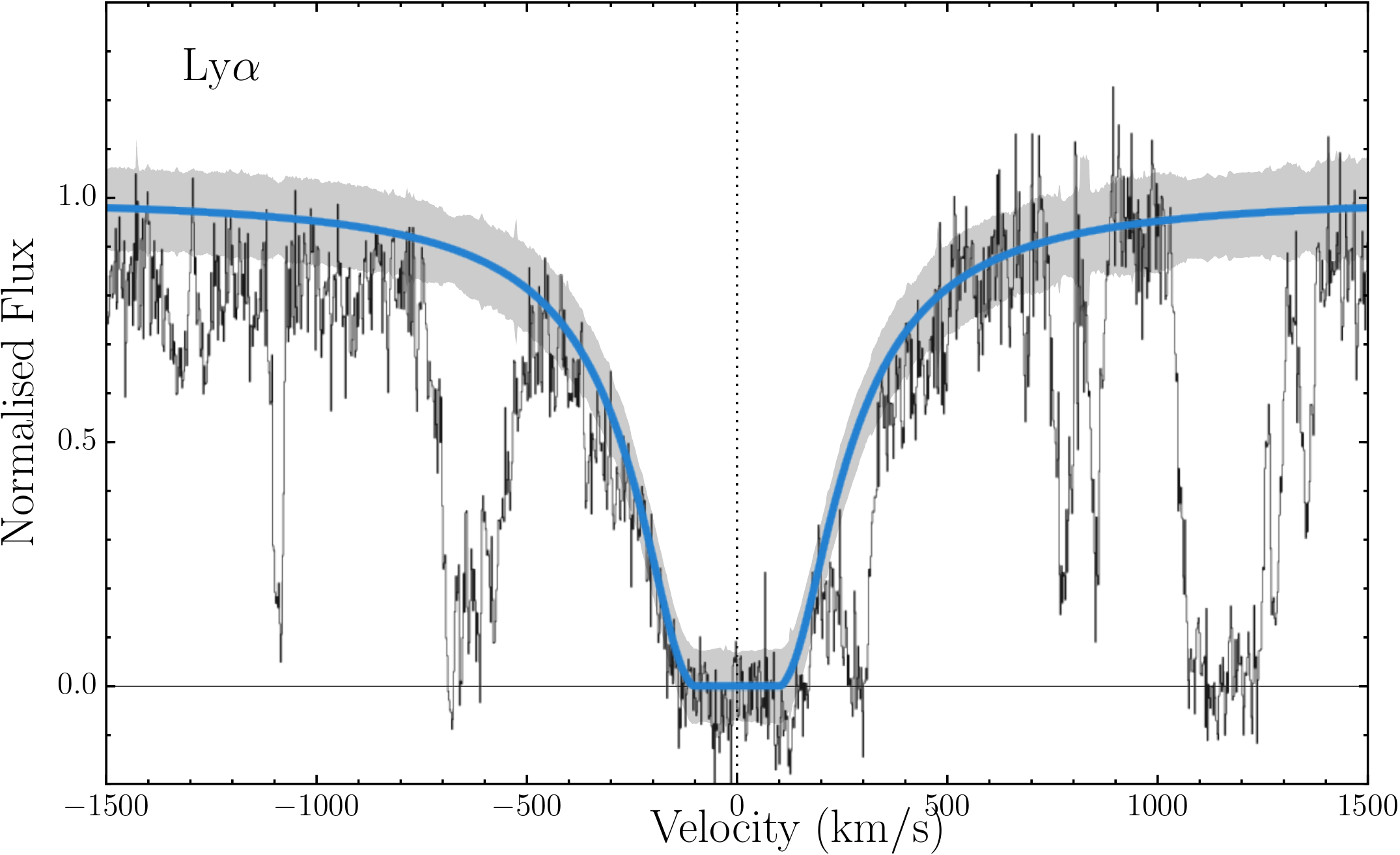}
\caption{Same as \Fref{f:lls0344_deut} but for the \lya\ line of LLS1156 centred at $\zab=2.616$ towards J1156$+$5513. The model (blue solid line) has $\lNHI=19.30 \pm 0.10$ and $b = 15$\,\kms.}
\label{f:lls1156_hi}
\end{figure}

Our fiducial \HI\ model is depicted in \Fref{f:lls1156_hi} (solid blue line) with $\lNHI=19.30 \pm 0.10$ and $b = 15$\,\kms. Having established the redshift of LLS1156 from the low-ionisation metal lines, we checked the corresponding Lyman series lines to assess \NHI. LLS1156's redshift value is low compared to the rest of our LLS sample, so we only have access to its associated \lya\ line. However, its damping wings are clearly seen in \Fref{f:lls1156_hi} and they directly constrain the neutral hydrogen column density to be $\lNHI = 19.30 \pm 0.10$. We also use $b = 15$\,\kms\ in our model but note that the precise value does not have an impact on the column density estimate. The uncertainty on \NHI\ was determined from the flux scatter in the HIRES spectrum and the uncertainties in the continuum placement, the latter being the dominant factor.

\subsubsection{Photoionisation modeling results}\label{para:lls1156}

We first construct a minimal photoionisation model of LLS1156 using the column densities listed in the third column of \Tref{t:logN}. Following the discussion above (\Sref{para:lls1156_metal}), we consider the bulk of \HI\ to be associated with the singly ionised metal lines (SiII\ and \AlII). Comparison of \textsc{cloudy}'s predictions with the measured column densities reveals no significant discrepancies, therefore hinting that [Al/Si] must be very close to the solar value. The initial metallicity estimate from this minimal model is $\textrm{[Si/H]} = -2.93 \pm 0.26$. While strictly higher than $-3$, this is still consistent with it being a ``near-pristine'' system compared to the higher metallicity LLSs described in \Aref{s:higher} where $\textrm{[Si/H]} \ga -2.5$, so it warrants further investigation.

\begin{figure}
\includegraphics[width=0.95\columnwidth]{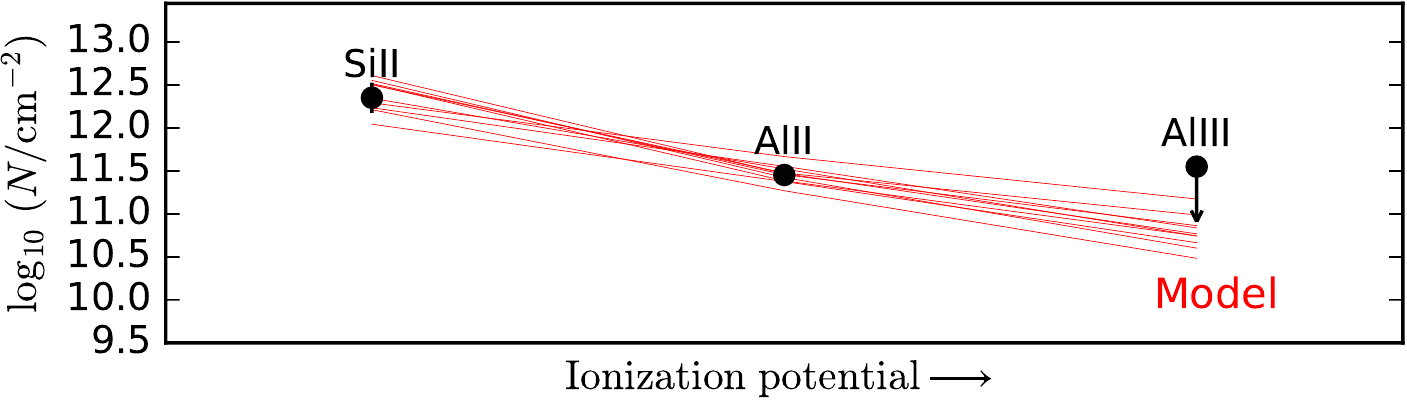}
\caption{\label{f:finer_fiducial_model_lls1156}Same as \Fref{f:minimal_model_lls0344} but for the fiducial model of LLS1156.}
\end{figure}

\begin{figure}
\includegraphics[width=0.95\columnwidth]{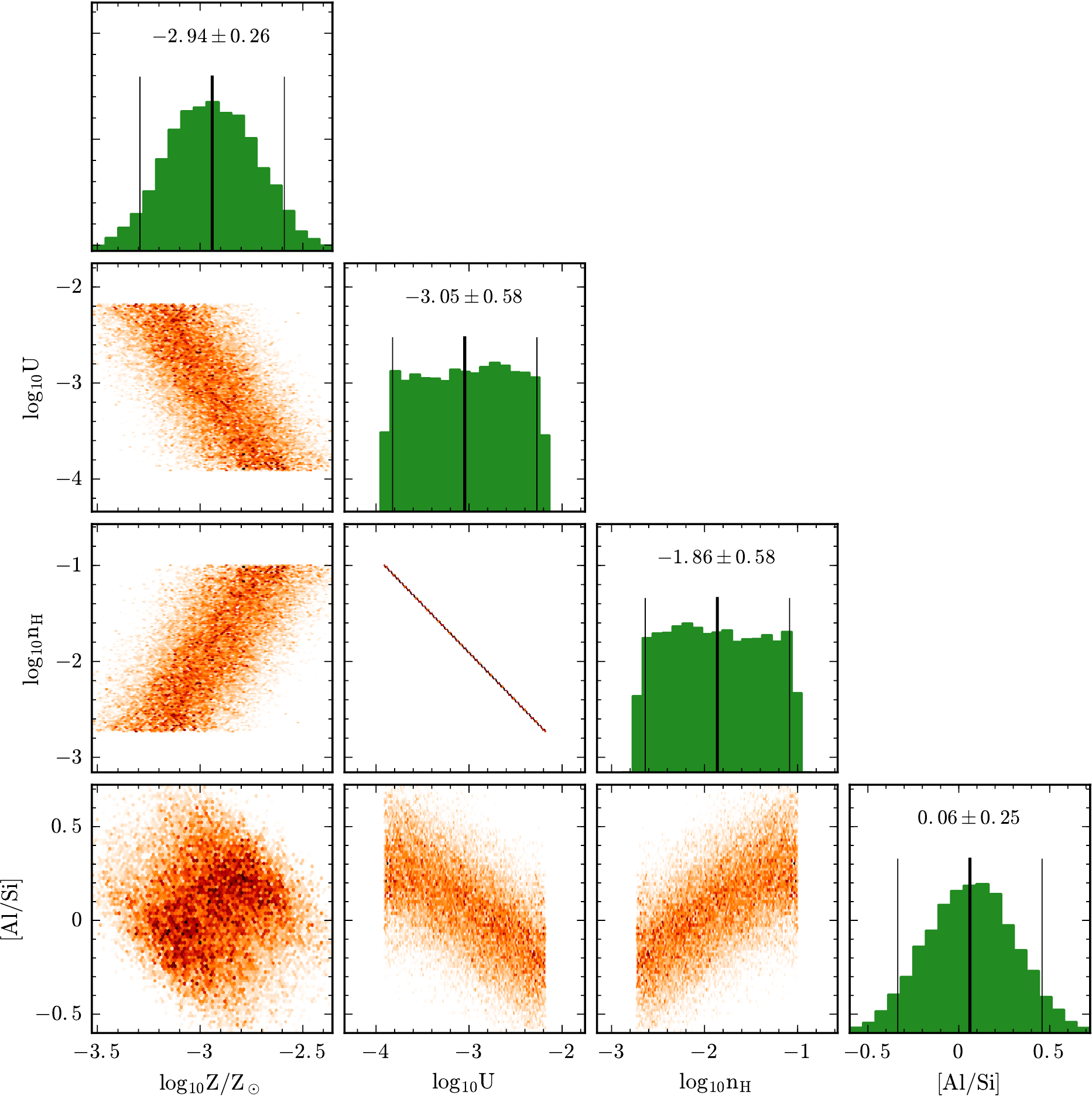}
\caption{\label{f:finer_fiducial_distribution_1156}Same as \Fref{f:distribution_0344_fiducial} but for the fiducial photoionisation model of LLS1156.}
\end{figure}

We therefore generated a finer grid of \textsc{Cloudy}'s models for LLS1156, with a fiducial model in which [Al/Si] is also a free parameter. However, given that even the minimal model above closely matched the observed column densities, and that we do not detect higher ions (\AlIII), we do not include the possibility of varying the UV background slope in the fiducial model. \Fref{f:finer_fiducial_model_lls1156} shows \textsc{cloudy}'s predictions compared to the measured column densities and \Fref{f:finer_fiducial_distribution_1156} shows the distributions of $Z/Z_\odot$, $U$, \nH\ and [Al/Si] from the MCMC sampling algorithm. The fiducial metallicity measurement is $\textrm{[Si/H]} = -2.94 \pm 0.26$, almost identical to the minimal model value. The fiducial model also confirms a near-solar [Al/Si] ratio of $0.06 \pm 0.25$. Unfortunately, this does not provide an important constraint on LLS1156's chemical abundance pattern; i.e.\ this single abundance ratio does not enable a detailed comparison with models of pristine gas enrichment.

\section{Discussion}
\label{s:discussion}

\begin{figure}
\includegraphics[width=0.95\columnwidth]{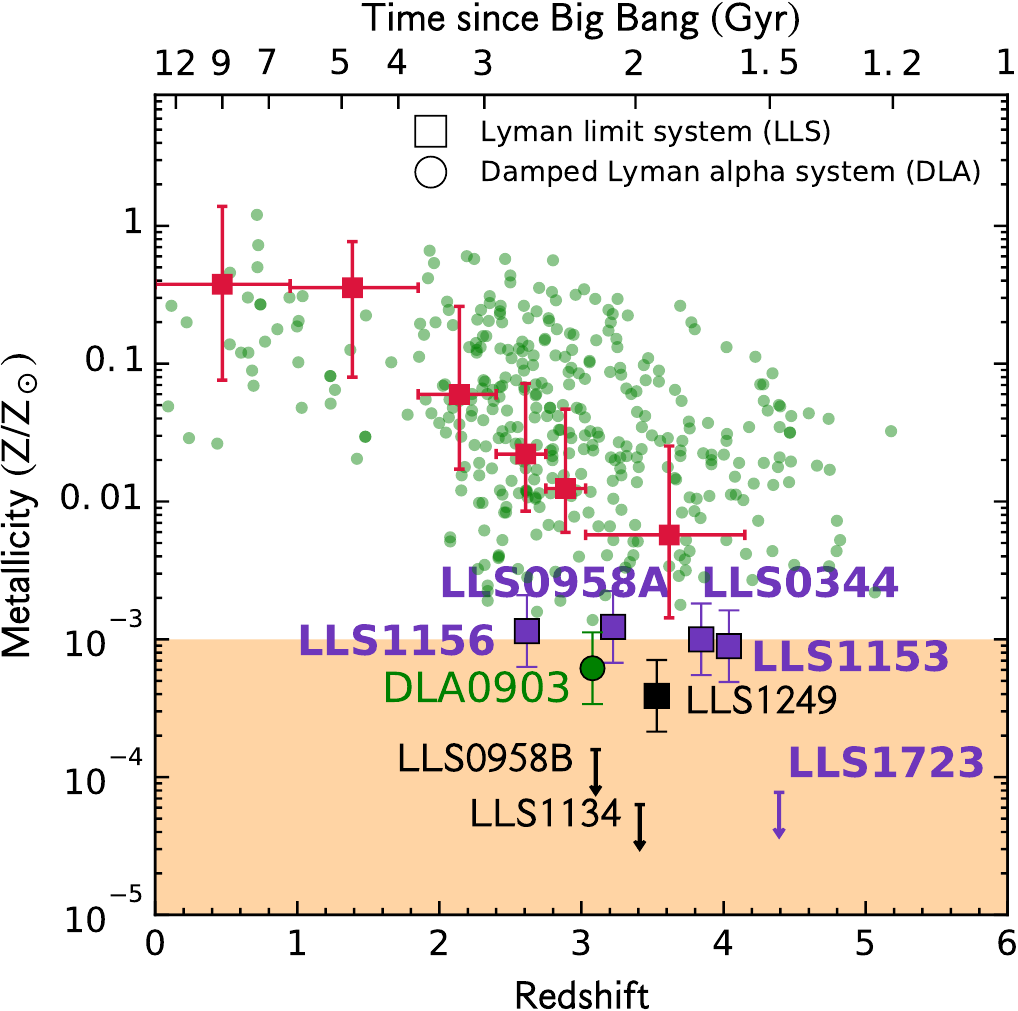}
\caption{Summary of the metallicity distribution of LLSs (squares) and DLAs (circles) in the literature, showing the three new near-pristine systems discovered in this work: LLS0344, LLS1153 and LLS1156, shown in purple. Our remeasurement of LLS0958A's metallicity is also shown in purple \citep[0.44\,dex higher than in][]{2016ApJ...833..283L}. The lowest metallicity measurement for a LLS from \citet{2016MNRAS.457L..44C}, LLS1249, is shown in black. All these near-pristine LLS metallicites are plotted with 95\% confidence error bars. The three known, apparently metal-free LLSs are indicated with arrows (2$\sigma$ upper limits): LLS0958B and LLS1134 from \citet{2011Sci...334.1245F} in black, and LLS1723 from \citet{2019MNRAS.483.2736R} in purple. The red squares and error bars represent the LLS sample of \citet{2016MNRAS.455.4100F} and show the median and 25--75\% range of the composite posterior probability density function in redshift bins containing at least 25 LLSs each. The green circles are the DLA samples of \citet{2011MNRAS.417.1534C}, \citet{2012ApJ...755...89R} and \citet{2013MNRAS.435..482J} [while more recent, dust-corrected DLA metallicity samples are available in \citet{2018A&A...611A..76D} and \citet{2020ARA&A..58..363P}, dust bias at metallicities $\la-2$ is negligible]. The lowest metallicity measurement ([Si/H]) for a DLA from \citet{2017MNRAS.467..802C}, DLA0903, is shown in green. The shaded orange region shows the expected metallicity range for gas enriched by PopIII supernovae from the simulations of \citet{2012ApJ...745...50W}.}
\label{f:LLS}
\end{figure}

In \Sref{s:interesting} we reported the discovery of 3 new near-pristine LLSs (LLS0344, LLS1153, LLS1156) with metallicity measurements of, respectively, $\textrm{[Si/H]}  = -3.00$, $-3.05$ and $-2.94$, all with 95\% confidence uncertainties of 0.26\,dex (dominated by the uncertainties in $\NHI$ which are estimated to be 0.1\,dex in all cases). \Fref{f:LLS} summarises the metallicity distribution at $z>2$ of LLSs and DLAs in the literature \citep[updated from][]{2019MNRAS.483.2736R}. It appears that the 3 new near-pristine LLSs reported here are separated from the bulk of the significant and representative portion of the LLS population shown on the figure. Indeed, the majority of LLSs generally have metallicities $\ga$0.5\,dex higher than those of the 3 ``near-pristine'' LLSs. It is important to note that the range of LLS metallicities in each redshift bin (red points), taken directly from \citet{2016MNRAS.455.4100F}, represent the composite probability distribution from the constituent absorbers. The red error bars in \Fref{f:LLS} indicate the 25--75\% range of these composite distributions, and this extends almost down to $\lmetal = -3$ for the $z=3$--4 bin. However, the low-metallicity tails of the distributions for individual LLSs in this bin are particularly broad, with no real lower limit, because a larger fraction of them ($\sim$30\%) have poorly constrained \HI\ column densities in the range $\lNHI \approx 17.9$--18.9 \citep[figure 1 of][i.e.\ damping wings were not detected but no flux was detected bluewards of the Lyman limit]{2016MNRAS.455.4100F}. Indeed, no individual LLS was measured to have $\lmetal \la -3$ in this sample. Furthermore, our survey targeted six of these absorbers, which had metallicity upper limits of $\approx-3$ from that study, but we found that three actually have considerably higher metallicities ($\textrm{[Si/H]} \ga -2.5$; see \Aref{s:higher}). That is, the binned LLS points in \Fref{f:LLS} are biased to low metallicities, particularly the lower error bars and especially for $z=3$--4, and include the LLSs we targeted, which all tend to artificially reduce the apparent gap between the near-pristine systems and the higher metallicity bulk of the LLS sample.

Regarding the metallicity distribution of LLSs at $\zab \geq 2$, it appears to be broadly unimodal, with a peak at $\textrm{[Si/H]} \sim -2$. This was found in both \citet{2016ApJ...833..283L} and \citet{2016MNRAS.455.4100F}, who also derived probabilities for finding LLSs with $\textrm{[Si/H]} < -3$ of $\sim$10 and $\sim$3--18\%, respectively, at $2 \leq \zab \leq 4.5$. In \citet{2019MNRAS.483.2736R}, we estimated that LLSs in this redshift range with metallicity $\textrm{[Si/H]} < - 4$ constitute $\sim$1.6\% of the population, drawing from an effective sample of $\sim$191 LLSs without strong metallicity biases. With the discovery of 3 new near-pristine LLSs, we now extend this simple estimate to 8 out of 191 systems with $\lmetal \leq -3$, a proportion of $\sim$4\%. This is substantially lower than the $\sim$10\% and $\sim$18\% upper bounds of \citet{2016ApJ...833..283L} and \citet{2016MNRAS.455.4100F}. And, while the latter's lower-bound of $\sim$3\% is close to our $\sim$4\% estimate, we must recall that theirs includes the low-metallicity tails of the probability distributions for many LLSs, none of which actually had measured metallicities 3\,dex (or more) lower than the solar value. That is, our new estimate appears to indicate that very metal-poor LLSs are rarer than previously estimated. Nevertheless, we note that our sample of 8 candidates was deliberately selected to have higher \NHI\ values, on average, than these previous studies, so it remains possible that a smaller fraction of higher \NHI\ LLSs have $\lmetal \la -3$ than lower \NHI\ systems. Naively, we would also expect that the cosmological abundance of $\lmetal \la -3$ LLS may evolve with redshift.

While the above results cannot be considered reliable estimates of the cosmological abundance of these systems, they nevertheless may still provide interesting insights. Indeed, the real shape of the LLS metallicity distribution at $z \geq 2$ is currently unclear. After all, we have been able to find 4 LLSs with $\lmetal \leq -3$ non-serendipitously in this paper and \citet{2019MNRAS.483.2736R}. If, as we estimate, they are apparently very rare, they may simply be examples of the very low metallicity tail of a broad unimodal metallicity distribution. On the other hand, if they are significantly more common than we estimate (due to unknown biases), they may instead constitute a second mode of the metallicity distribution, possibly with a different origin to the higher-metallicity LLSs. Clarifying this issue would help infer the origins of very metal-poor LLSs. Measuring the redshift evolution of the cosmological abundance of $\lmetal \leq -3$ LLSs, and its dependence of \NHI, would be particularly important in this context. In \Sref{s:origin}, we consider the possible origin scenarios for the near-pristine LLSs in a similar way to LLS1723 in \citet{2019MNRAS.483.2736R}. The fact that our new near-pristine LLSs have detected metal lines, unlike LLS1723, leaves the question of their enrichment histories more open. Then, to refine the current probability estimates of finding very metal-poor LLSs, we discuss in \Sref{s:strategy} how we can better identify very metal-poor candidates in future searches.

\subsection{Possible origins of the 3 new near-pristine LLSs}
\label{s:origin}

The abundance pattern of a given absorption system should theoretically exhibit the chemical fingerprint left by the contributing stars that ended their life as SNe. Establishing its chemical profile offers the possibility of investigating the properties of the progenitor stars such as their number, mass distribution and SN explosion energies. Perhaps more importantly, it would determine in which stellar generation the progenitor stars were formed. If the progenitors could be confidently identified as being PopIII stars, this would represent an important, relatively direct observational probe of the properties of the first stars. A similar approach has been employed for very metal-poor local stars in the Milky Way \citep[e.g.][]{2002Natur.419..904C,2014Natur.506..463K} which may have been created from the remnants of PopIII stars. However, the interpretation of stellar abundances is neither direct nor simple \citep[e.g.][]{2005ARA&A..43..481A,2005ARA&A..43..531B}. For instance, the abundance pattern will change substantially over time as the interior is polluted by a star's environment and its internal nuclear burning. Compared to very metal-poor local stars, the large reservoirs of hydrogen in LLSs and DLAs may be less complex to model and could, in principle, have abundance patterns subject to fewer sources of metal enrichment. So far, this approach has been mostly applied to DLAs \citep[e.g.][]{2008MNRAS.385.2011P,2010ApJ...721....1P,2011MNRAS.412.1047C}, and offers promising results in some cases \citep[e.g.][]{2015ApJ...800...12C,2017MNRAS.467..802C}. However, as can be seen from \Fref{f:LLS}, LLSs appear to be more metal-poor than DLAs, on average, at $z>2$. This may indicate that LLSs have a greater chance of pollution by fewer generations of stars -- possibly PopIII stars -- or even remaining metal-free as has been suggested for LLS1134 and LLS0958B by \citet{2011Sci...334.1245F} and LLS1723 by \citet{2019MNRAS.483.2736R}.

Generally speaking, the enrichment level of a gas cloud through the explosion of SNe depends on many properties, such as the form of the initial mass function (IMF) of the first stars, their explosion energies, and the level of mixing between stellar layers. Therefore, the fate of these remnants and their interactions with their environment spans many orders of magnitude in length scale: from sub-pc to Mpc. This is a complex process and key point for numerical simulations if they are to provide clues for the enrichment history of LLSs. In \citet{2019MNRAS.483.2736R}, we discussed the possible origins of LLS1723, an apparently metal-free system. Briefly, it may be a cold gas stream accreting towards a galaxy, or it may represent a high-density portion of the intergalactic medium containing either pristine gas (unpolluted by stellar debris for $1.4$ Gyr after the Big Bang) or the remnants of low-energy supernovae from (likely low-mass) PopIII stars.  For the 3 new near-pristine LLSs, we may reject the pristine gas possibility because we clearly detected metal absorption. The other two possibilities -- cold streams and PopIII remnants -- are considered further below.

Before discussing the different possible origin scenarios for the 3 new near-pristine LLSs, we note that, before this work, only two others had been found. \citet{2016MNRAS.457L..44C} and \citet{2016ApJ...833..283L} analysed the properties of LLS1249 and LLS0958A, respectively, both with metallicities\footnote{\label{fn:LLS0958}In our reassessment of LLS0958A's metallicity in \Aref{s:lls0958A} we find $\lmetal\approx-2.9$. Our estimate differs from \citet{2016ApJ...833..283L} as we include metal absorption over the full velocity structure. However, this does not change the fact that LLS0958A may be in an isolated, intergalactic-like environment, as discussed further below.}  $\lmetal \sim -3.4$. For both, only a [C/Si] ratio could be determined, and therefore distinguishing between PopIII and PopII/I scenarios for the progenitor star(s) was not possible. One of the main motivations for our survey was to find more such systems, but with higher column densities, to increase the possibility of establishing more than a single abundance ratio. However, only two of the 3 new near-pristine systems provided a potentially useful ratio: a $\textrm{[C/Si]} = 0.14 \pm 0.17$, plus an upper-limit on $\textrm{[Al/Si]} \leq -0.35$ (95\% confidence) for LLS0344, and $\textrm{[Al/Si]} = 0.06 \pm 0.25$ for LLS1156. Unfortunately, these are again not enough to single out a particular nucleosynthetic model. Indeed, using the online fitting tool \textsc{starfit}\footnote{See \urlstyle{rm}\url{https://2sn.org/starfit/}} \citep{2019ApJ...871..146F} we could not conclude which model best-fitted our measured ratios in LLS0344: enrichment coming from a 140--260 \msun\ PopIII progenitor, a $\leq 100$\,\msun\ PopIII progenitor, or a PopII/I progenitor gave equally acceptable fits. Therefore, without more detailed abundance ratio constraints coming from their weak metal lines, the possible origins for the 3 new near-pristine LLSs becomes similar to that discussed for the apparently metal-free LLS1723, though with their higher metallicities taken into account.

A first possible origin scenario for the 3 new near-pristine LLSs (LLS0344, LLS1153, LLS1156) is that they could be intergalactic clouds that have been enriched by PopIII stars. Indeed, the fact that these 3 LLSs have $\lmetal \leq -3$ is reminiscent of the metallicity floor found in different numerical simulations \citep{2007MNRAS.382..945T,2011MNRAS.414.1145M,2012ApJ...745...50W,2018MNRAS.475.4396J}. In \citet{2012ApJ...745...50W}, a single pair-instability supernova in the mass range 140--260\,\msun\ can enrich its host halo and surrounding $\sim 5$\,kpc to such a metallicity as early as $z=13$--16. Such a scenario would leave near-pristine LLSs as isolated portions of the IGM that have not been polluted any further by $z \sim 3$. It may also be possible for intergalactic LLSs to have been enriched to a similar level by less massive PopIII stars. A less massive progenitor, with a typical mass $\leq 40$\,\msun, means less metallic ejecta to enrich the surrounding IGM. Therefore, the enrichment level necessary to form PopII/I stars would not be reached immediately after the death of these less massive PopIII stars, i.e.\ in contrast to a single very massive PopIII star, many $\leq 40$\,\msun\ PopIII progenitors would be needed to reach an enrichment level of $\lmetal \sim -3$.

Similar conclusions were reached by \citet{2019MNRAS.487.3363W} who consider the possible origins of the 11 most metal-poor DLAs known.  DLA0903, represented in \Fref{f:LLS}, is the most metal-poor DLA from this sample, by a considerable margin, with $\lmetal = -3.21 \pm 0.05$\ (95\%-confidence), which is very similar to the near-pristine LLSs. One of \citeauthor{2019MNRAS.487.3363W}'s main conclusions is that the ``typical'' metal-poor DLA has been enriched by $\leq 72$ massive stars with masses $\leq 40$\,\msun.  Moreover, the simulations of \citet{2012ApJ...745...50W} suggest that haloes in which one or many of these low mass PopIII stars explodes would tend to keep the metal ejecta within them. Such self-enrichment would allow reaching the metallicity level necessary to form PopII/I stars. It would then be interesting to consider how portions of the surrounding IGM could end up enriched by low mass PopIII remnants, and stay isolated until $z \sim 3$. \citet{2008ApJ...682...49W} considered the evolution of different types of PopIII SNe for haloes of different masses. For 15--40\,\msun\ stars, their explosions can disrupt haloes with $\leq 10^7$\,\msun\ but not more massive haloes where the ejecta stays contained and self-enrichment can occur. From the same work, PopIII stars with masses in the range 140--260\,\msun\ easily disrupt haloes more massive than $10^7$\,\msun\ when exploding as pair-instability supernovae, and it can be approximated that any PopIII supernova explosion can disrupt a halo less massive than $10^7$\,\msun. Therefore, to have a portion of the IGM polluted by the remnants of $\leq 40$\,\msun\ PopIII stars, the progenitor stars' birth site would need to be within a small halo with $\leq 10^7$\,\msun.

While still assuming that the near-pristine LLSs arise in the intergalactic medium, we can instead consider a PopII/I enrichment scenario. However, following a similar argument about the apparently metal-free LLS1723 in \citet{2019MNRAS.483.2736R}, this seems unlikely. Firstly, it would require prior enrichment by PopIII stars which, as already discussed, already raises the metallicity of pristine gas to $\lmetal \sim -3$. The additional enrichment from one or more PopII/I supernovae will then increase the metallicity significantly above what is observed in the near-pristine LLSs. Furthermore, because most PopII/I stars would be less massive than PopIII stars, the PopII/I supernovae ejecta is more likely to be contained within the host halo. That is, the halo would self-enrich rather than lead to pollution of the IGM with $\lmetal \sim -3$.

Given that the hydrogen column densities for the new 3 near-pristine absorbers are all $\lNHI > 19$, it may be more likely that they arise in a circumgalactic region, possibly as part of a cold stream of gas. This scenario is a well-known prediction of cosmological simulations where the streams of gas being accreted into galactic haloes have column densities consistent with LLSs \citep[e.g.][]{2011MNRAS.418.1796F,2011MNRAS.412L.118F,2012MNRAS.421.2809V}. Nonetheless, the origin of their metal enrichment is unclear and may arise in several different ways. Indeed, they could have been enriched by PopIII or PopII/I remnants before the redshift at which they are observed (\zab). The gas would have then remained unaffected, without further metal pollution, and become part of a cold stream observed at \zab. On the other hand, the LLSs could have been polluted for the first time at \zab\ by metals with a PopII/I origin, coming from a recent SN event in the galaxy/galaxies they surround. Indeed, as pointed out above with the work of \citet{2008ApJ...682...49W}, while remnants of relatively low-mass PopII/I stars are unlikely to be able to reach the IGM, it is possible they may reach a circumgalactic cold stream of gas close to the host-galaxies. The added enrichment from PopII/I explosions would likely exceed or stay close to $\lmetal \sim -3$, being consistent with the metallicities derived for the 3 new near-pristine LLSs discovered in \Sref{s:interesting}. In the absence of detailed chemical profiles (i.e.\ several different abundance ratios) to compare to predictions of theoretical nucleosynthetic yields, both the above options are possible for the source of enrichment in the circumgalactic gas stream scenario.

Finally, we note that, with the advent of optical integral field spectrographs such as Keck/KCWI and VLT/MUSE, it is now possible to observe the galactic environment of metal-poor LLSs relatively efficiently. This is a promising approach that is likely to help better understand the origins of these systems. For instance, \citet{2016MNRAS.462.1978F} used $\sim$5\,hrs of VLT/MUSE observations to map the \lya-emitting\ galaxy environment down to luminosities $L_{\lya} \ge 3\times10^{41}$\,erg\,s$^{-1}$ ($0.1L^*_{\lya}$) in a $\sim$160\,kpc radius around LLS0958A and LLS0958B, with metallicities $\lmetal = -3.35 \pm 0.05$ (though see footnote \ref{fn:LLS0958}) and $\leq -3.8$, respectively. Interestingly, this work reports two different environments for these possibly near-pristine and apparently metal free LLSs. On the one hand, the near-pristine LLS0958A shows no nearby galaxies while the apparently metal-free LLS0958B is surrounded by five \lya\ emitters at its redshift with impact parameters in the range 77--312\,kpc. Three of the emitters appear aligned in projection and may indicate a filamentary intergalactic structure. Interestingly, LLS0958A may then correspond to the case of an isolated, enriched portion of the IGM as we described above, while LLS0958B may be akin to a portion of a circumgalactic cold stream that has remained free of metal enrichment. By contrast, recently \citet{2020MNRAS.491.2057L} studied the galactic environment of the first near-pristine system, LLS1249, with MUSE and discovered three \lya-emitting galaxies within 185\,kpc (projected) of the absorber.
 
While it is promising that the different origin scenarios for metal-poor LLSs described in this section and in \citet{2019MNRAS.483.2736R} can be observed, the initial results above from \citet{2016MNRAS.462.1978F} and \citet{2020MNRAS.491.2057L} above highlight the complexity of metal enrichment at $z>2$. Indeed, the ``naive'' expectation that the most metal-poor LLSs should reside in the least galaxy-rich environment, similar to the IGM, appears to be contradicted by the examples of LLS0958A and B, but perhaps confirmed in the case of LLS1249. However, it is important to consider that LLS0958A may still arise close to a galaxy, or galaxies, which simply fall below to detection limits of the VLT/MUSE observations \citep{2016MNRAS.462.1978F}. In any case, it is clear that to refine our discussion of the nature of metal-poor LLSs, building a larger, statistical sample is much needed (ideally, combining absorption and integral field spectroscopy). In that regard, we discuss in \Sref{s:strategy} how our strategy for identifying new metal-poor LLSs can be improved.

\subsection{Strategies for identifying new metal-poor LLSs}
\label{s:strategy}

In this work we targeted with Keck/HIRES 8 LLSs without existing high-SNR, high-resolution spectra. These systems were selected because their metallicity upper limits were found to be $\lmetal \leq -3$ by \citet{2015ApJ...812...58C} or \citet{2016MNRAS.455.4100F}. Nevertheless, 4 LLSs turned out to be more metal-rich than expected, with $\lmetal \ga -2.5$ for three of them (LLS0952, LLS1304, LLS2241) and one not a LLS at all (LLS0106) -- see \Aref{s:higher}. We do not discuss their possible origins since they have metallicities not too different from the vast majority of LLSs already studied in the literature \citep[e.g.][]{2016MNRAS.455.4100F,2016ApJ...833..283L,2017ApJ...837..169P}. Nonetheless, as emphasised above, new observations of metal-poor LLSs are clearly needed, given the variety of possible origin cases for the (so-far) small sample of systems with a robustly measured low metallicity. Therefore, we consider the 4 rejected LLSs to help establish a better strategy for identifying future new metal-poor systems.

Among the 4 rejected LLSs, the case of LLS0106 is the simplest. At redshift $\zab=4.172$, \citet{2015ApJS..221....2P} estimated $\lNHI=19.05 \pm 0.20$ (1$\sigma$) based on visual inspection of theoretical Voigt profiles superimposed on the \lya\ line. When selecting our metal-poor candidates, we confirmed these values, and the non-detections of strong metal lines, using the same MIKE spectrum as \citet{2015ApJS..221....2P}. However, some remaining flux bluewards of LLS0106's Lyman limit can, in fact, be observed in this spectrum, implying an upper-limit on \NHI\ of $\lNHI \leq 17$. Even with no secure metal line detection, the very low \NHI\ would only provide a very high -- and, in this context, not useful -- upper limit on the metallicity of this LLS, so it would not have been selected for Keck/HIRES follow-up. This illustrates the point that establishing the different properties of LLSs in large surveys is, currently, still very dependent on manual, human interaction which, occasionally, is prone to error. More systematic checks of key observables (in this case the existence of detectable flux bluewards of the Lyman limit) should be undertaken to avoid similar, erroneous LLS identifications.

Low spectral resolution can also lead to underestimating the metallicity of an absorber. Moderate resolution ($\sim 40 \kms$) spectra, such as those used in \citet{2015ApJ...812...58C} and \citet{2015ApJS..221....2P}, may not allow metal column density measurements accurate enough, in some cases, to be sure a system is truly a very metal-poor candidate. For the 4 rejected LLSs in \Aref{s:higher}, we note that for important transitions such as \tran{Si}{ii}{1260} or \tran{C}{ii}{1334} only upper-limits on their column densities were provided. Clearly, it is not possible to unambiguously detect weak, narrow ($b \sim 5$\,\kms) metal lines in such low-to-medium resolution spectra. We therefore recommend careful assessment of whether the column density measurements and upper limits derived using such spectra confidently constrain the metallicity to $\lmetal \leq -3$ in each individual absorber.

On a similar note, more detailed information about the metallicity upper limit should be used to select metal-poor candidate LLSs. Our candidates in \Tref{t:log_obs_lls} were selected using estimates of their metallicity upper limits from \citet{2016ApJ...833..283L} and \citet{2015ApJ...812...58C}. These were derived with a method similar to that described in \Sref{s:interesting}, i.e.\ a MCMC sampling to compare observed metal column densities to the predictions of a grid of \textsc{cloudy} photoionisation models. In general (except for LLS1304), the resulting metallicity posterior distribution was very broad for each candidate, so the metallicity upper limit was poorly defined. Also, as stated above, the low-to-medium resolution spectra used in these studies mostly provided upper-limits on the metal column densities, and this strongly impacted the shape of the metallicity posterior distribution. Therefore, to more securely select very metal-poor candidates from low-to-medium resolution spectra, we recommend running a MCMC sampling of a full suite of photoionisation models which cover all conditions found in known LLSs \citep[e.g.][]{2016ApJ...833..283L,2016MNRAS.455.4100F}. Follow-up spectroscopy could then only be conducted if its metallicity distribution shows, with good confidence, an upper-limit consistent with $\lmetal \leq -3$.

Finally, it is also important that the wavelength coverage of a LLS survey is tailored not only to the redshift range, but also the column density range targeted. For systems with $\lNHI \leq 19.0$, the high-order Lyman series and Lyman limit must be covered to ensure the best constraints on \NHI, while the \lya\ line should be sufficient to cover for higher-\NHI\ systems because its damping wings strongly constrain \NHI. Clearly, the SNRs required in the spectral regions of these features also needs careful consideration. The SNR and wavelength coverage requirements for a LLS survey are also impacted by the \HI\ \lya\ forest. Indeed, depending on a LLS's redshift, many important metal lines can fall inside the forest. Therefore, \HI\ blends might prevent estimation of a metal transition's column density at low-to-medium resolution. Furthermore, while high resolution can allow deblending a narrow metal line from a \HI\ forest feature, it can also fail depending on how broad and saturated the \HI\ feature is. To reduce this problem, observing a wavelength range that allows different metal lines to be observed simultaneously is recommended. Obviously, one can select metal transitions that are redwards of the \lya\ forest so they are less likely to be contaminated by unrelated features. While it is common practice to use the strongest available transition of an ion to measure its column density, the weaker ones can also be useful. They have some chance of being less affected by \HI\ forest blends but may also be redwards of it. Nonetheless, for transitions that are much weaker than their ion's strongest one (e.g.\ \tran{Si}{ii}{1526} and $\lambda$1260, respectively), it might not be feasible to obtain a detection or a useful upper-limit when $\lmetal < -3$. For example, detecting \tran{Si}{ii}{1526} in LLS1304 would require $\textrm{SNR} \sim 300$ at 6620\,\AA\ with MagE ($\sim$187 for ESI). Such high SNRs are clearly not reachable in most surveys. For comparison, only $\textrm{SNR} \sim 40$ would be required for HIRES. 

To summarise the discussion above, the following are important building blocks for a survey to identify very metal-poor candidates with low-to-medium resolution spectra:
\begin{itemize}
\item Resolution high enough, for the SNR targeted, to allow column density measurements precise enough (and low enough upper limits for metals) for a maximum metallicity of $\lmetal \leq -3$ to be determined through photoionisation modeling. What constitutes ``enough'' for these various survey parameters needs to be carefully modelled, with simulated spectra, when initially considering a survey, depending on the available telescopes and spectrographs.
\item MCMC sampling of a full suite of photoionisation models of very metal-poor candidates, and to conduct a high-resolution spectroscopic follow-up of the candidates whose metallicity distribution shows an upper-limit consistent with $\lmetal \leq -3$ at high confidence.
\item Regardless of the spectral quality of the data, a reliable estimate of \NHI\ is important.
\end{itemize}
Regarding the high-resolution follow-up observing strategy, we recommend:\begin{itemize}
\item Wavelength coverage of the strong metal transitions (e.g. \tran{Si}{ii}{1260}) and of weaker representatives of the same ions;
\item Wavelength coverage of at least the Lyman limit and \lya\ line for LLSs with $\lNHI \leq 19.0$\ and $\geq 19.0$, respectively;
\item High enough SNR and resolving power for these regions (SNR $\geq 25$--30 and $R \sim 35000$) to robustly measure the column density of absorption lines and resolve important velocity structure.
\end{itemize}

As a final note, we add that it is also important to apply consistent methodologies to establish a large sample of near-pristine absorbers. In \Sref{s:discussion} we mentioned that only two other near-pristine LLSs had previously been found: LLS1249 and LLS0958A by \citet{2016MNRAS.457L..44C} and \citet{2016ApJ...833..283L}, respectively, who measured their metallicities to be $\lmetal\approx-3.4$. Due to the availability of a high-resolution HIRES spectrum of LLS0958A in the KODIAQ sample of \citet{2015AJ....150..111O}, we conducted a reassessment of its properties, particularly its metallicity, with the same methodology we applied to the other 8 LLSs described in this work. Details are provided in \Aref{s:lls0958A}. We find $\lmetal = -2.91 \pm 0.26$ for LLS0958A, which is substantially higher than \citet{2016ApJ...833..283L}'s value of $-3.35 \pm 0.05$. However, LLS0958A's low metallicity is still intriguing given that it could be in an isolated, intergalactic-like environment in \citet{2016MNRAS.462.1978F}, as we discussed in \Sref{s:origin}.

\section{Conclusions}\label{s:conclusion}

In this work we described a dedicated search for near-pristine LLSs in which we selected a total of 8 LLSs from large, low-to-medium resolution spectroscopic surveys \citep{2015ApJ...812...58C,2015ApJS..221....2P}. The 8 candidates exhibited no, or only weakly detected metal absorption lines, suggesting a very low metallicity, $\lmetal < -3$, derived in the photoionisation modelling of \citet{2015ApJ...812...58C} and \citet{2016MNRAS.455.4100F}. To confirm their very metal-poor nature, we observed them in a Keck/HIRES high-resolution spectroscopy campaign from 2016 to 2017. One of these 8 LLSs, LLS1723, did not display metal lines in the follow-up spectra, as presented in \citet{2019MNRAS.483.2736R}. All the remaining 7 LLSs, presented in this work, exhibited detectable, albeit weak, metal absorption lines in the Keck/HIRES spectra.

To make more accurate metallicity measurements for these 7 near-pristine candidates, we first derived metal and hydrogen column densities from the Keck/HIRES spectra using multi-component Voigt profile fits. These were compared to a grid of \textsc{cloudy} photoionisation models, sampled via an MCMC algorithm, to measure the metallicities. This confirmed the discovery of 3 new near-pristine LLSs -- LLS0344, LLS1153, LLS1156 -- with final metallicities $\textrm{[Si/H]} = \lmetal = -3.00 \pm 0.26$, $ -3.05  \pm 0.26$ and $ -2.94 \pm 0.26$ (95\% confidence), respectively. Unfortunately, only two of these systems exhibited more than one detected metal ion species: LLS0334's [C/Si] ratio is consistent with solar, while its [Al/Si] ratio is constrained to be less than 0.14\,dex below solar; for LLS1156, the [Al/Si] ratio is consistent with solar. These constraints on the nucleosynthetic profile of these LLSs are not detailed enough to distinguish between PopIII and later population progenitor models. Even the low [Al/Si] ratio for LLS0344 did not strongly constrain the progenitor mass within the PopIII scenarios explored. For the remaining four LLSs, we found that the previous metallicity estimates, prior to this work, were underestimated; our Keck/HIRES spectra and photoionisation modelling strongly suggest metallicities $\lmetal \ga -2.5$ -- see \Aref{s:higher}.

With the discovery of 3 new systems via our dedicated search, and the first two serendipitous discoveries (LLS1249 by \citealt{2016MNRAS.457L..44C} and LLS0958A by \citealt{2016ApJ...833..283L}, noting our reassessment of the latter in \Aref{s:lls0958A}), there are now 5 LLSs with metal detections known to be consistent with $\lmetal \leq -3$. \Fref{f:LLS} summarises their metallicities in the context of the general LLS population. The near-pristine systems have clearly lower metallicities than the vast majority of LLSs and are relatively rare. Given the samples from which these 5 systems and the 3 apparently metal-free LLSs \citep{2011Sci...334.1245F,2019MNRAS.483.2736R} were drawn, we estimated that systems with $\lmetal \leq -3$ constitute only $\sim$4\% of LLSs. However, it is not clear from the metallicity distribution alone whether they are truly a separate population, with a different origin, to the higher metallicity LLSs. We considered the possible origins for near-pristine LLSs, based on results from existing simulations in the literature, in \Sref{s:origin}. Systems arising in the high-density regions of the IGM (i.e.\ not near galaxies) may well be the chemical remnants of PopIII stars with no contamination from later generations. However, if the near-pristine LLSs arise in circumgalactic environments, as seems most likely for the new systems, given their higher column densities [i.e.\ $\lNHI > 19$], then their metallicites are consistent with being polluted by PopIII or PopII/I stars at very early times, or by PopII/I supernovae from the nearby galaxy/galaxies, at or just prior to, the epoch at which we observe them. Therefore, the origin of the 3 new near-pristine systems remains unclear.

Simulations may be able to help clarify the origins of near-pristine LLSs. The major challenge in this context is to resolve very low-density gas. \citet{2019ApJ...873..129P} made concerted progress in this direction for circumgalactic gas, but simulating the IGM at high-enough mass resolution will be substantially more difficult. Testing the predictions of these simulations will also require a larger survey for near-pristine and apparently metal free LLSs, with well-controlled selection effects. For example, a basic test would be to compare the redshift and column density dependence of the number density of such LLSs with simulations of different origin scenarios. Similarly, discovering near-pristine LLSs with a suitably large variety of detected metal species to allow detailed nucleosynthetic modelling, will also likely require a significantly larger sample. Our success in discovering 3 near-pristine LLSs in a targeted way opens the possibility for creating a well-defined, larger statistical sample in the future to conduct such tests. We recommended observational strategies for future surveys in \Sref{s:strategy}.

\section*{Acknowledgements}

We thank Ryan Cooke for assisting with aspects of the data reduction. PFR acknowledges supports through a Swinburne University Postgraduate Research Award (SUPRA) PhD scholarship, and travel support through the International Telescopes Support Office. MTM thanks the Australian Research Council for \textsl{Discovery Project} grant DP130100568 which supported this work. MF acknowledges funding from the European Research Council (ERC) under the European Union's Horizon 2020 research and innovation programme (grant agreement No 757535) and funding support by Fondazione Cariplo, grant No 2018-2329.

The data presented herein were obtained at the W.\ M.\ Keck Observatory, which is operated as a scientific partnership among the California Institute of Technology, the University of California and the National Aeronautics and Space Administration. The Observatory was made possible by the generous financial support of the W.\ M.\ Keck Foundation. The authors wish to recognise and acknowledge the very significant cultural role and reverence that the summit of Maunakea has always had within the indigenous Hawaiian community. We are most fortunate to have the opportunity to conduct observations from this mountain. Australian access to the W.\ M.\ Keck Observatory has been made available through Astronomy Australia Limited via the Australian Government's National Collaborative Research Infrastructure Strategy, via the Department of Education and Training, and an Australian Government astronomy research infrastructure grant, via the Department of Industry, Innovation and Science. This research has made use of the Keck Observatory Archive (KOA), which is operated by the W.\ M.\ Keck Observatory and the NASA Exoplanet Science Institute (NExScI), under contract with the National Aeronautics and Space Administration. Our analysis made use of \textsc{astropy} \citep{2013A&A...558A..33A}, \textsc{matplotlib} \citep{Hunter2007}, and \textsc{barak} (\url{https://github.com/nhmc/Barak}).

\section*{Data Availability}

All raw exposures of all quasars studied in this work are publicly
available in the Keck Observatory Archive (KOA). The final, combined
spectrum of each quasar is available on request to the corresponding
author (MTM).


%\bibliographystyle{mnras}
%\bibliography{biblio}{}

%%%%%%%%%%%%%%%%%%%%%%%%%%%%%%%%%%%%%%%%%%%%%%%%%%
%%%%%%%%%% SUPPORTING INFORMATION %%%%%%%%%%%%%%%%

\section*{Supporting Information}

\noindent Supplementary figures are available at MNRAS online.\\

\noindent\textbf{LLS0952\_column\_densities.pdf}\\
\textbf{LLS0952\_parameter\_distributions.pdf}\\
\textbf{LLS1304\_parameter\_distributions.pdf}\\
\textbf{LLS1304\_column\_densities.pdf}\\
\textbf{LLS2241\_column\_densities.pdf}\\
\textbf{LLS2241\_parameter\_distributions.pdf}\\

\noindent Please note: Oxford University Press is not responsible for the content or functionality of any supporting materials supplied by the authors. Any queries (other than missing material) should be directed to the corresponding author for the article.

%%%%%%%%%%%%%%%%%%%%%%%%%%%%%%%%%%%%%%%%%%%%%%%%%%
%%%%%%%%%%%%%%%%% APPENDICES %%%%%%%%%%%%%%%%%%%%%
\appendix

\section{Higher metallicity Lyman Limit Systems in the sample}\label{s:higher}

As explained in \Sref{s:cloudy_general}, only 3 of the near-pristine candidates have measured metallicities consistent with $\lmetal \leq -3$: LLS0344, LLS1153, LLS1156. Among the remaining 4, LLS0106 has \lNHI\ well below the normal LLS definition of $17.2$, and the others (LLS0952, LLS1304, LLS2241) are consistent with $\lmetal \ga -2.5$. The analysis of their absorption features is described very briefly below for completeness. Plots of the photoionisation results for each system (akin to Figs.\ \ref{f:fiducial_model_lls0344} and \ref{f:distribution_0344_fiducial} for LLS0344) are available as supporting information online. \Tref{t:logNhigher} summarises the metal and \HI\ column density measurements and upper limits, and metallicity measurements for each system. These four systems were considered in \Sref{s:strategy} when discussing how to improve searches for very metal-poor candidates in future surveys. Finally, for LLS2241, we find that its \HI\ column density is $\lNHI \approx 20.3$, consistent with the normal DLA definition, so this system may be better considered as a very low metallicity DLA rather than a LLS.

\newcommand\oldtabcolsep{\tabcolsep}
\setlength{\tabcolsep}{0.55em}
\begin{table}
\caption{Same as \Tref{t:logN} but for the 4 LLSs found to have higher metallicities than near-pristine systems ($\lmetal \leq -3$). \textsuperscript{c}We did not calculate a metallicity for LLS0106 because its \NHI\ was below that of the normal definition for a LLS; see \Sref{s:lls0106}.}
\label{t:logNhigher}
\begin{center}
\begin{tabular}{l|cccc}\hline
 Ion                                       & \multicolumn{4}{c}{$\log_{10} (N/\mathrm{cm^{-2}})$} \\
 \hline
                                           & LLS0106                & LLS0952                & LLS1304               & LLS2241 \\
 \ion{Si}{ii}                              & $11.75 \pm 0.15      $ &  $ 13.55 \pm 0.04    $ & $ 12.55 \pm 0.01    $ &$ 13.22 \pm 0.05$  \\
 \ion{Si}{iv}                              & -                      & -                      & $ 13.01 \pm 0.02$     &-                  \\
 \ion{C}{ii}                               & -                      & $ 14.41 \pm 0.17     $ & $ 13.30 \pm 0.03    $ & -                 \\
 \ion{C}{iv}                               & -                      & -                      & $ 13.15 \pm 0.06    $ & -                 \\
 \ion{Fe}{ii}                              & -                      & $13.59 \pm 0.05$       & -                     & -                 \\
 \ion{Al}{ii}                              & -                      & -                      & -                     & $ 12.08 \pm 0.04$ \\
 \ion{Al}{iii}                             & -                      & -                      & $\leq 11.79$          & $ \leq 12.35$     \\
 \ion{O}{i}                                & -                      & $14.99 \pm 0.18      $ & -                     & -                 \\
 \HI                                       & $\leq 17$              & $19.80 \pm 0.10$        & $17.90 \pm 0.10$     & $20.3 \pm 0.10$   \\
\hline
 [Si/H]                                    & -\textsuperscript{c}   & $ -1.85 \pm 0.26$      &$-2.30 \pm 0.26$       & $-2.75 \pm 0.26$  \\
\hline
\end{tabular}
\end{center}
\end{table}
\setlength{\tabcolsep}{\oldtabcolsep}

\subsection{LLS0106}\label{s:lls0106}

\begin{figure}
\centering
\includegraphics[width=0.95\columnwidth]{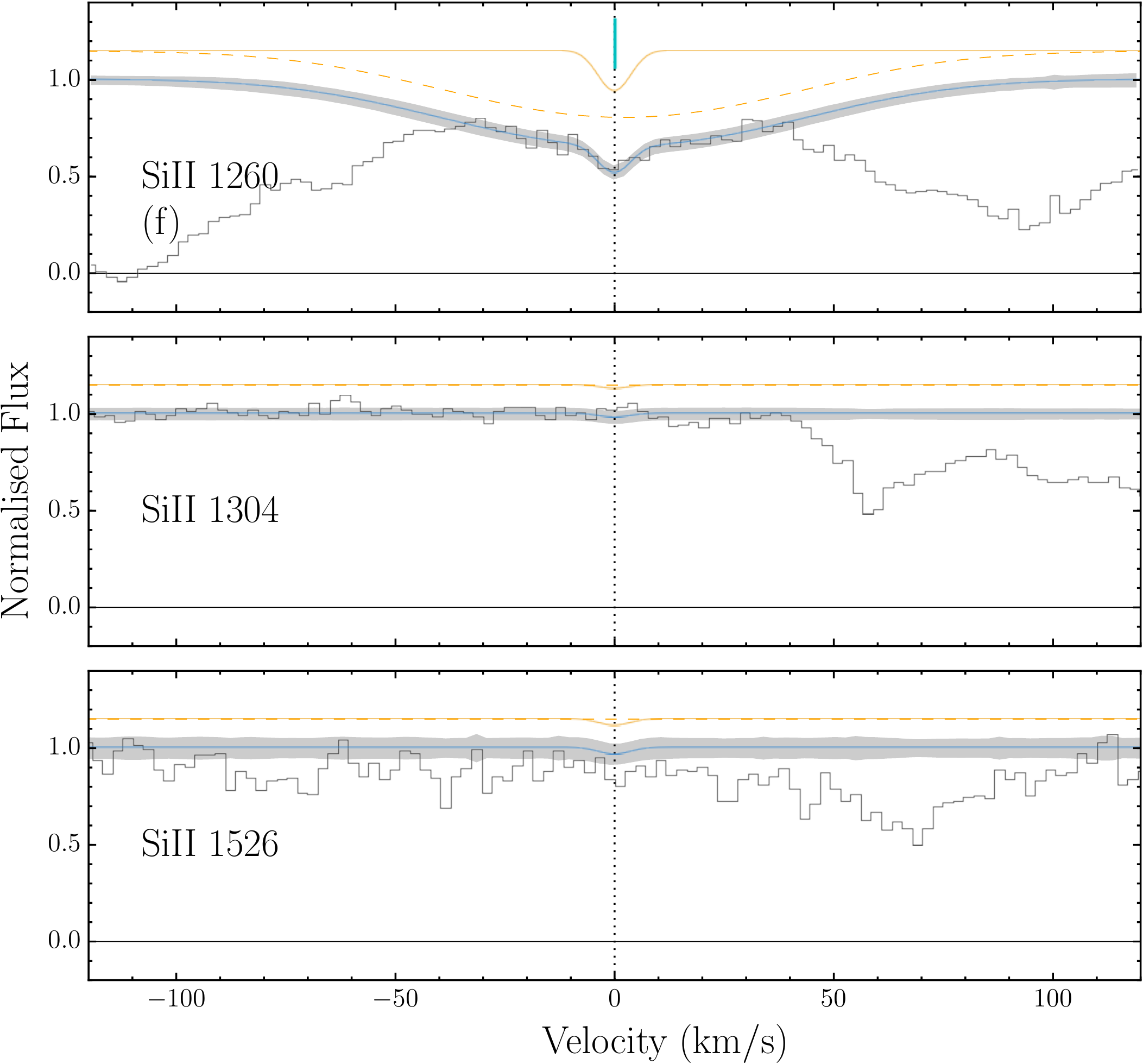}
\caption{Same as \Fref{f:lls0344_metals} but for J0106$+$0048 and LLS0106 with the zero velocity redshift set at $\zab=4.172$. For \tran{Si}{ii}{1304} and \tran{Si}{ii}{1526}, the column densities used for the Voigt profile fits are the same used for \tran{Si}{ii}{1260}, the strongest \SiII\ transition.}
\label{f:lls0106_metals}
\end{figure}

LLS0106 was first identified by \citet{2015ApJS..221....2P} towards the $\zem=4.43$ quasar SDSS J010619.20$+$004823.3 (hereafter J0106$+$0048), based on absorption features at redshift $\zab=4.172$, using a Magellan/MIKE spectrum. At this redshift \NHI\ was estimated at $\lNHI=19.05 \pm 0.20$ along with $b=30$\,\kms\ from the \lya\ and \lyb\ lines. \citet{2015ApJS..221....2P} did not detect any metal lines, finding upper limits for \doublet{C}{iv}{1548/1550}, \tran{Si}{ii}{1304}, \doublet{Si}{iv}{1393/1402}\ and \tran{O}{i}{1302}. In our HIRES spectrum, we tentatively detect \tran{Si}{ii}{1260}, the strongest \SiII\ line, in a broad \lya\ forest feature and provide the corresponding column density \Tref{t:logNhigher}. \Fref{f:lls0106_metals} depicts this \tran{Si}{ii}{1260} feature and the weaker (smaller oscillator strength) \tran{Si}{ii}{1304} and \tran{Si}{ii}{1526} transitions for comparison. The weaker \SiII\ lines were also used to constrain $N(\textrm{Si})$ in the \textsc{vpfit} modelling.

\begin{figure}
\centering
\includegraphics[width=0.95\columnwidth]{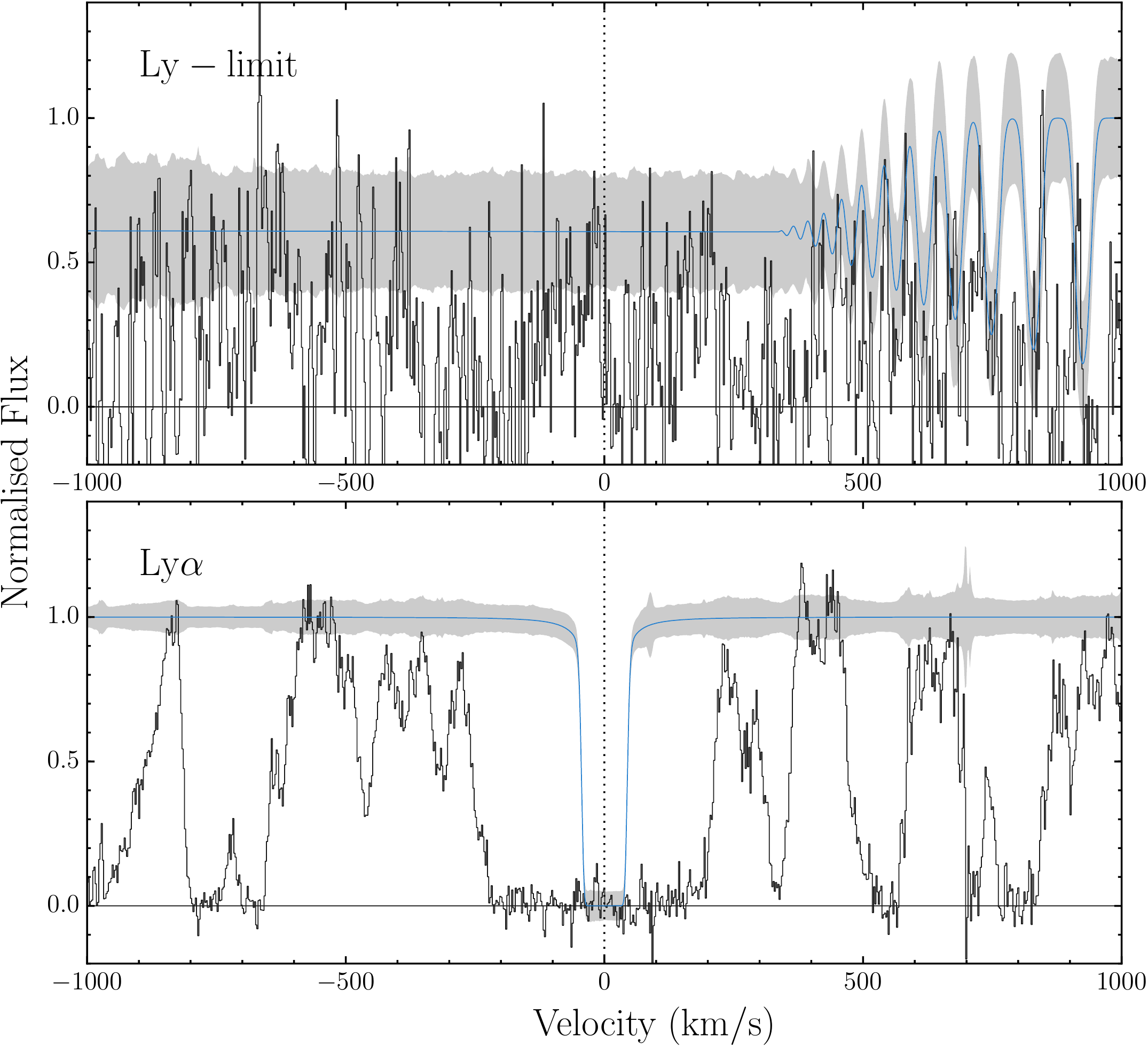}
\caption{Same as \Fref{f:lls0344_deut} but for J0106$+$0048 and LLS0106 with the zero velocity redshift set at $\zab=4.172$ and profile model (blue solid line) parameters $\lNHI = 17$ and $b = 15$\,\kms.}
\label{f:lls0106_hi}
\end{figure}

Our fiducial \HI\ model is depicted in \Fref{f:lls0106_hi}: $\lNHI = 17$ and $b = 15$\,\kms. Despite the previous estimate of $\NHI\geq 19$ \citep{2015ApJS..221....2P}, we find that there is significant flux bluewards of the Lyman limit, as can be seen in \Fref{f:lls0106_hi}. The nominal upper limit value on \NHI\ of 
$\lNHI \leq 17$ is therefore conservative in this case, demonstrating that this system is almost certainly not a true LLS considering the conventional threshold of $\lNHI \geq 17.2$. Therefore, we do not proceed to attempt a metallicity measurement for LLS0106.

\subsection{LLS0952}

\begin{figure}
\centering
\includegraphics[width=0.95\columnwidth]{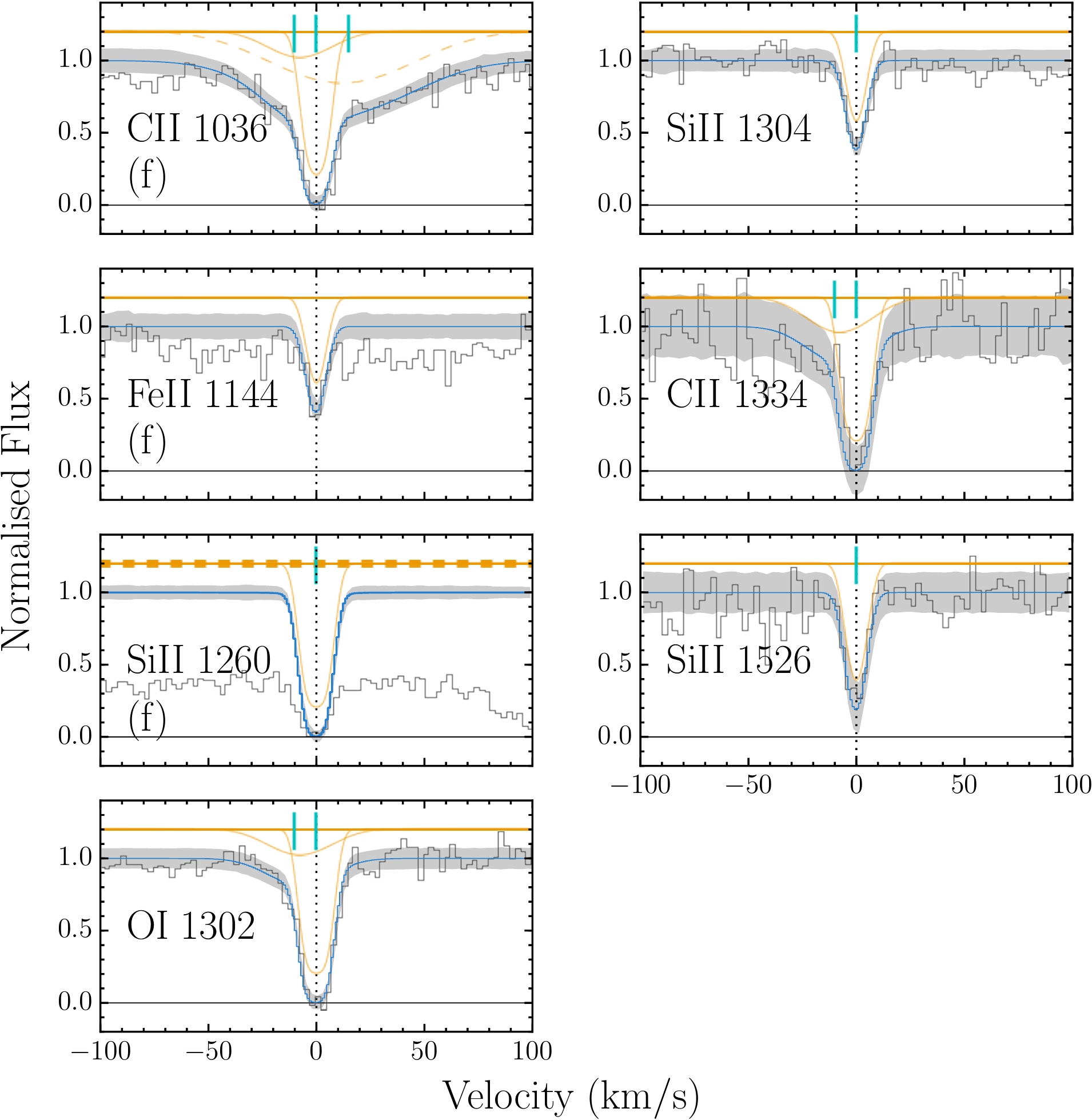}
\caption{Same as \Fref{f:lls0344_metals} but for J0952$+$3329 and LLS0952 with the zero velocity redshift set at $\zab=3.213$.}
\label{f:lls0952_metals}
\end{figure}

LLS0952 was first identified by \citet{2015ApJS..221....2P} towards the $\zem=3.396$ quasar SDSS J095256.41$+$332939.0 (hereafter J0952$+$3329), based on absorption features at redshift $\zab=3.212$, using a Keck/ESI spectrum. \NHI\ was estimated at $\lNHI=19.90 \pm 0.20$ (1$\sigma$) along with $b=30$\,\kms\ from the \lya\ line, the main constraint coming from its damping wings. \citet{2015ApJS..221....2P} reported detections of \tran{C}{ii}{1334}, \doublet{Si}{iv}{1393/1402}\ and \tran{O}{i}{1302}, with only upper limits determined for other singly-ionised species (\ion{Si}{ii}, \ion{Al}{ii}, \ion{Fe}{ii}) and higher ions (\ion{Al}{iii} and \ion{Si}{iv}). Our Keck/HIRES spectrum provides strong detections of \ion{C}{ii}, \ion{O}{i}, \ion{Si}{ii} and even \ion{Fe}{ii}, with column density measurements listed in \Tref{t:logNhigher}, but we do not detect the high ions (\ion{C}{iv} and \ion{Si}{iv}). \Fref{f:lls0952_metals} depicts the strongest transitions of the most abundant metal species in this LLS at $\zab=3.213$, clearly showing a low-ionisation phase for LLS0952: the different transitions depicted are aligned in velocity space and comprise the same, simple velocity structure. Therefore, for the fiducial model of LLS0952, all the transitions depicted are considered to be produced by the same phase as the \HI\ absorption.

\begin{figure}
\centering
\includegraphics[width=0.95\columnwidth]{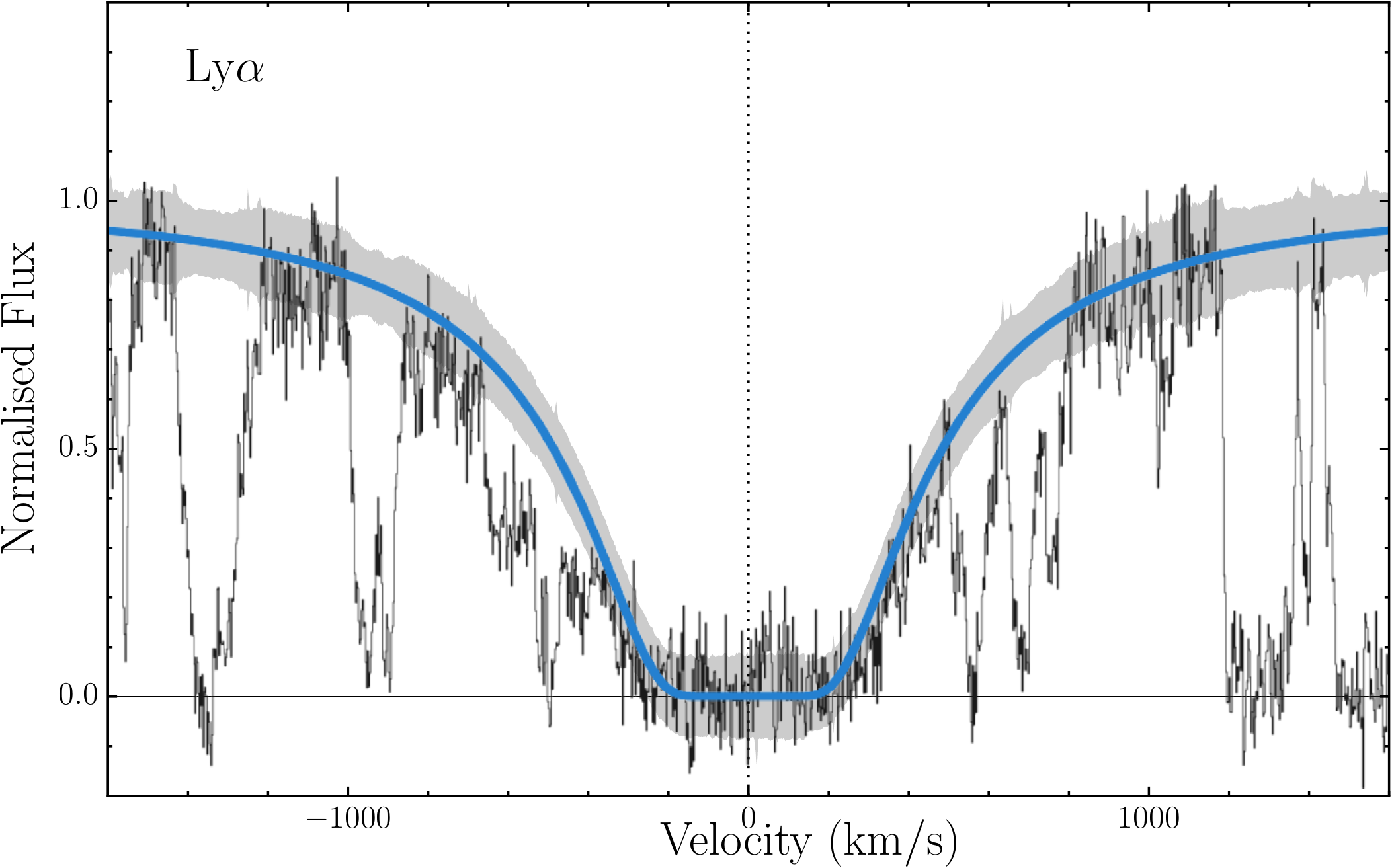}
\caption{Same as \Fref{f:lls0344_deut} but for J0952$+$3329 and LLS0952 with the zero velocity redshift set at $\zab=3.213$ and profile model (blue solid line) parameters $\lNHI=19.80 \pm 0.10$, and $b = 15$\,\kms.}
\label{f:lls0952_hi}
\end{figure}

Our fiducial \HI\ model is depicted in \Fref{f:lls0952_hi} with $\lNHI=19.80 \pm 0.10$ and $b = 15$\,\kms. \Fref{f:lls0952_hi} clearly shows that the \lya\ transition has damping wings and will therefore strongly constrain the column density. The uncertainty in \NHI\ is determined in a similar way to that for LLS0344, taking into account the scatter in flux and uncertainty in the continuum placement.

\subsection{LLS1304}

\begin{figure}
\centering
\includegraphics[width=0.95\columnwidth]{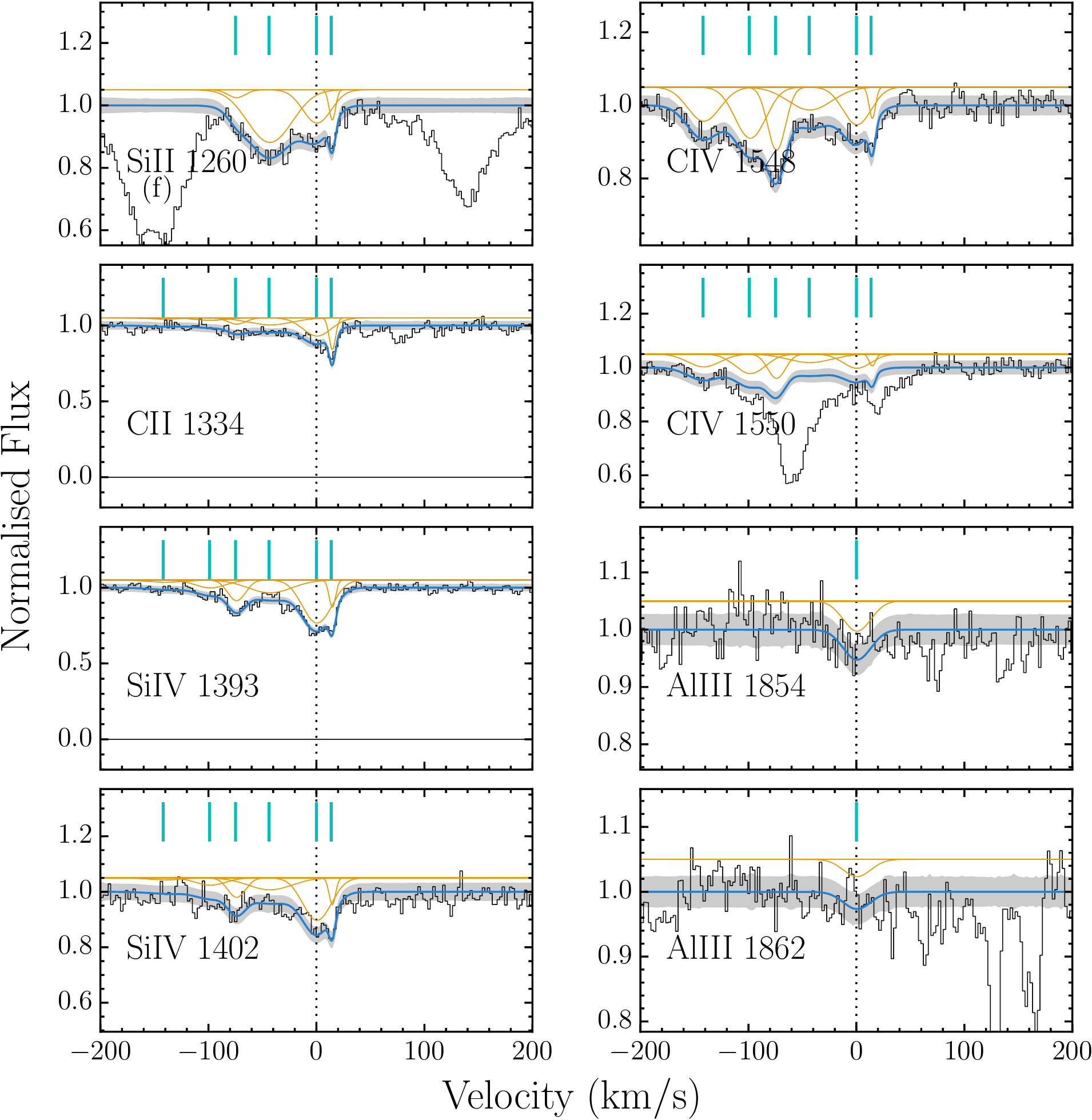}
\caption{Same as \Fref{f:lls0344_metals} but for J1304$+$0239 and LLS1304 with the zero velocity redshift set at $\zab = 3.3369$. For \doublet{Al}{iii}{1854/1862}, the column densities used for the Voigt profile fits correspond to a 2$\sigma$ upper limit derived using the apparent optical depth method. Note that \tran{C}{iv}{1550} is blended with \tran{C}{iv}{1548} absorption from a higher redshift $\zab =3.3433$ system, the absorption profiles of which are not shown.}
\label{f:lls1304_metals}
\end{figure}

LLS1304 was first identified by \citet{2015ApJ...812...58C} towards the $\zem=3.651$ quasar SDSS J130452.57$+$023924.8 (hereafter J1304$+$0239), based on absorption features at redshift $\zab=3.3369$, using a Magellan/MagE spectrum. A plausible range of values for \NHI\ was estimated based on the absence of detected flux bluewards of the Lyman limit and the apparent lack of \lya\ damping wings: $17.9<\lNHI<18.7$. \citet{2015ApJ...812...58C} reported detections of the high ions (\ion{C}{iv} and \ion{Si}{iv}) and non-detections (upper limits) of \ion{Si}{ii} and \ion{Al}{iii}. Our HIRES spectrum provided detections of two low ions (\ion{C}{ii}, \ion{Si}{ii}) and the same high ions, as shown in \Fref{f:lls1304_metals}. \tran{Si}{ii}{1260} is in the Lyman forest of J1304$+$0239, between two \HI\ blends, and this almost certainly biases the \ion{Si}{ii} column density estimate to higher values. We therefore treated it as an upper limit in the photoionisation modelling. The low and high ions have very similar velocity structures, suggesting they may be part of the same phase. However, the metallicity obtained from the photoionisation modelling does not depend strongly on this assumption.

\begin{figure}
\centering
\includegraphics[width=0.95\columnwidth]{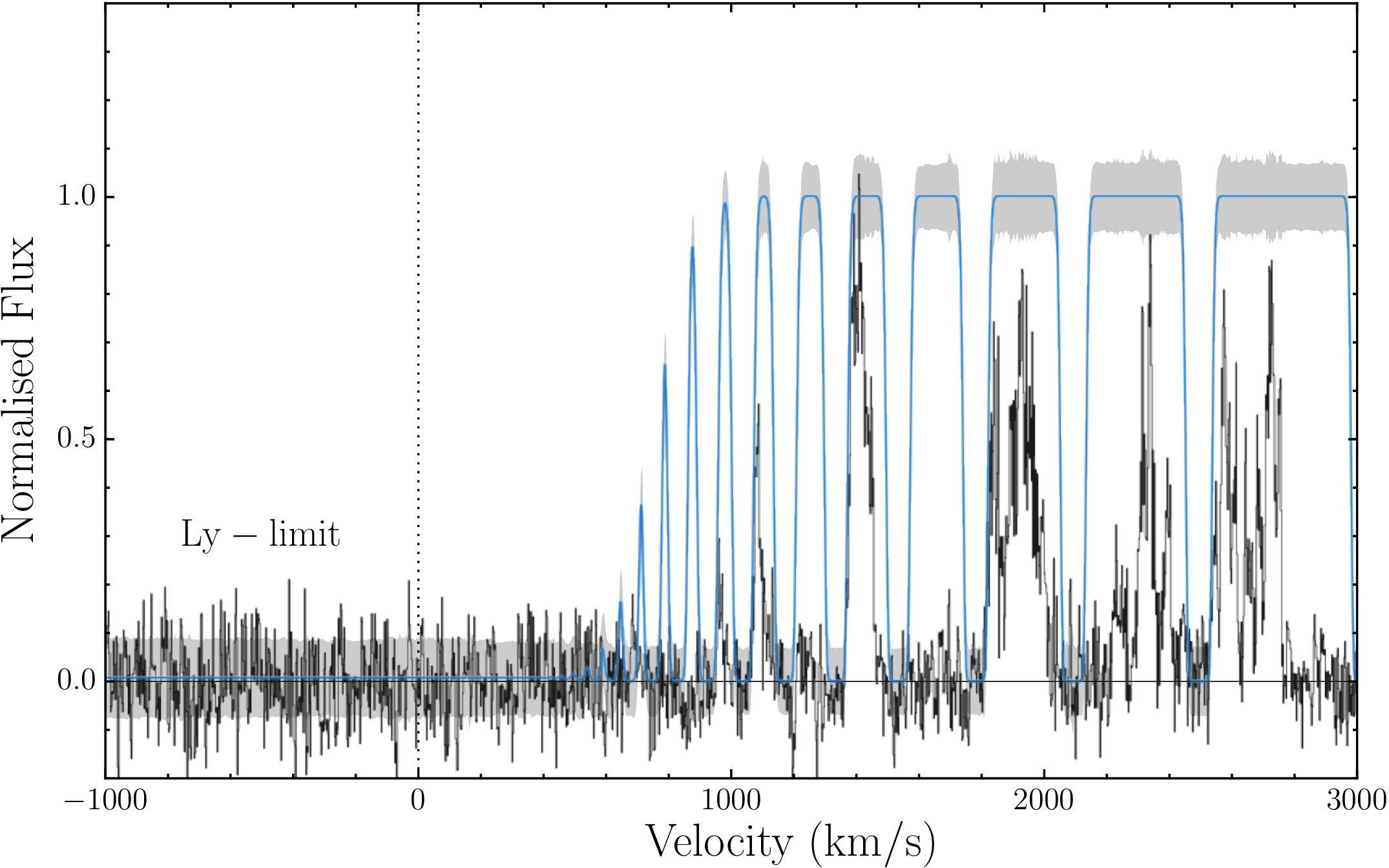}
\caption{Same as \Fref{f:lls0344_deut} but for J1304$+$0239 and LLS1304 with the zero velocity redshift set at $\zab = 3.33695$ and profile model (blue solid line) parameters $\lNHI=17.90 \pm 0.10$, and $b = 20$\,\kms.}
\label{f:lls1304_hi}
\end{figure}

\Fref{f:lls1304_hi} shows our fiducial \HI\ model with $\lNHI = 17.90 \pm 0.10$ and $b = 20$\,\kms. There is no apparent flux bluewards of the Lyman limit so we proceed by following \citet{2015ApJ...812...58C} to establish a plausible \NHI\ range. The upper limit is defined by the \lya\ and $\beta$ lines, which show no apparent damping wings. The lower limit on \NHI\ (from the Lyman limit), $\lNHI = 17.90 \pm 0.10$, provides the highest (i.e.\ most conservative) estimate of LLS1304's metallicity. The uncertainty on \NHI\ is determined by taking into account the scatter in flux in the HIRES spectrum, and the continuum placement.

\subsection{LLS2241}

\begin{figure}
\centering
\includegraphics[width=0.95\columnwidth]{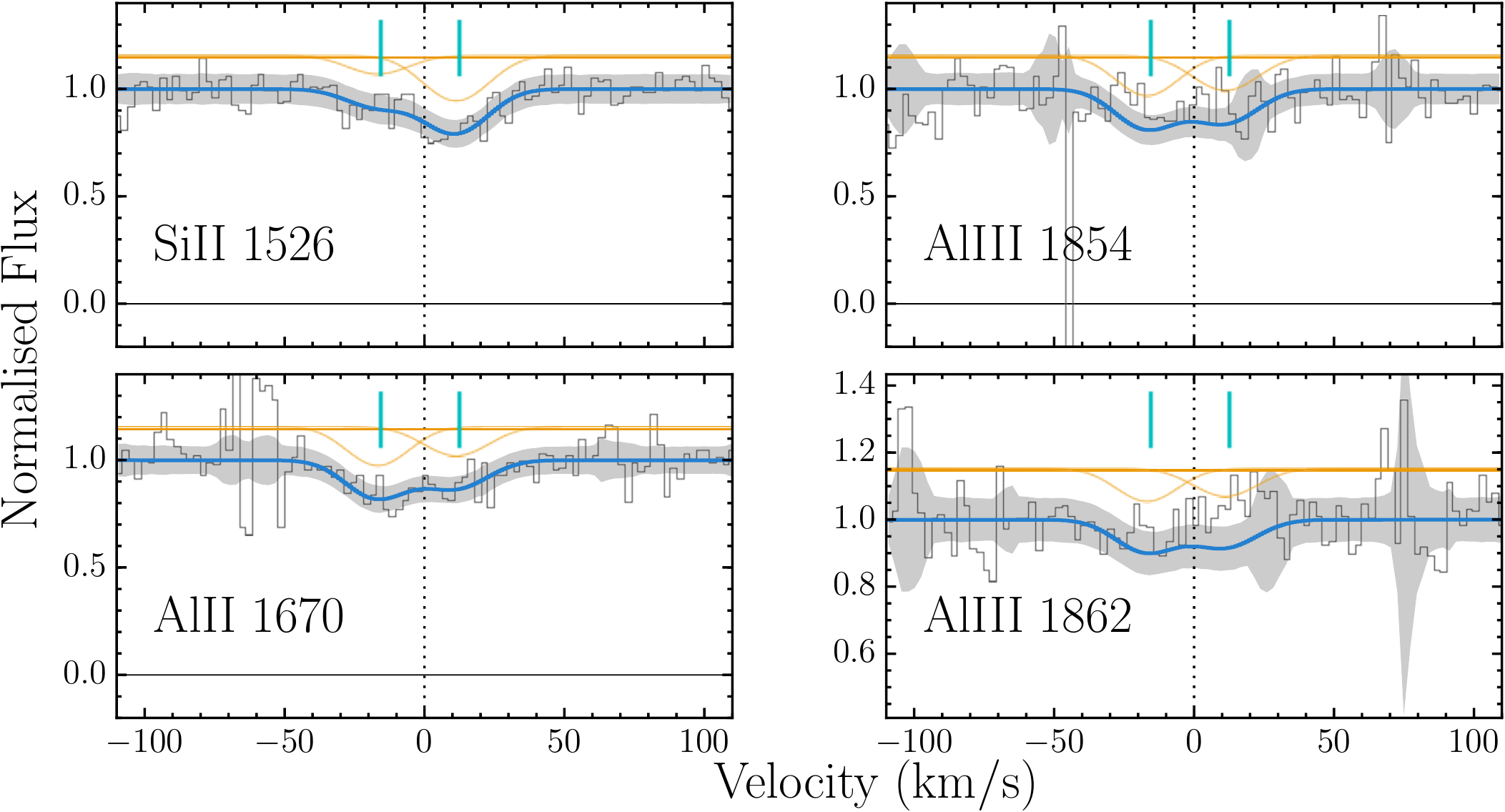}
\caption{Same as \Fref{f:lls0344_metals} but for J2241$+$1352 and LLS2241 with the zero velocity redshift set at $\zab = 3.65492$. For \doublet{Al}{iii}{1854/1862}, the column densities used for the Voigt profile fits correspond to a 2$\sigma$ upper limit derived using the apparent optical depth method.}
\label{f:lls2241_metals}
\end{figure}

LLS2241 was first identified by \citet{2015ApJS..221....2P} towards the $\zem=4.470$ quasar SDSS J224147.70$+$135203.0 (hereafter J2241$+$1352), based on absorption features at redshift $\zab=3.65393$, using a Keck/ESI spectrum. Damping wings in the \lya\ profile indicated $\lNHI=20.20 \pm 0.20$ with $b=30$\,\kms. \citet{2015ApJS..221....2P} reported no metal detections, with upper limits determined for \ion{Si}{ii}, \ion{Al}{ii}, \ion{Fe}{ii}, \ion{Al}{iii} and \ion{C}{iv}. We detect only low ions \ion{Si}{ii} and \ion{Al}{ii} in our Keck/HIRES spectrum, with an upper limit determined for \ion{Al}{iiI}, as detailed in \Tref{t:logNhigher}, and depicted in \Fref{f:lls2241_metals}.  \doublet{C}{iv}{1548/1550}\ falls in a region affected strongly by telluric absorption, while the wavelengths where \doublet{Si}{iv}{1393/1402}\ fall are heavily crowded with \lya\ forest lines. We associate the detected singly-ionised metal species with the \HI\ content to estimate the metallicity.

\begin{figure}
\centering
\includegraphics[width=0.95\columnwidth]{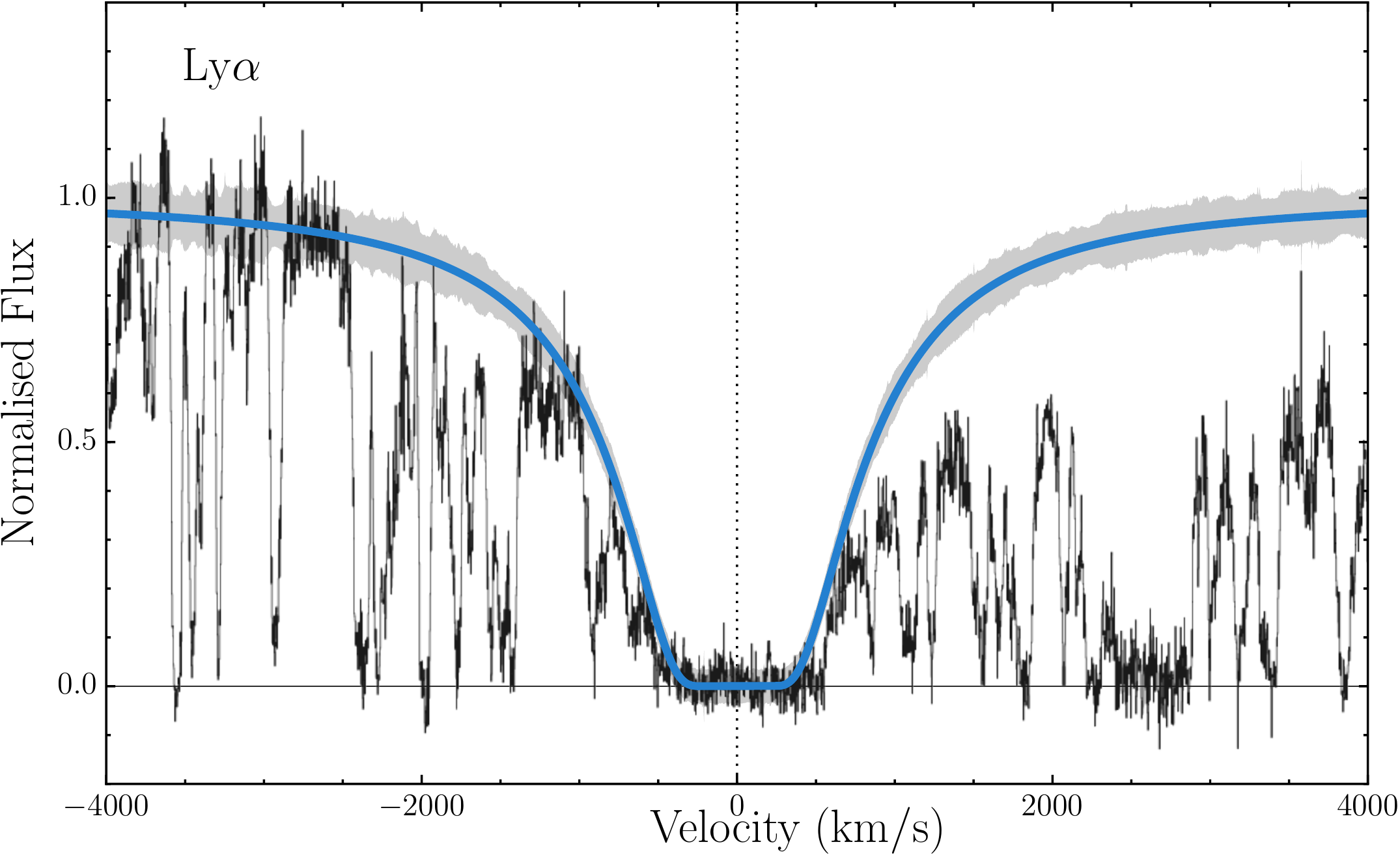}
\caption{Same as \Fref{f:lls0344_deut} but for J2241$+$1352 and LLS2241 with the zero velocity redshift set at $\zab = 3.65492$ and profile model (blue solid line) parameters $\lNHI=20.30 \pm 0.10$, and $b = 30$\,\kms.}
\label{f:lls2241_hi}
\end{figure}

\Fref{f:lls2241_hi} depicts our fiducial \HI\ model with $\lNHI=20.30 \pm 0.10$ and $b = 30$\,\kms. The \lya\ damping wings are very prominent in LLS2241, and these strongly constrain \NHI. The uncertainty on \NHI\ is determined by taking into account the scatter in flux in the HIRES spectrum and the continuum placement.

\section{Reassessment of LLS0958A's metallicity}
\label{s:lls0958A}

Here we measure the metallicity of LLS0958A, which was previously identified as a near-pristine absorption system by \citet{2016ApJ...833..283L}, using the same approach as used for our other LLSs in this work.

\subsection{Metal line column densities}

LLS0958A was first identified by \citet{1990ApJS...74...37S} towards the $\zem=3.297$ quasar Q095852.3$+$120243 (hereafter Q0958$+$1202), based on absorption features at redshift $\zab=3.2228$, using a Hale Telescope/Double Spectrograph \citep{1982PASP...94..586O} spectrum. High-resolution spectroscopy with HIRES was later obtained by \citet{2011Sci...334.1245F}, establishing $\zab=3.223194 \pm 0.000002$, $\lNHI=17.36 \pm 0.05$ and $b=20.4$\,\kms. These values were estimated from the flux decrement at the Lyman limit. In the same spectrum, \citet{2016ApJ...833..283L} detected \tran{Si}{ii}{1206} and 1260, and the high ion doublets (\ion{C}{iv} and \ion{Si}{iv}), and determined upper limits for \tran{C}{ii}{1334} and \tran{Al}{ii}{1670}. The column densities were measured using the apparent optical depth method over a velocity range of $-15$--$+30$\,\kms\ which included only the strongest metal absorption features; a range $-15$--$+15$\,\kms\ was used to determine the upper limits for non-detections.

\begin{figure}
\centering
\includegraphics[width=0.95\columnwidth]{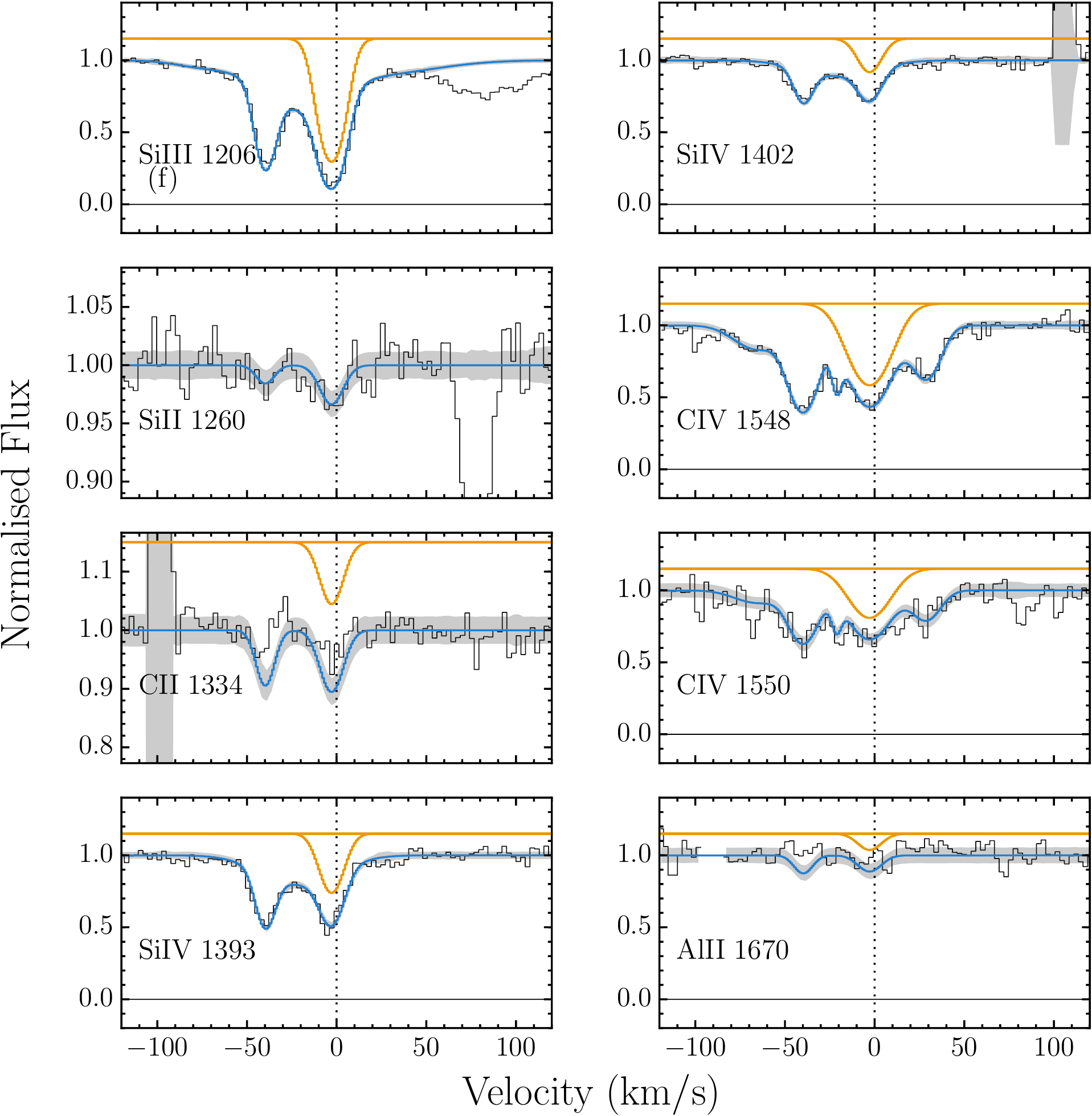}
\caption{Similar to \Fref{f:lls0344_deut} but for Q0958$+$1202 and LLS0958A with the zero velocity redshift set at $\zab = 3.223194$. However, in this case the blue solid lines here represent our fit to the entire metal absorption observed, with total column densities provided in the first column of \Tref{t:logN0958}, while the orange lines show just the component at 0\,\kms\ in our model, with column densities in the second column of \Tref{t:logN0958}. For \tran{Al}{ii}{1670}, the column density used for the profile fit corresponds to a 2$\sigma$ upper limit derived using the apparent optical depth method.}
\label{f:lls0958_metals}
\end{figure}

We confirmed the metal detections and non-detections of \citet{2016ApJ...833..283L} in the HIRES spectrum of LLS0958A from the KODIAQ sample \citep{2015AJ....150..111O}. \Fref{f:lls0958_metals} depicts the strongest transitions of the most abundant metal species at $\zab = 3.223194$, and we list the detections and upper limits in the first column of \Tref{t:logN0958}. \Fref{f:lls0958_metals} illustrates how our reassessment of LLS0958A's metallicity differs from \citeauthor{2016ApJ...833..283L}'s: consistent with our approach to other LLSs in this work, we included all of the detected metal absorption (blue curves); by contrast, \citeauthor{2016ApJ...833..283L} effectively included only the strongest metal component in their narrower velocity range. Given the lack of information about the hydrogen velocity structure offered by the Lyman series lines (i.e.\ they are too broad), we did not find evidence that the \HI\ content could be associated with only one of the main metal-line components. Obviously, including all the metal absorption implies larger total metal column-densities: the other main absorption feature, at $-40$\,\kms\ in \Fref{f:lls0958_metals}, clearly contains a comparable column density as the main feature at 0\,\kms. This is evident in \Tref{t:logN0958} which compares our measured column densities with those of \citeauthor{2016ApJ...833..283L}. Our larger metal column densities imply a higher metallicity estimate for LLS0958A, as calculated below. \Tref{t:logN0958} also provides the column density contained in the main component at 0\,\kms\ in our model, as depicted by the orange profiles in \Fref{f:lls0958_metals}. These values are much closer to the those derived from \citeauthor{2016ApJ...833..283L} using the apparent optical depth method.

Our fiducial photoionisation model of LLS0958A assumes that only the low ions in \Fref{f:lls0958_metals} are associated with the bulk of the \HI\ content. This is due to the apparent difference in velocity structure displayed by \ion{C}{iv} compared to the low ions. We note that \citeauthor{2016ApJ...833..283L} assumed that both the low and high ions are in the same phase and are all associated with the \HI\ content. Indeed, this may be justified because the \ion{Si}{iv} velocity structure appears more similar to that of the low ions. Therefore, we test the effect of including the high ions in our photoionisation modelling below (\Sref{ss:lls0958_phot}) and find the metallicity is fairly insensitive to this assumption.

\subsection{\HI\ column density}

\begin{figure}
\centering
\includegraphics[width=0.95\columnwidth]{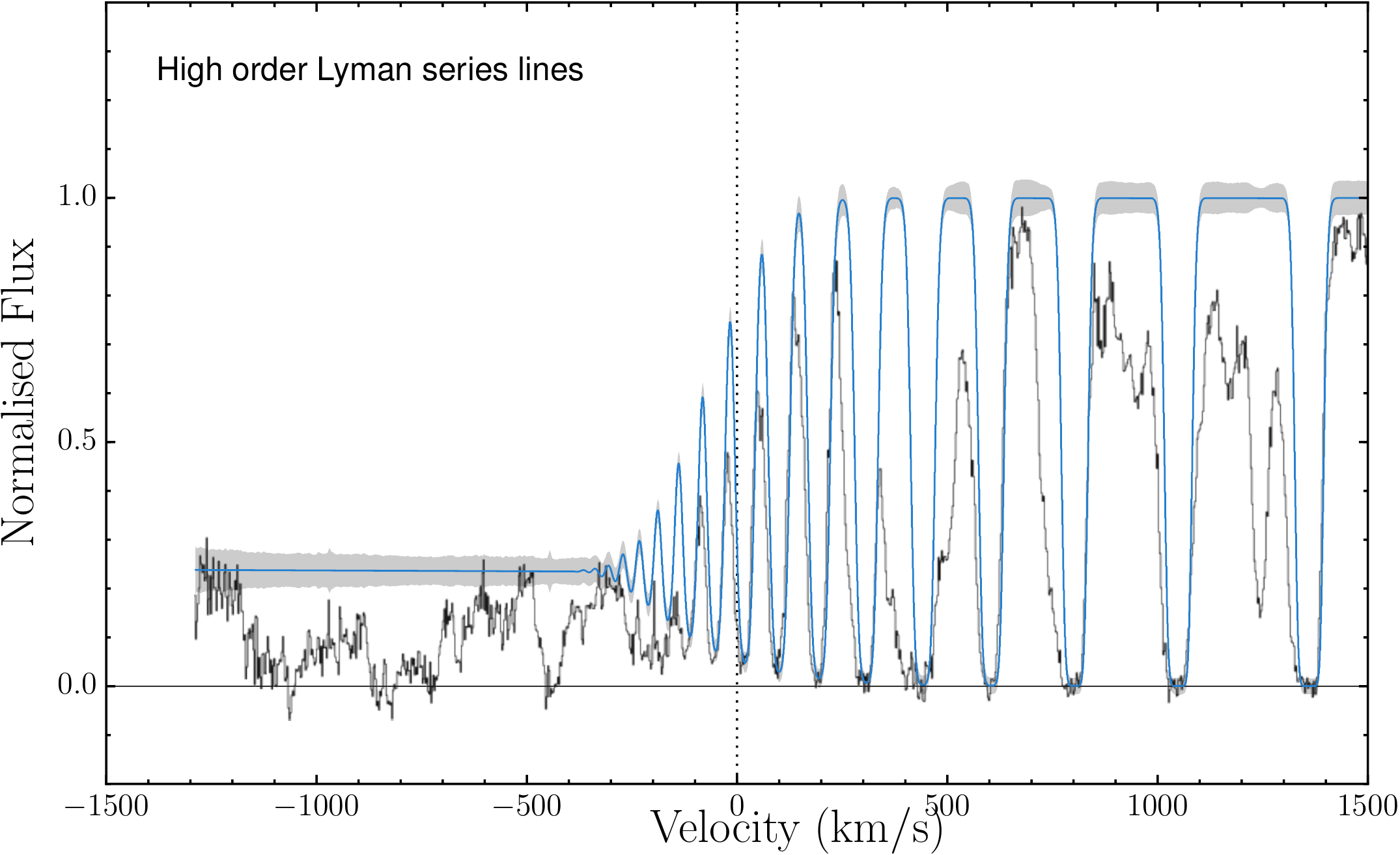}
\caption{Continuum normalised flux of the Lyman limit and other high order Lyman series lines used to constrain \NHI\ for LLS0958A in the Q0958$+$1202 HIRES spectrum (black histogram). The zero velocity redshift is set at $\zab=3.223194$ for Ly-19 (at 914.039\,\AA\ in the rest frame). The blue solid lines corresponds to \citet{2016ApJ...833..283L}'s fiducial model of LLS0958A with $\lNHI=17.36 \pm 0.05$, $b = 20.4$\,\kms\ at $\zab=3.223194$. The grey shading shows the 1$\sigma$ uncertainty in the flux.}
\label{f:lls0958_HI}
\end{figure}

Our fiducial \HI\ model is depicted in \Fref{f:lls0958_HI} which shows
the high-order Lyman series and limit for LLS0958A. We find that the
\NHI\ value derived by \citeauthor{2016ApJ...833..283L} is entirely
appropriate, even though we took a very different approach to
determining the continuum in the Lyman limit region of the
spectrum. To form their combined spectrum,
\citeauthor{2016ApJ...833..283L} first fitted continua to the spectra
from each echelle order in each exposure \citep[as
in][]{2015AJ....150..111O}. The continuum-normalised orders were then
averaged to form a normalised, combined spectrum. While this approach
will work well for orders containing only a small number of narrow
absorption features (e.g.\ redwards of the \lya\ emission line), it is
unlikely to work well for orders spanning the Lyman limit, like in
\Fref{f:lls0958_HI}. Our approach was the same as for the rest of this
paper: i.e.\ reduce the original exposures with \textsc{makee} and
combine them with \popler. We obtained all available exposures of
Q0958$+$1202, and appropriate calibration exposures, from the Keck
Observatory Archive
(KOA)\footnote{\urlstyle{rm}\url{https://www2.keck.hawaii.edu/koa/public/koa.php}}. \popler\ scales individual order spectra to optimally match each other before
averaging them and fitting a continuum to the final, combined
spectrum. The result is the Lyman limit spectrum shown in
\Fref{f:lls0958_HI} overlaid with the \HI\ model of
\citeauthor{2016ApJ...833..283L}: $\lNHI=17.36 \pm 0.05$ (1$\sigma$)
and $b=20.4$\,\kms. Clearly, this model adequately matches our
combined spectrum in this region, so we adopt this \NHI\ value without
alteration.

\begin{table}
\caption{Same as \Tref{t:logN} but for LLS0958A. For the fiducial model in the first column, the total column densities, from all fitted velocity components in \Fref{f:lls0958_metals}, are listed. Those for the main component at $v = 0$\,\kms\ are provided in the second column for the best comparison with the apparent optical depth values in the third column, as derived by \citet{2016ApJ...833..283L} using only the $-15$--$+30$\,\kms\ velocity range. The last row provides the metallicity measurements inferred from the corresponding photoionisation analyses.}
\label{t:logN0958}
\begin{center}
\begin{tabular}{l|ccc}\hline
 Ion             & \multicolumn{3}{c}{$\log_{10} (N/\mathrm{cm^{-2}})$} \\
 \hline
                 & Fiducial               & Main component        & \citeauthor{2016ApJ...833..283L} \\
 \hline
 \ion{Si}{ii}    & $ 11.40 \pm 0.10     $ & $ 11.09 \pm 0.18    $ & $ 11.26 \pm 0.10     $ \\
 \ion{Si}{iii}   & $ 13.10 \pm 0.12     $ & $ 12.77 \pm 0.54    $ & $ 12.85 \pm 0.01     $ \\
 \ion{Si}{iv}    & $ 13.16 \pm 0.01     $ & $ 12.65 \pm 0.15    $ & $ 12.87 \pm 0.01     $ \\
 \ion{C}{ii}     & $ \leq 12.60         $ & $ \leq 12.60        $ & $ \leq 12.07         $ \\
 \ion{C}{iv}     & $ 13.76 \pm 0.20     $ & $ 13.32 \pm 0.14    $ & $ 13.55 \pm 0.01     $ \\
 \ion{Al}{ii}    & $ \leq 11.40         $ & $ \leq 11.40        $ & $ \leq 11.11         $ \\
 \HI             & \multicolumn{3}{c}{$17.36 \pm 0.03$}                                    \\
 \hline
 [Si/H]          & $-2.91 \pm 0.26$       & $ -3.25 \pm 0.26$     & $-3.35 \pm 0.05      $ \\
\hline
\end{tabular}
\end{center}
\end{table}

\subsection{Photoionisation modelling results}\label{ss:lls0958_phot}

\begin{figure}
\includegraphics[width=0.95\columnwidth]{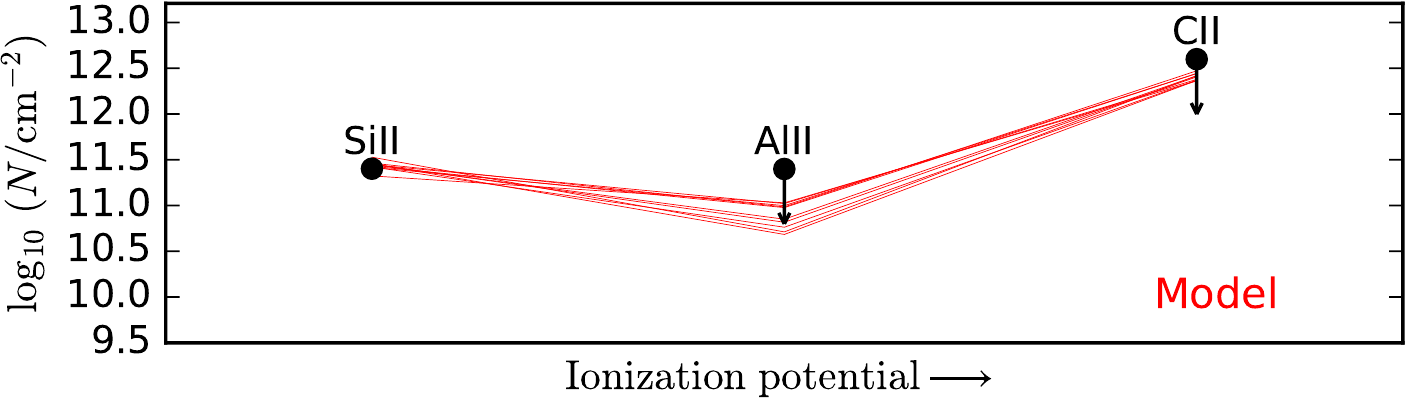}
\caption{\label{f:fiducial_model_lls0958} Same as \Fref{f:distribution_0344_fiducial} but for LLS0958A's fiducial model where only the singly ionised species listed in \Tref{t:logN0958} are associated with the \HI.}
\end{figure}

We use the same photoionisation modelling approach for LLS0958A as the
other systems in this work -- see Sections \ref{s:cloudy_general} and
\ref{s:interesting}. As discussed above, our fiducial model assumes
that the bulk of the \HI\ is associated with the low-ions only:
\ion{C}{ii}, \ion{Si}{ii}, \ion{Al}{ii}. As only \ion{Si}{ii} is
regarded as a detection, we assume a non-tilted HM12 UV background
(i.e.\ we set \auv $=0$) and solar values of [C/Si] and
[Al/Si]. \Fref{f:fiducial_model_lls0958} shows the comparison between
\textsc{cloudy}'s predictions for the metal column densities and the
values listed in the first column of \Tref{t:logN0958}. This simple
fiducial model is clearly consistent with the data. The corresponding
distributions of $Z/Z_\odot$, \nH, $U$ were derived by the MCMC
sampling algorithm as usual, providing a metallicity measurement of
$\lmetal = -2.91 \pm 0.26$ (95\% confidence). This is $0.44$\,dex
higher than \citeauthor{2016ApJ...833..283L}'s value of
$-3.35 \pm 0.05$.

\begin{figure}
\includegraphics[width=0.95\columnwidth]{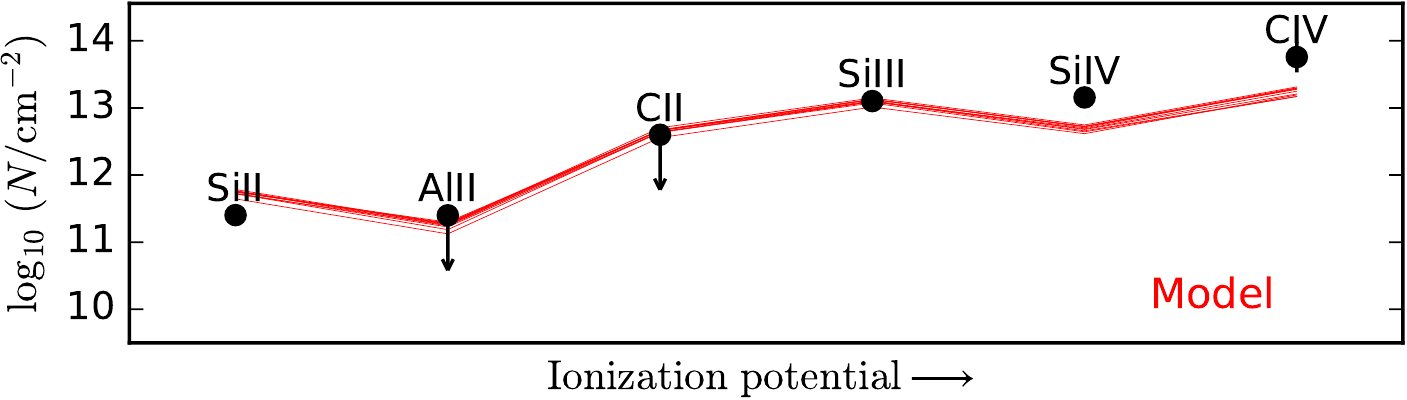}
\caption{\label{f:test_model_lls0958}Same as \Fref{f:fiducial_model_lls0958} but where all metallic species listed in \Tref{t:logN0958} are associated with the \HI.}
\end{figure}

\Fref{f:test_model_lls0958} compares the fiducial column density
measurements with a \textsc{Cloudy} model where all metal ions are
assumed to trace the observed \HI. In this case, there are clear
mismatches between the data and model: the model cannot simultaneously
fit the low and high ions (\ion{Si}{ii}/\textsc{iv} and
\ion{C}{ii}/\textsc{iv}). This remains the case even when allowing a
variable slope for the UV background and non-solar value for [C/Si]. In
both cases, the metallicity remains $\geq-3$:
$\lmetal = -2.89 \pm 0.26$ and $-2.60 \pm 0.26$, respectively. The
former, simpler case reflects most closely the assumptions made by
\citeauthor{2016ApJ...833..283L} in their photoionisation analysis. In
this respect, it appears that the assumption about which ions trace
the \HI\ -- low-ions only, or all ions -- makes little difference to
the metallicity estimate.

\begin{figure}
\includegraphics[width=0.95\columnwidth]{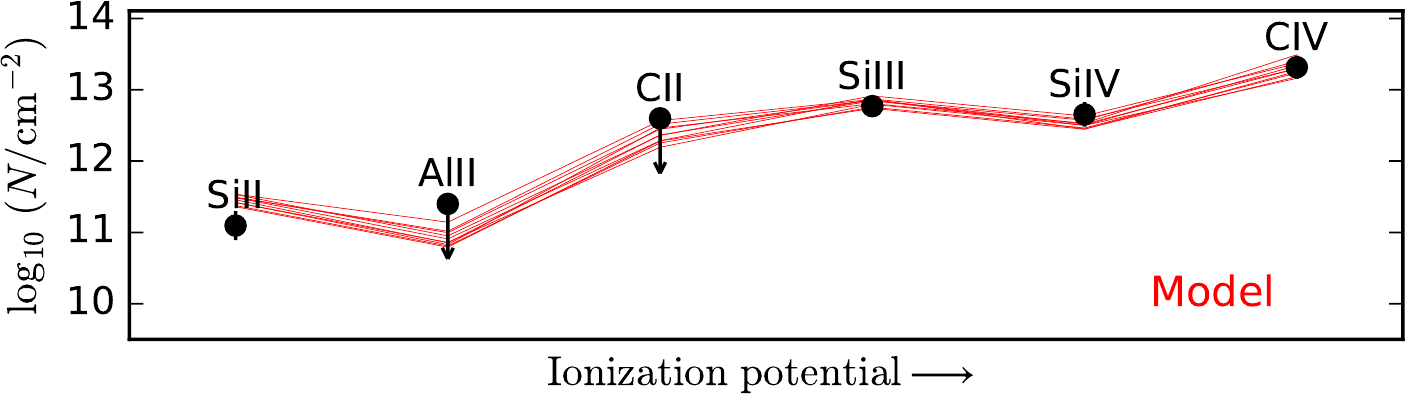}
\caption{\label{f:lehner_like_model_lls0958} Same as \Fref{f:fiducial_model_lls0958} but using only the main metal component's column densities (second column of \Tref{t:logN0958}) and the same photoionisation modelling assumptions as \citet{2016ApJ...833..283L}.}
\end{figure}

Instead, our metallicity estimate is 0.44\,dex higher than \citeauthor{2016ApJ...833..283L}'s predominantly because we include all the metal absorption detected, rather than that found within a small velocity range. To illustrate that this is not adequately explained by differences in our modelling approaches, we also analysed LLS0958A replicating \citeauthor{2016ApJ...833..283L}'s assumptions as closely as possible. We used only the metal column densities from the main, $v=0$\,\kms\ component in \Fref{f:lls0958_metals}. These compare closely with those derived using the apparent optical depth method, in a restricted velocity range around that component, by \citeauthor{2016ApJ...833..283L}, as can be seen in the second and third columns of \Tref{t:logN0958}. Again, we assume that all metal ions are associated with the \HI\ and we use the same UV background in our \textsc{Cloudy} model as \citeauthor{2016ApJ...833..283L}, i.e.\ HM05 -- a revised version of that originally published by \citet{1996ApJ...461...20H}. Finally, a non-solar [C/Si] value is allowed as a free parameter. \Fref{f:lehner_like_model_lls0958} compares this model with the metal column densities from the main component. There are some mismatches between the model and data; in particular, the model does not simultaneously reproduce the column densities of \ion{Si}{ii}, \textsc{iii} and \textsc{iv}. The distributions of $Z/Z_\odot$, \nH, $U$ and [C/Si] were derived by the MCMC sampling algorithm in the same way as the other absorbers in this paper, with the most likely metallicity being $\lmetal = -3.25 \pm 0.26$. This is only $0.1$\ dex higher than \citeauthor{2016ApJ...833..283L}'s value of $-3.35$, demonstrating that differences in photoionisation modelling are not responsible for the higher metallicity we find for LLS0958A.

%%%%%%%%%%%%%%%%%%%%%%%%%%%%%%%%%%%%%%%%%%%%%%%%%%

% Don't change these lines
\bsp	% typesetting comment
\label{lastpage}
\end{document}